# Detuning Choice for solving MIS and MWIS

Sem Saada Khelkhal
ssksem015@gmail.com

Louis Barcikowsky
louisbarcikowsky@gmail.com

## Abstract

In this article, we focus on the realization of a Maximum Weighted Independent Set (MWIS) as well as a quantum Maximum Independent Set, while respecting the constraints imposed by Pasqal's neutral atom processor: limited qubit number, bounds on $\mathbf{\Omega}$ and $\mathbf{\Delta}$, sequence durations, confinement space, minimum interatomic distances, and parasitic interactions.

Our goal was to obtain results that are directly usable with current technology, on asymmetric graphs whose size is only limited by the QPU's capabilities, within the bounds of the topological constraints.

To achieve these results, we introduce an innovative method for computing detuning that departs from traditionally accepted approaches. Indeed, for arbitrary graphs, parasitic interactions between nearby but unconnected atoms strongly bias the results, making the usual detuning bounds insufficient.

We thus propose three approaches that implement our detuning calculation, adapted to different levels of hardware maturity:
(I) a speculative approach where each atom is treated individually (local detuning) to demonstrate the application of pure theory;
(II) an approach using Detuning Map Modulation (DMM), which approximates our theory as closely as possible, with a view toward future implementation on the QPU;
(III) a method based on global pulses and frequency shifts, which deviates significantly from our theoretical framework but is directly applicable today, providing good results under stricter conditions.

Our strategies were tested against experimental realities using Pasqal's emulators on various graphs of up to 30 qubits.

Our approach places particular emphasis on respecting all current QPU constraints, in order to propose a solution that can be immediately transferred to practical applications.

# 1. Introduction

Combinatorial problems represent one of the most promising domains for demonstrating a practical quantum advantage. Among them, the Maximum Independent Set (MIS) and its weighted generalization, the Maximum Weighted Independent Set (MWIS), occupy a central place due to their numerous applications in optimization, complex networks, resource allocation, and scheduling. These problems, however, are notoriously difficult to solve classically, especially for large graphs. Classical methods typically rely on optimization procedures that yield results whose validity may be questionable as the problem size increases.

Analog quantum processors based on neutral atoms, such as those developed by Pasqal, provide a particularly suitable platform for tackling these problems via the Ising model, in which independent graph configurations can be natively encoded. Nevertheless, implementing this approach faces several physical constraints: a limited number of qubits, bounds on the control parameters $\mathbf{\Omega}$ and $\mathbf{\Delta}$, maximum sequence durations, atomic confinement space, minimum interatomic distances, and the presence of parasitic interactions between unconnected atoms.

These limitations make it necessary to develop new strategies that reconcile experimental feasibility with theoretical fidelity. Current detuning-tuning approaches often prove insufficient, particularly for arbitrary graphs where residual interactions distort the expected outcomes.

In this context, our work proposes an innovative method for computing detuning and explores three complementary approaches reflecting different levels of hardware maturity: a speculative approach based on local detuning, an intermediate approach relying on Detuning Map Modulation (DMM), and an operational approach applicable today using global pulses. These methods are tested against the current constraints of Pasqal's QPU and evaluated experimentally using its emulators, on asymmetric graphs of up to 30 qubits.

The objective of this paper is twofold: first, to demonstrate the validity of our approach on concrete cases, and second, to pave the way for practical quantum MIS/MWIS implementations on current and future analog processors.

The Figure 1 concretely illustrates the construction of the graph associated with a set of rubidium atoms confined in optical tweezers. Each green point $q_i$ represents an atom and corresponds to a vertex of the graph. The circle surrounding each atom denotes the Rydberg blockade radius $R_b$, defined by the relation

$$\frac{C_6}{R_b^6} = \Omega(t)$$

Two atoms are connected by an edge when their interatomic distance is smaller than $R_b$, reflecting the mutual exclusion of simultaneous Rydberg excitations. Thus, the image provides a direct representation of the graph $\mathcal{G} = (\mathcal{V}, \mathcal{E})$, where quantum exclusion links naturally arise from the physical properties of the atoms.

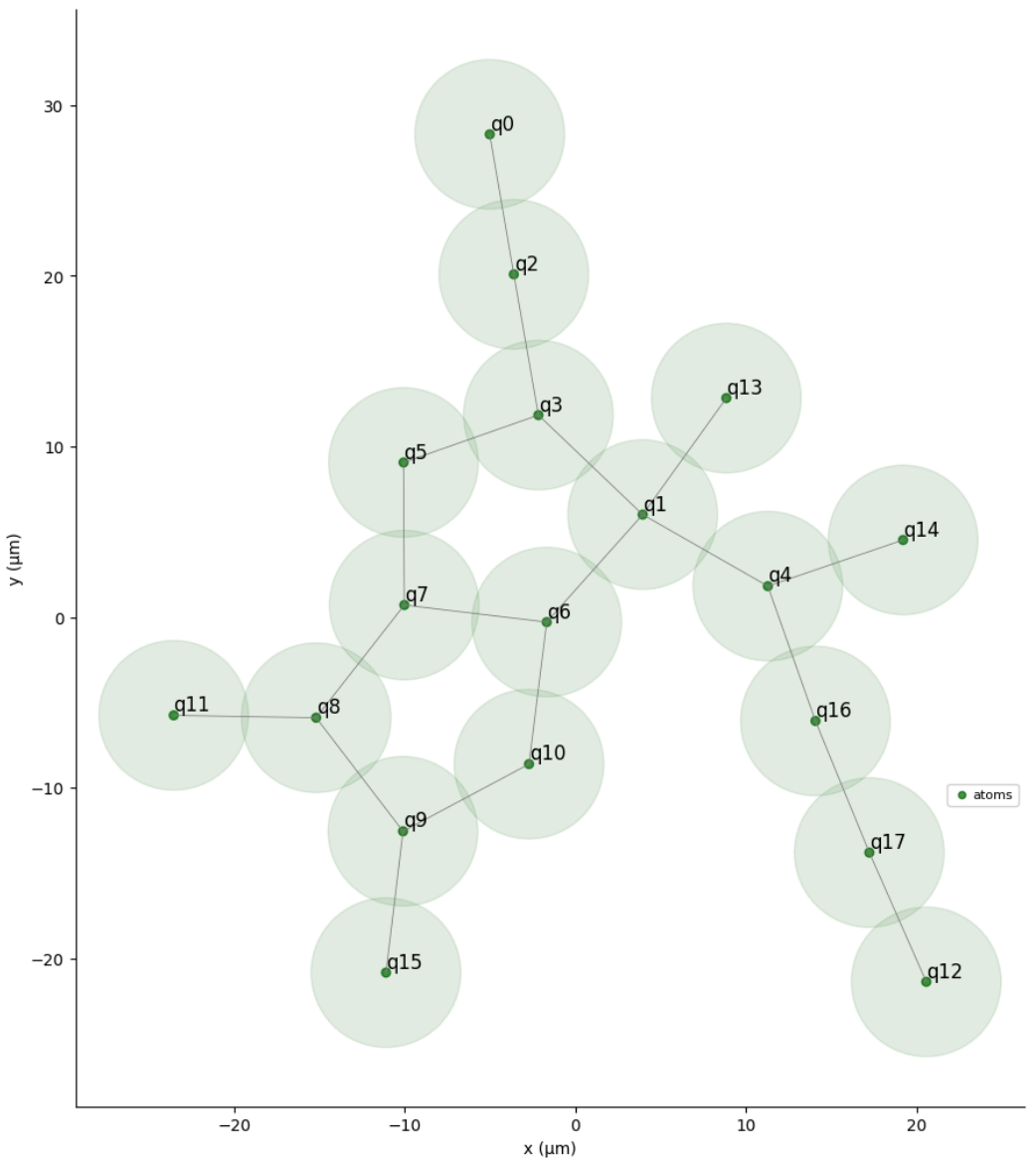

Figure (1): Example of register mapping of a graph showcasing how the blockad is used to determine connections

# 2. Context

An *Independent Set* (IS) in a graph $\mathcal{G} = (\mathcal{V}, \mathcal{E})$ is a subset of vertices $\tilde{\mathcal{V}} \in \mathcal{V}$, so that no pair of element from $\tilde{\mathcal{V}}$ is connected by an edge. Then, the *Maximum Independant Set* (MIS) is the largest subset of vertices $\tilde{\mathcal{V}} \in \mathcal{V}$. [1]

$$\mathbf{y} = \mathbf{x}^T Q \mathbf{x} = \sum_{i,j} q_{ij} x_i x_j$$

To find the MIS, we assign to each vertex $i$ a value $x_i$ that is 0 (excluded) or 1 (included) [2]. We then minimize the cost function which counts the included vertices via $-\sum_{i=1}^{\mathcal{V}} x_i$ and penalizes undesirable adjacencies with $U \sum_{(i \neq j) \in \mathcal{E}} x_i x_j$, where $U$ an arbitrary coefficient to penalize unfeasible solutions [2]. The MIS problem can be encoded with the following cost function:

$$-\sum_{i=1}^{\mathcal{V}} x_i + U \sum_{(i \neq j) \in \mathcal{V}} x_i x_j$$

Finding the MIS in a graph is often a complex problem, especially in large graphs, because it requires examining many possible combinations of vertices [2].

A weighted graph $\mathcal{G}_{\mathcal{W}} = (\mathcal{V}_{\mathcal{W}}, \mathcal{E})$ is a graph where each vertex $i$ is associated with a weight $w_i$ [2].

It follows that the *Maximum Weighted Independent Set* (MWIS) is the *Independent Set* (IS) of vertices $\tilde{\mathcal{V}}_{\mathcal{W}} \in IS(\mathcal{G}_{\mathcal{W}})$ which the sum of the weights is the maximum, i.e :

$$\max_{\tilde{\mathcal{V}}_{\mathcal{W}} \in IS(\mathcal{G}_{\mathcal{W}})} \sum_{i \in \tilde{\mathcal{V}}_{\mathcal{W}}} w_i$$

This makes the problem even more complex, as it's not just about finding the largest number of independent vertices, but also considering the associated weights to maximize the total sum [2].

In this study, we explore the neutral-atom quantum computer using rubidium atoms confined in optical tweezer arrays. Each atom is described by two electronic states. These atoms serve as qubits. A laser is used to illuminate them.

The volume of the electron cloud around the atom is determined by the Rabi frequency $\Omega(t)$ of the laser. The radius of the cloud in the Rydberg state is called the **Rydberg blockade radius** $R_b$. Considering a set of atoms $0, ..., \mathcal{V}$ and their positions $p_0, ..., p_{\mathcal{V}}$, we define a graph $\mathcal{G} = (\mathcal{V}, \mathcal{E})$ where each atom represents a vertex.

$C_6$ characterizes the strength of the van der Waals interaction between atoms. This force manifests if two atoms have an interatomic distance $\| p_i - p_j \|$ smaller than $R_b$. This interaction is defined by:

$$C_{-}\frac{6}{R}_{-} b^6 = (t)$$

“”

Since $C_6$ is constant, $R_b$ varies according to $\Omega(t)$. If two atoms are closer than $R_b$, they mutually block each other from occupying the Rydberg state. The associated vertices are then considered connected.

We can therefore define a function to quantify the Rydberg interaction energy by summing the interactions between pairs of atoms. We use $\hat{n}_{i/j} = |r\rangle\langle r|_{i/j}$ to project the state of atom $i/j$ onto $| \ r\rangle$, preserving this state if the atom is already in the Rydberg state, or setting it to zero otherwise:

$$\sum_{i \neq j}^{\mathcal{V}} \frac{C_6}{| \ p_i - p_j \ |^6} \hat{n}_i \hat{n}_j$$

The sixth power in the denominator highlights the similarity to $U \sum_{i \neq j}^{\mathcal{V}} z_i z_j$, as both functions heavily penalize nearby vertices.

However, the system's energy also depends on the laser-atom interaction, modeled by the sum:

$$\hbar \sum_{i=1}^{\mathcal{V}} \left( \frac{\Omega(t)}{2} \hat{\sigma}_i^x - \Delta(t) \hat{n}_i \right)$$

Here, $\hbar$ is the reduced Planck constant. $\Omega(t)$ determines the intensity of the laser-atom interaction, with $\hat{\sigma}_i^x$ being the Pauli-X operator representing the qubit state flip. $\Delta(t)$ represents the energy detuning, linked to an intentional offset of the laser frequency relative to the atomic transition, and $\hat{n}_i$ projects onto the excited state.

With negative $\Delta(t)$, minimizing the energy implies placing most atoms in the $|g\rangle$ state, as $\hat{n}_i$ cancels non-excited states. This corresponds to the initial state of a quantum computer where all atoms are in $|g\rangle$. With positive $\Delta(t)$, minimizing energy requires most atoms to be in the $|r\rangle$ state, as $\hat{n}_i$ preserves the excited states.

By following an adiabatic evolution with constant $\Omega(t)$ and $\Delta(t)$ varying from negative to positive, the system evolves from a low-energy initial state with all atoms in $| \, g\rangle$ to a low-energy final state with the majority of atoms in $| \, r\rangle$. The system thus seeks to maximize the number of excited atoms, which resembles what $-\sum_{i=1}^{\mathcal{V}} z_i$ does by selecting the maximum number of vertices in the graph.

The system Hamiltonian can therefore be represented as the sum of these interaction energies:

$$\hbar \sum_{i=1}^{\mathcal{V}} \left( \frac{\Omega(t)}{2} \hat{\sigma}_i^x - \Delta(t) \hat{n}_i \right) + \sum_{i \neq j}^{\mathcal{V}} \frac{C_6}{| \, p_i - p_j \, |^6} \hat{n}_i \hat{n}_j$$

In the case of adiabatic evolution, with constant $\Omega(t)$ and $\Delta(t)$ going from negative to positive, this closely resembles:

$$-\sum_{i=1}^{\mathcal{V}} z_i + U \sum_{i \neq j}^{\mathcal{V}} z_i z_j$$

Under this evolution, the neutral-atom quantum computer naturally solves an MIS problem on a graph where each atom is a vertex and a link exists between atoms separated by distance $R_b$.

It then remains only to correctly choose $\Omega(t)$ and $\Delta(t)$, respecting the physical limitations of current QPUs, so that the system yields this result.

# 3. Current Approach

The standard method for selecting the detuning parameter is based on imposing a lower bound that ensures unwanted long-range interactions do not affect the ground state. As formulated in recent work [3], [4], this condition is expressed as

$$\max_{\{(i,j) \notin E\}} V_{\{i,j\}} < \min_k \Delta_k$$

where $V_{i,j}$ denotes the interaction between atoms $i$ and $j$, and $\Delta_k$ is the boundary value of the detuning function $\delta_{k(t)}$. Typically, the detuning function is chosen to vary linearly in time, consistent with the adiabatic theorem [5], with boundary conditions $\delta_{i(0)} = -\Delta_i$ and $\delta_{i(t_f)} = \Delta_i$. The purpose of this schedule is to gradually increase the energy cost of remaining in the ground state, thereby favoring excitation to the Rydberg state.

An alternative but equivalent way to express this condition on a per-vertex basis is:

$$\Delta_i > \max_{(i,j) \notin E} V_{i,j}$$

which requires that the detuning applied to each atom exceed the strongest unwanted interaction involving that atom. Both formulations aim to prevent unconnected atoms from interfering with each other's activation by ensuring that the detuning dominates the corresponding residual couplings.

However, this strategy suffers from two key limitations. First, it assumes that it is sufficient to consider only the largest unconnected interaction. In practice, a single atom may be subject to several strong unconnected interactions. In such cases, the cumulative effect may suppress activation, even though the bound above is satisfied. Second, the method does not impose an upper bound on the detuning. If $\Delta_i$ is chosen too large, it may allow connected atoms to be simultaneously excited, thereby violating the independent set constraint.

To illustrate the first limitation, consider the embedding shown in Figure 2. None of the four atoms are connected by an edge, so all should in principle be eligible for activation. However, applying the current method to determine the detuning results in the central atom q0 being deactivated. This occurs because q0 experiences three unconnected interactions of similar strength. Although each is individually bounded by the condition above, their combined effect overwhelms the detuning and prevents q0 from being excited. Figure 3 shows the corresponding measurement statistics, where the dominant outcome excludes q0 despite the absence of graph edges.

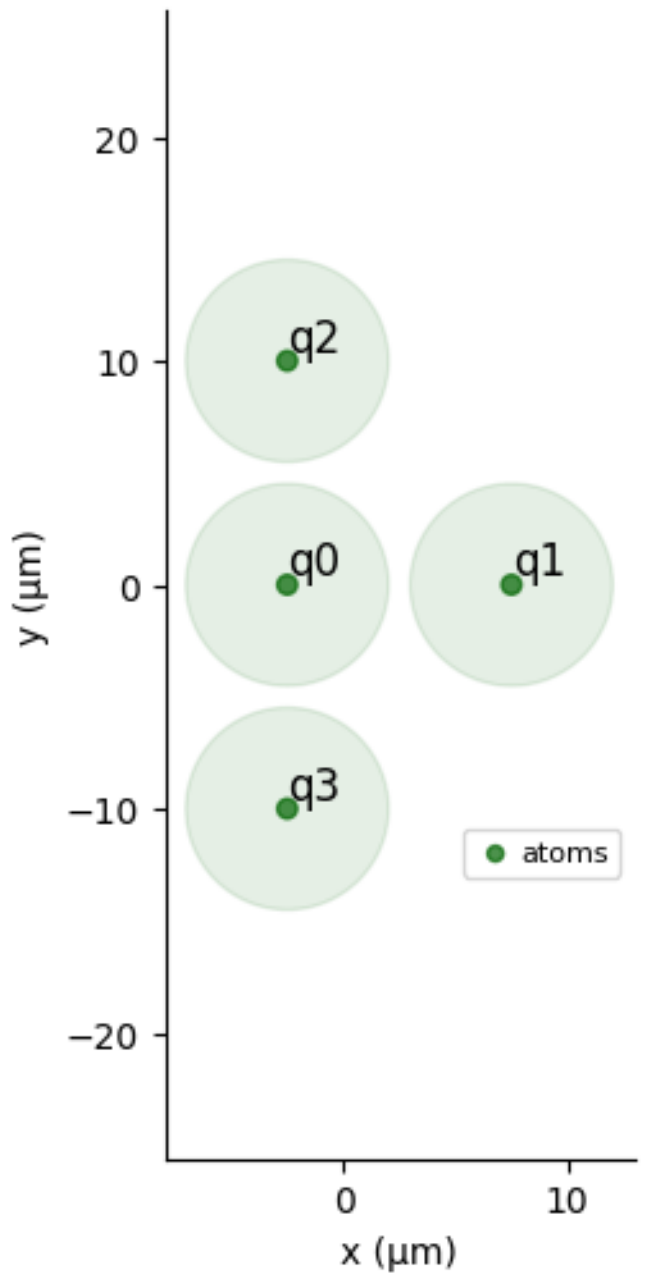

Figure (2): Atom layout used in the example. All atoms are unconnected.

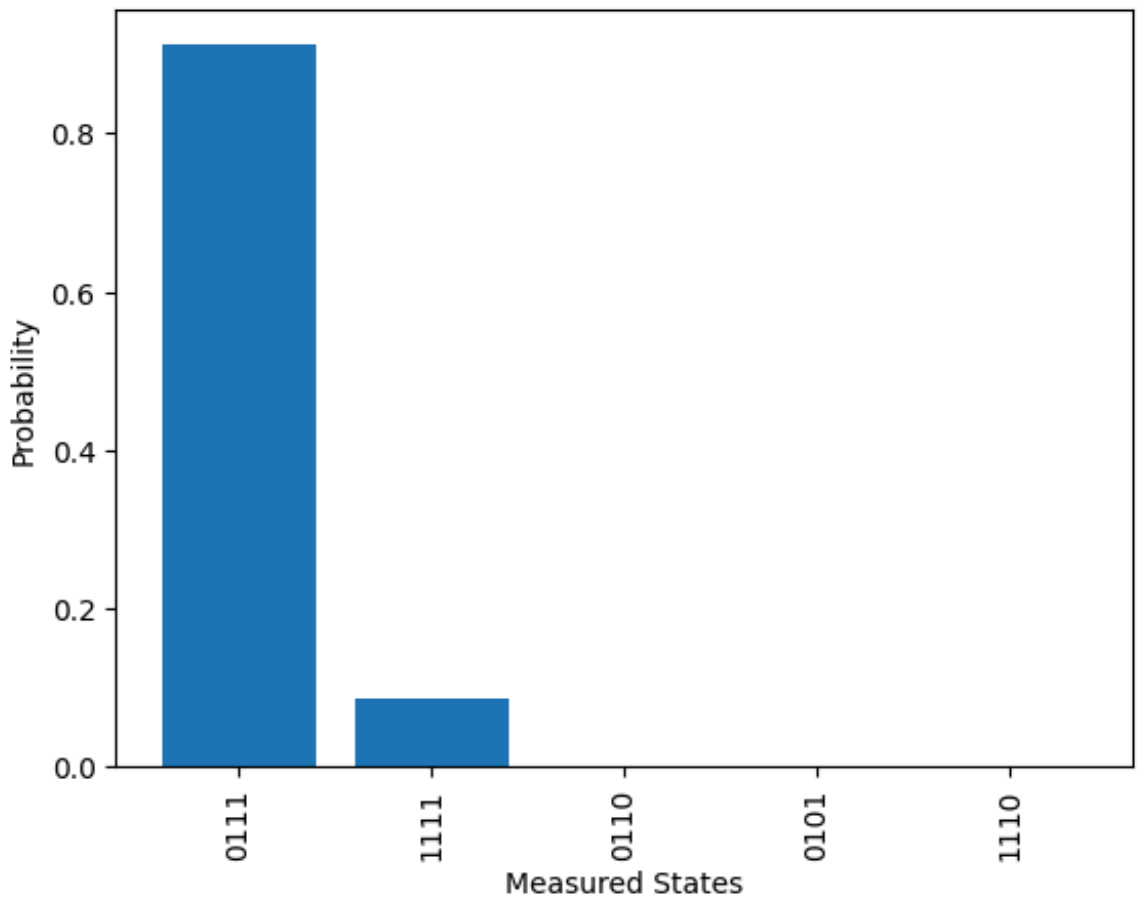

Figure (3): Measured probability distribution showing that q0 is suppressed despite being unconnected.

## 4. New Method for Determining the Detuning

In existing approaches, the detuning parameter is often chosen by identifying the maximum unconnected interaction and selecting a value slightly larger. However, as we have shown in the previous section, this strategy can lead to unwanted activations and deactivations of atoms. To address this issue, we propose an alternative method for computing the detuning, defined as

$$\Delta_i = \sum_{(i,j)\notin E} V_{i,j} + \min_{(i,j)\in E} V_{i,j} \cdot \tau$$

where $0 < \tau < 1 \in \mathbb{R}$

This formulation ensures that the detuning assigned to each atom exceeds the contributions from unconnected interactions while remaining below the weakest connected interaction. As a result, the method more reliably produces correct maximum independent sets (MIS) across a wide range of topologies.

Moreover, this choice naturally adapts to topologies in which unconnected interactions are negligible: in such cases, the lower bound on $\Delta_i$ approaches zero (a short proof may be provided here). Importantly, the computational complexity of this approach remains unchanged compared to the state of the art. Both the summation and the minimization can be evaluated in $O(n^2)$, and thus the overall asymptotic cost is unaffected [6].

To illustrate the effect of the new method, we apply it to the same four-atom layout considered in Figure 2. Unlike the previous approach, the central atom q0 is now correctly activated alongside the others, the probability distribution in Figure 4 confirms that all atoms can be simultaneously excited, consistent with the absence of edges in the graph.

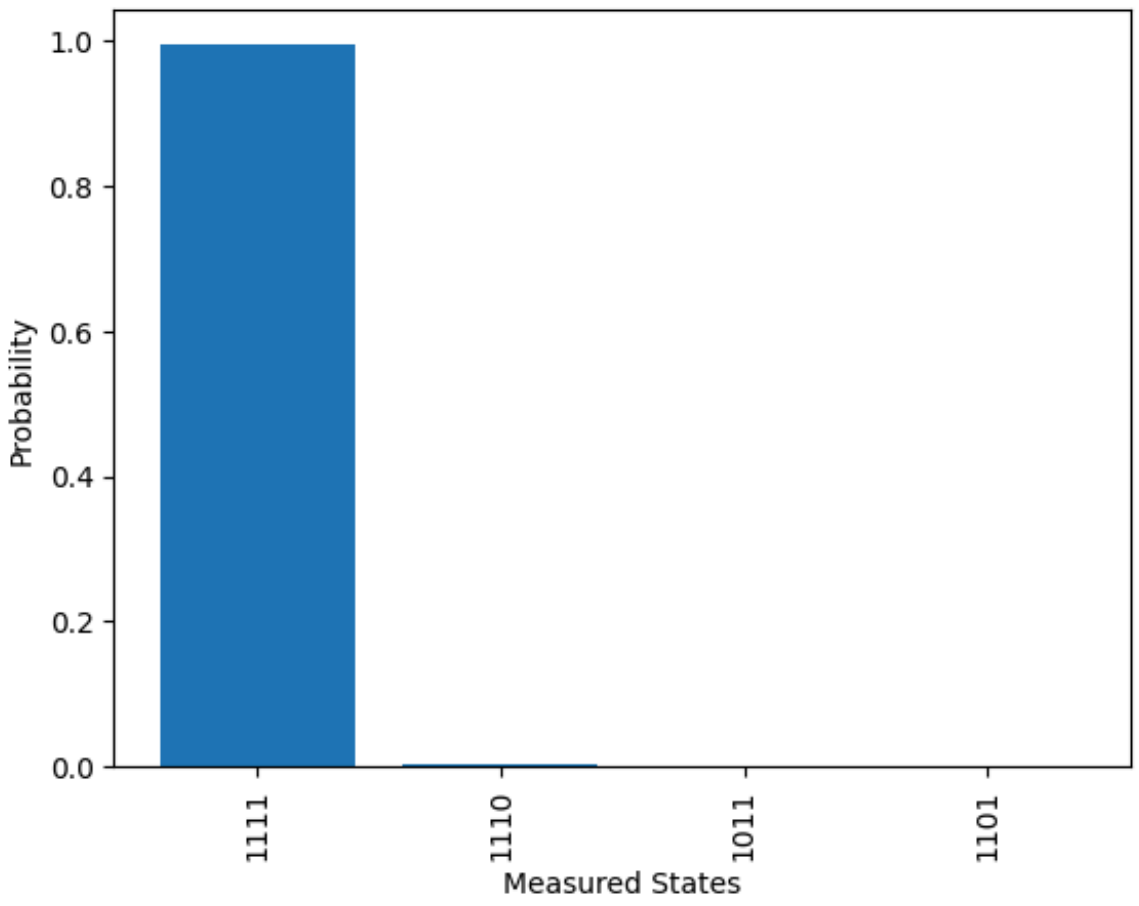

Figure (4): Measured probability distribution using the new method, showing correct activation of all atoms.

It should be noted that this formulation, in its current form, applies only to the MIS problem. We will show in the next section how it can be adapted to handle the maximum weight independent set (MWIS). Another caveat is that our method assumes the availability of local detuning within the quantum processing unit (QPU), which is not yet supported by current hardware. To address this, we propose two adaptation strategies: one compatible with current devices, and another anticipating forthcoming technological improvements.

# 5. Extension to the Maximum Weight Independent Set (MWIS)

To extend our method to the maximum weight independent set (MWIS) problem, we modify the detuning assignment as follows:

$$\Delta_i = \sum_{(i,j)\notin E} V_{ij} + \min_{(i,j)\in E} V_{ij} \cdot \tau_i$$

where the scaling factor $tau_i$ is defined by

$$\tau_i = L(w_i, 0.1, 0.9, w_{max}, w_{min})$$

and $L(w, min, max, w_{max}, w_{min})$ denotes the linear interpolation of $x$ within the interval $[0.1, 0.9]$

he purpose of introducing $\tau_i$ is to bias the detuning values in proportion to the vertex weights, thereby promoting the activation of nodes with larger weights. This scaling ensures that heavier nodes are preferentially stabilized, which naturally guides the system toward MWIS solutions.

We deliberately employ interpolation rather than normalization, since the boundary values 0 and 1 carry special meaning in this context. Specifically, they may be used to enforce activation or deactivation of certain nodes. This flexibility allows for the encoding of more complex behaviors —for example, partitioning the graph into cliques that can then be solved independently.

# 6. Pasqal QPU Constraints

In this study, we strictly adhere to the physical limitations of the neutral-atom quantum processors developed by Pasqal.

The atomic register is constrained to a two-dimensional arrangement, with a maximum of 80 atoms, each confined within a radius of 38 $\mu$m around the origin, and separated by at least 5 $\mu$m from their nearest neighbors. The device operates using Rydberg states with a principal quantum number of 60, characterized by an Ising interaction coefficient C equal to 865723.02.

The dynamical evolution is further limited by a maximum sequence duration of 6000 ns and a maximum of 500 experimental runs.

The laser control parameters are also bounded, with a detuning constrained to $\mid \Delta(t) \mid \leq 48.6947$ and a Rabi frequency amplitude limited to $\Omega(t) \leq 12.5664$.

These experimental constraints are essential to ensure that the proposed protocols remain feasible on the currently available QPU hardware.

# 7. DMM Formulation

An alternative approach to local addressing, currently under development by Pasqal, is the **Detuning Map Modulator** (DMM). This technique modifies the laser–atom interaction part of the Hamiltonian as follows:

$$H(t) = \sum_{i=1}^{V} \left[ \frac{\Omega(t)}{2} \hat{\sigma}_i^x - (\Delta(t) + \epsilon_i \Delta_{\text{DMM}}(t)) \hat{n}_i \right]$$

Here, each $\epsilon_i \in [0,1]$ denotes a local scaling factor, and $\Delta_{\text{DMM}}(t)$ is an additional global detuning function (negative-valued) modulated locally by $\epsilon_i$ [7]. The key idea is that all atoms follow a shared detuning schedule but are gradually biased toward their target theoretical values.

To set the parameters, we impose:

$$\epsilon_i = \frac{\Delta_i}{\Delta_{\text{max}}}$$
$$\Delta_{\text{DMM}}(0) = 0, \quad \Delta_{\text{DMM}}(t_f) = -\Delta_{\text{max}}$$
$$\Delta(0) = -\Delta_{\text{max}}, \quad \Delta(t_f) = \Delta_{\text{max}}.$$

This ensures that all atoms start with the same detuning, and as time progresses, their effective detunings converge smoothly to the ideal theoretical values at $t_f$.

For the Maximum Weight Independent Set (MWIS), we further introduce a weight-dependent bias:

$$\epsilon_i = \frac{\Delta_i}{\Delta_{\max}} \cdot L(w_i, 0.1, 0.9; w_{\max}, w_{\min}),$$

where $L$ denotes the linear interpolation described in Section [New Method]. This bias promotes the activation of heavily weighted nodes while preserving the correctness of the embedding.

To illustrate, Figure 5 shows the graph we are working with, its embedding remains the same as that of Figure 1, Figure 6 shows the time evolution of $\Omega(t)$, $\Delta(t)$, and $\Delta_{\text{DMM}}(t)$ under this formulation. Figure 7 presents the distribution of $\epsilon_i$ values mapped to the register layout, highlighting the spatial bias introduced by the DMM scheme. Figure 8 reports the effective detuning values obtained after applying the DMM correction, which explicitly shows the per-node detuning landscape. The corresponding MWIS instance, and the measured probability distribution are shown in Figure 5 and . Finally, Figure 10 depicts the top activation pattern obtained, highlighting the vertices included in the MWIS solution, giving us a total weight of 5.7713 which is also the solution found using a deterministic classical solver

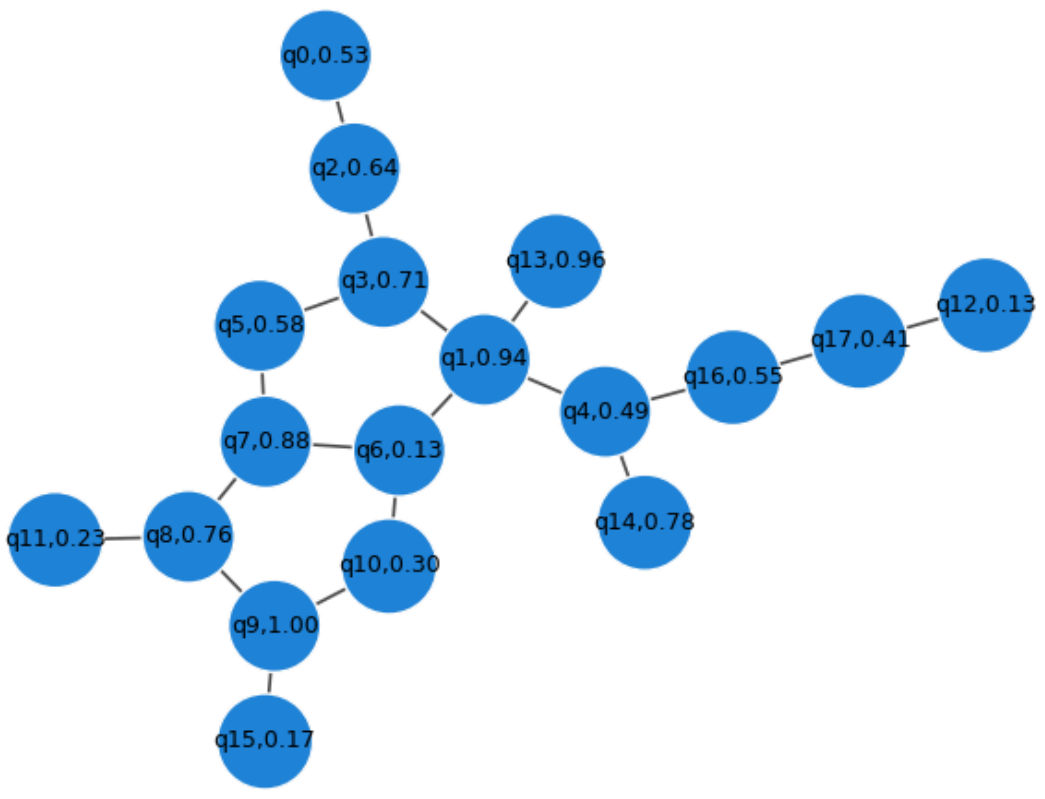

Figure (5): Graph instance and its embedding on the atomic array.

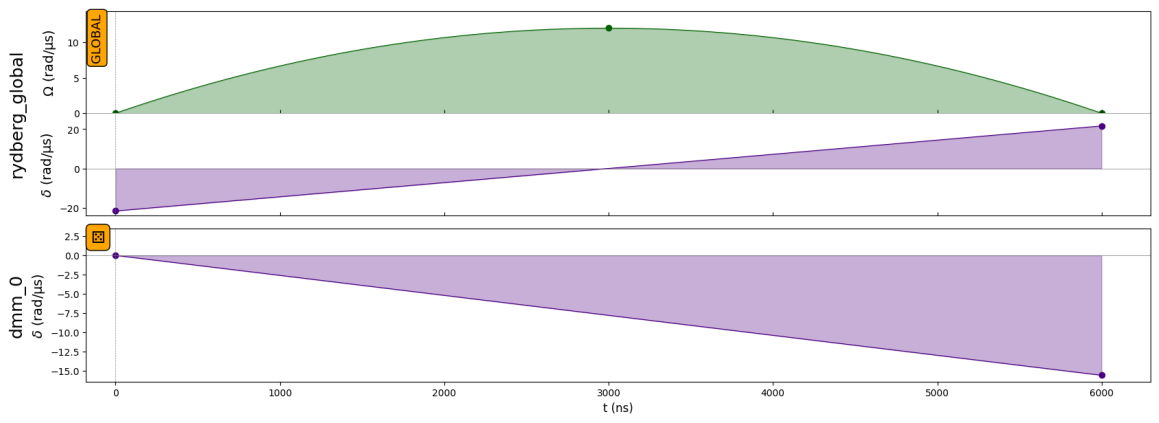

Figure (6): Time evolution of $\Omega(t)$, $\Delta(t)$, and $\Delta_{\text{DMM}}(t)$.

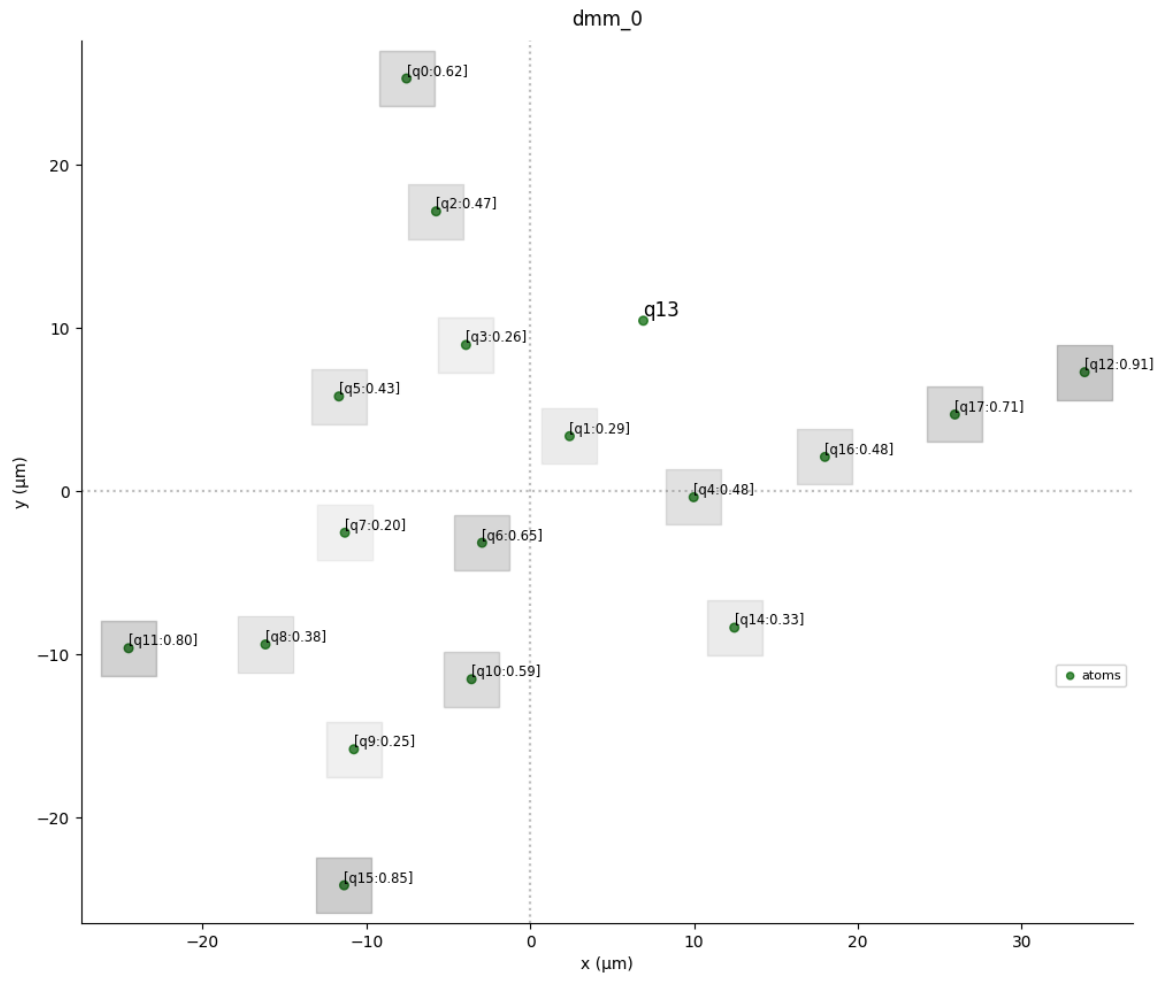

Figure (7): Distribution of $\epsilon_i$ values mapped to the register layout.

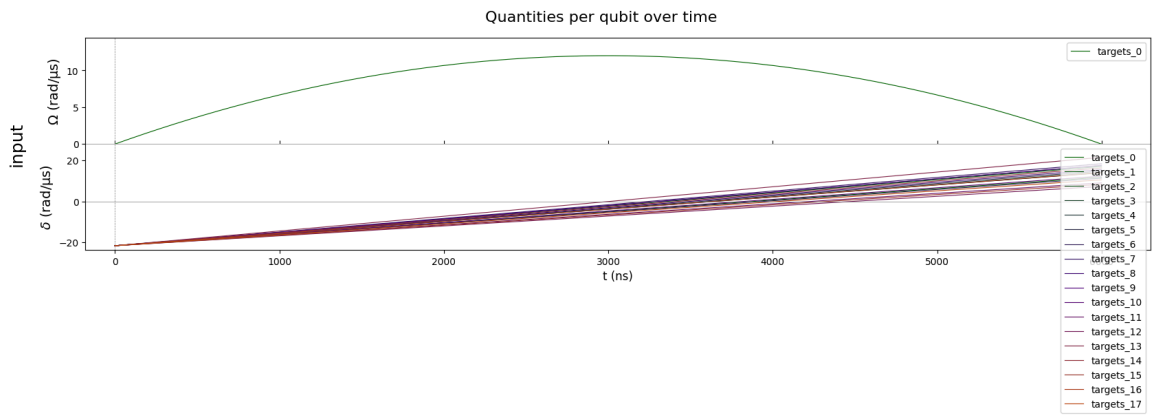

Figure (8): Effective detuning values per atom after combining global detuning and DMM contributions.

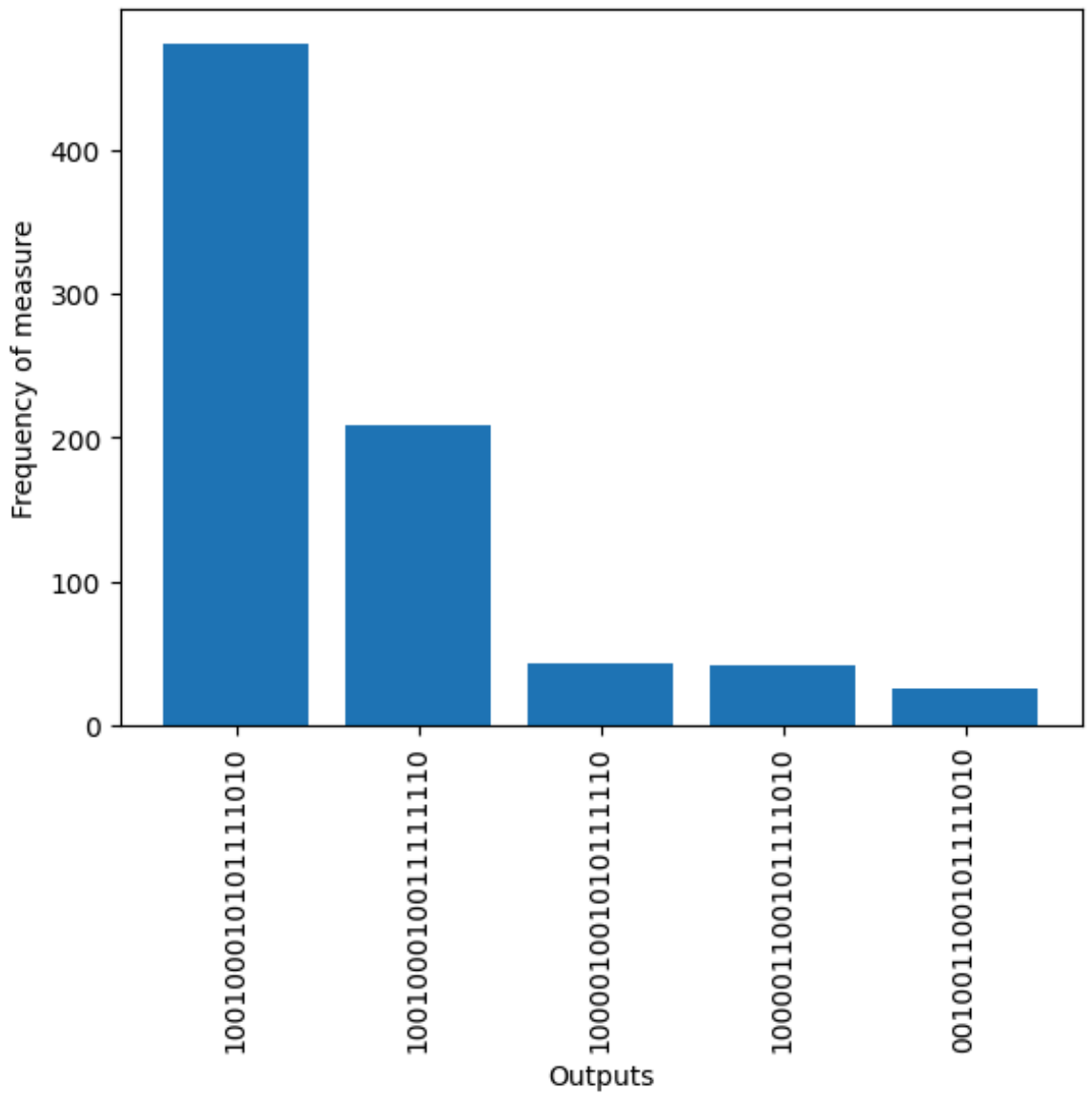

Figure (9): Measured probability distribution for the MWIS instance using the DMM scheme.

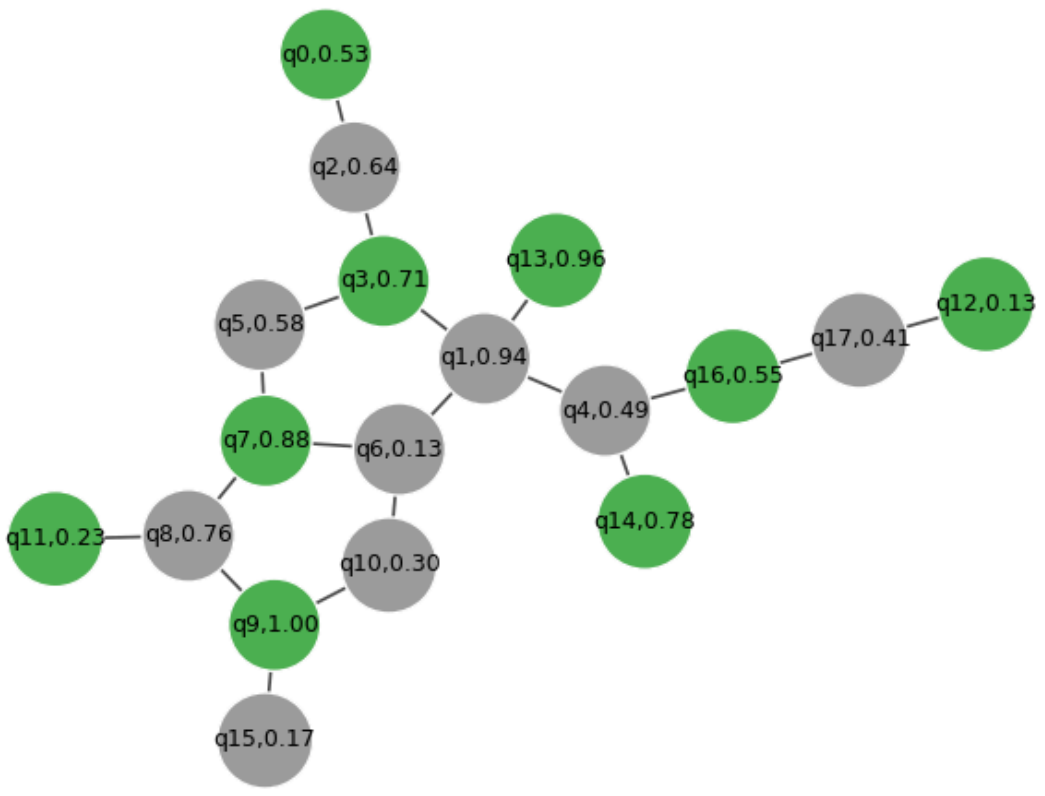

Figure (10): Top activation pattern represented on the graph, corresponding to the MWIS solution.

# 8. Global Detuning

in the case where having local addressability is impossible getting an MIS will be very topology dependent or impossible for the MWIS without breaking other constraints such as atom numbers or minimal distance now in order to achieve an MIS with global detuning the embedding of the graph must be constructed in a way such that :

$$\text{forall atom } \; i, j \left| \frac{\Delta_i - \Delta_j}{\Delta_{\text{max}}} \right| < \rho$$

here $\rho$ denotes the error level you are allowing on your results in our tests $\rho \leq 0.2$ has give good results but more that that has shown that the bias of interraction was too strong

# 9. Scalability

To demonstrate the scalability of our approach, we present the results obtained on a 32-node graph. The simulations were performed with a fixed duration of 6000 ns, a Rabi frequency set to $\Omega = 12$, and a detuning constrained to $\Delta < 48$.

Due to atomic confinement constraints, it was not possible to test graphs beyond 32 nodes. Indeed, extending to larger instances would have required exceeding the maximum allowed distances from the center of the atomic register.

The experiment was carried out using both the *local detuning* method and the *Detuning Map Modulation* (DMM) with very close output. In contrast, the *global detuning* approach could only be applied to less complex cases, and give us nothing interessting in this exemple.

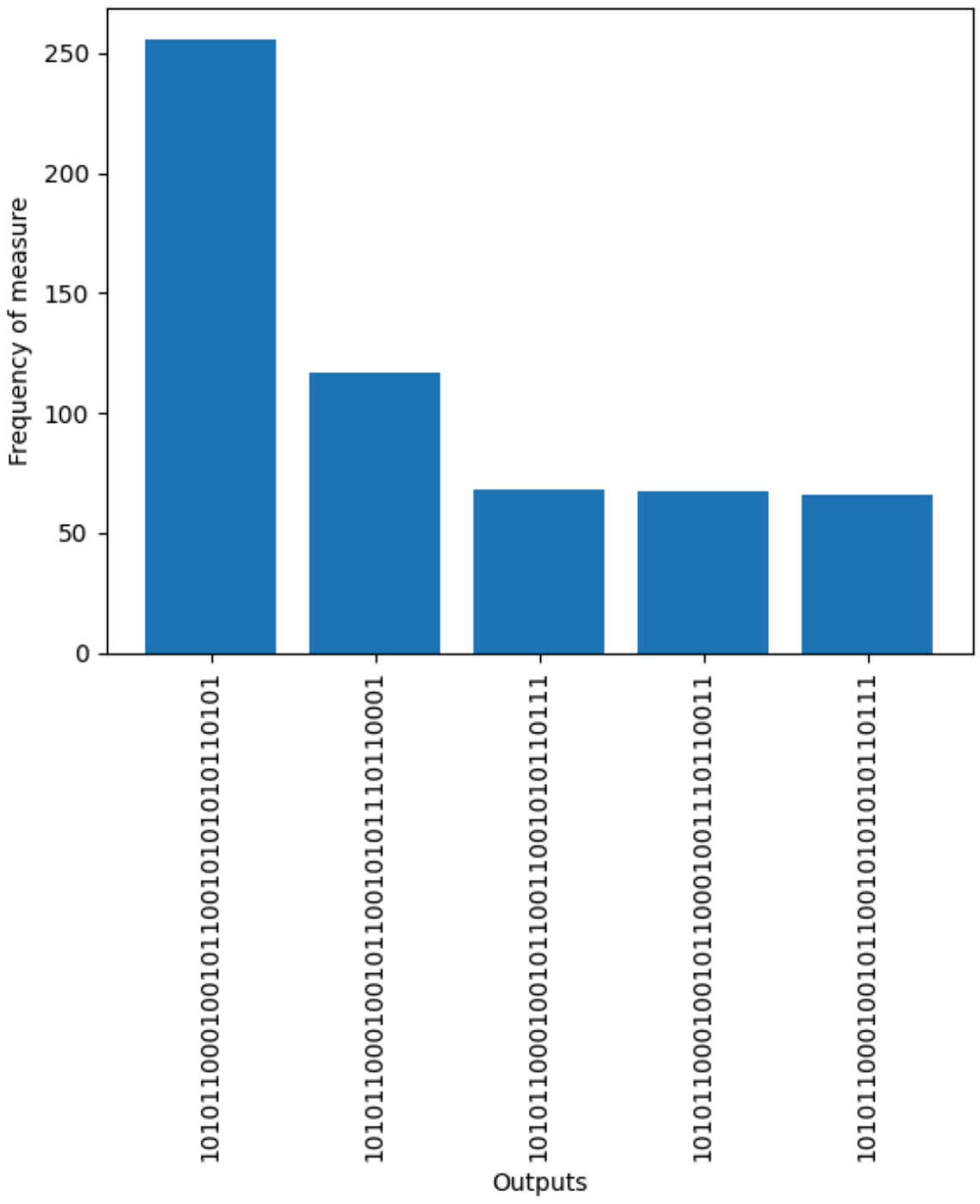

,

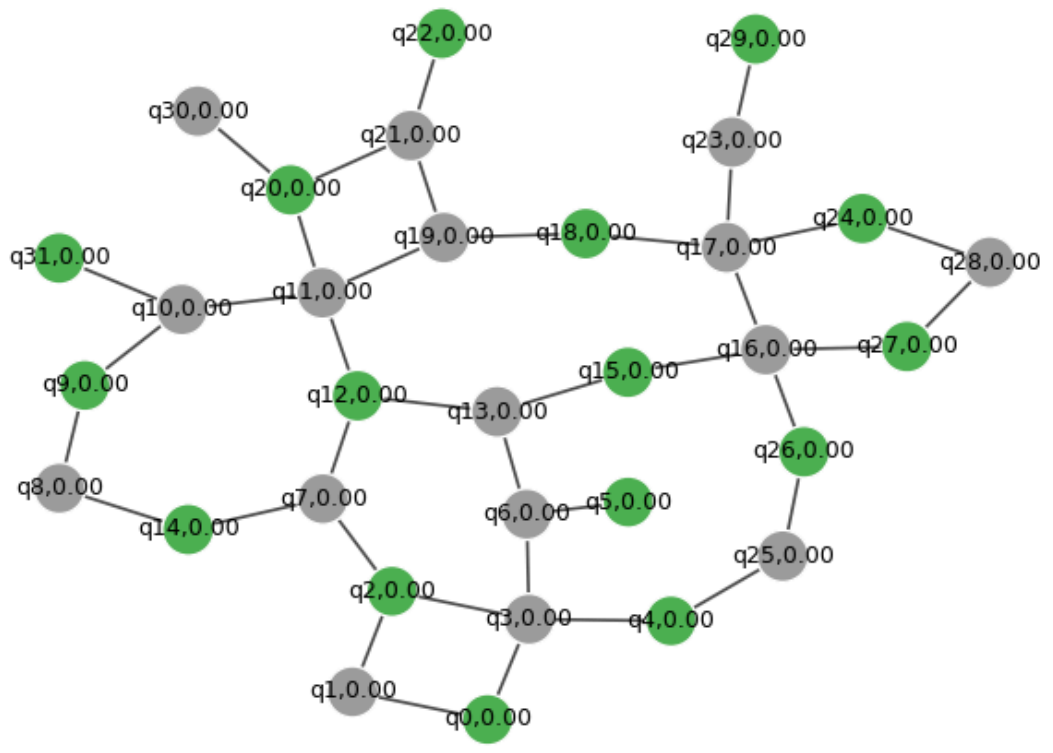

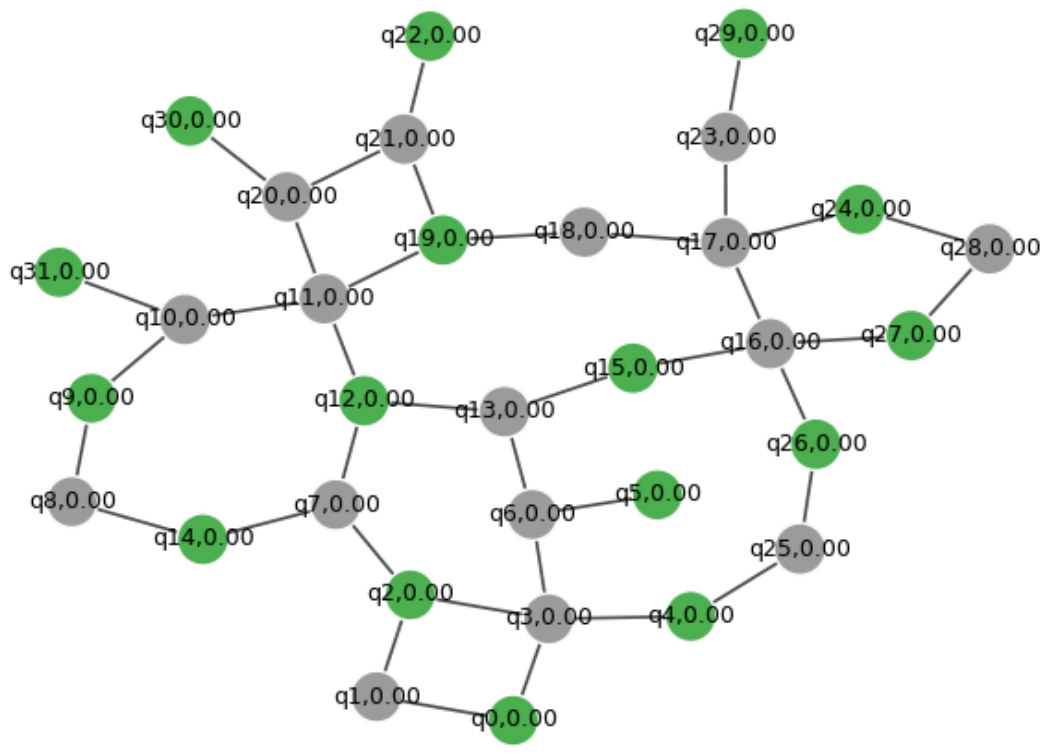

,

Figure (11): Top: Histogram of measurement outcomes, 32 nodes, local detuning. Middle: Graph representation of the most frequently measured configuration. Bottom: classical deterministic solution.

## 10. Conclusion

In this work, we have proposed a new method for computing detuning, enabling more reliable solutions to Maximum Independent Set (MIS) and Maximum Weighted Independent Set (MWIS) problems on neutral atom quantum processors. By accounting for parasitic interactions and the physical constraints of Pasqal's hardware, our approach strikes a balance between experimental feasibility and theoretical fidelity. We have shown that this method can be implemented in several variants adapted to different levels of technological maturity: a speculative local detuning, a Detuning Map Modulation (DMM) anticipating future capabilities, and a global strategy directly applicable on current devices.

Simulations performed on the emulator confirmed the relevance of our approach on asymmetric graphs of up to 30 qubits, while strictly respecting the instrumental limitations (number of qubits, minimum distances, maximum sequence durations, and bounds on $\Omega(t)$ and $\Delta(t)$). These results pave the way for practical use of analog QPUs in solving complex combinatorial problems.

The current limitation in sequence duration does not allow a thorough exploration of the system's evolution over longer timescales, restricting dynamic analysis. Moreover, the simulations performed are idealized as they do not account for the intrinsic noise of quantum processors and therefore do not fully reflect real experimental conditions. Implementation on physical hardware will be essential to confirm the robustness of our approach and to draw more definitive conclusions regarding the quality and reliability of the obtained results.

# 11. Acknowledgments

The authors would like to thank Reply for hosting this research project and for providing support throughout its development. We are also grateful to Pasqal for developing and maintaining the Pulser library used in our simulations, as well as for the technical support provided, and to OQI for their collaboration and partnership in this work. We additionally wish to thank Maëlle Toy-Riont–Le Dosseur and QUERELLA Laurent for their invaluable assistance and contributions to the research.

# 12. Appendix

## 12.1. Comparatif Example

# 13. Comparative Example

In this section, we compare the different methods for solving the Maximum Independent Set (MIS) on the same graph instance. All simulations are performed using the Pasqal simulator library **Pulser** with $\Omega = 12$ and a total evolution time $T = 6000ns$. For reference, we also include the classically computed MIS solution in Figure 12 on the same graph.

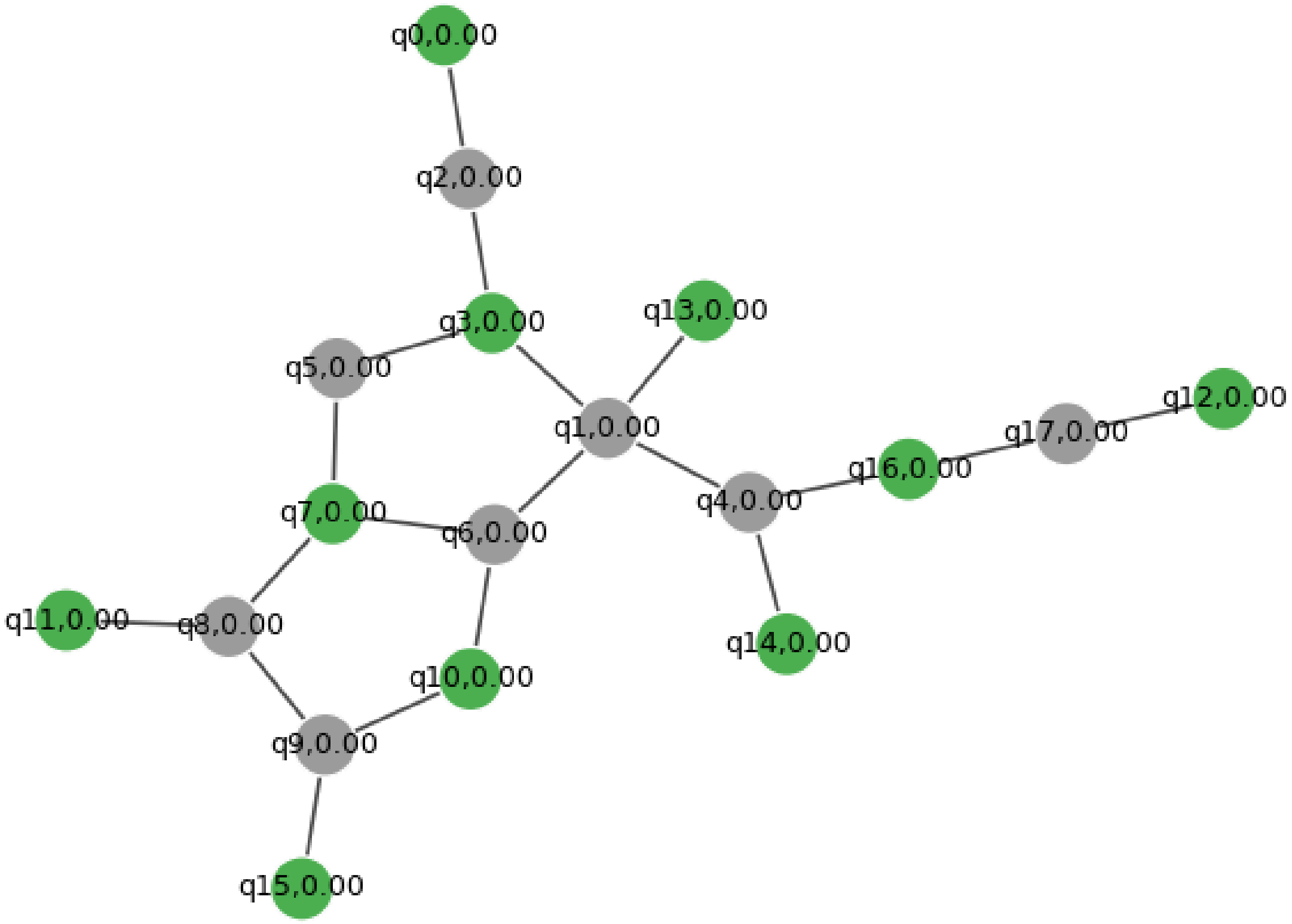

Figure (12): Classical MIS solution obtained using a deterministic solver.

### 13.0.1. Current Approach

The baseline approach, based on setting the detuning above the strongest unconnected interaction, yields suboptimal results. As illustrated in Figure 13, most frequently measured outcome does not correspond to the correct MIS, demonstrating the limitations of this strategy.

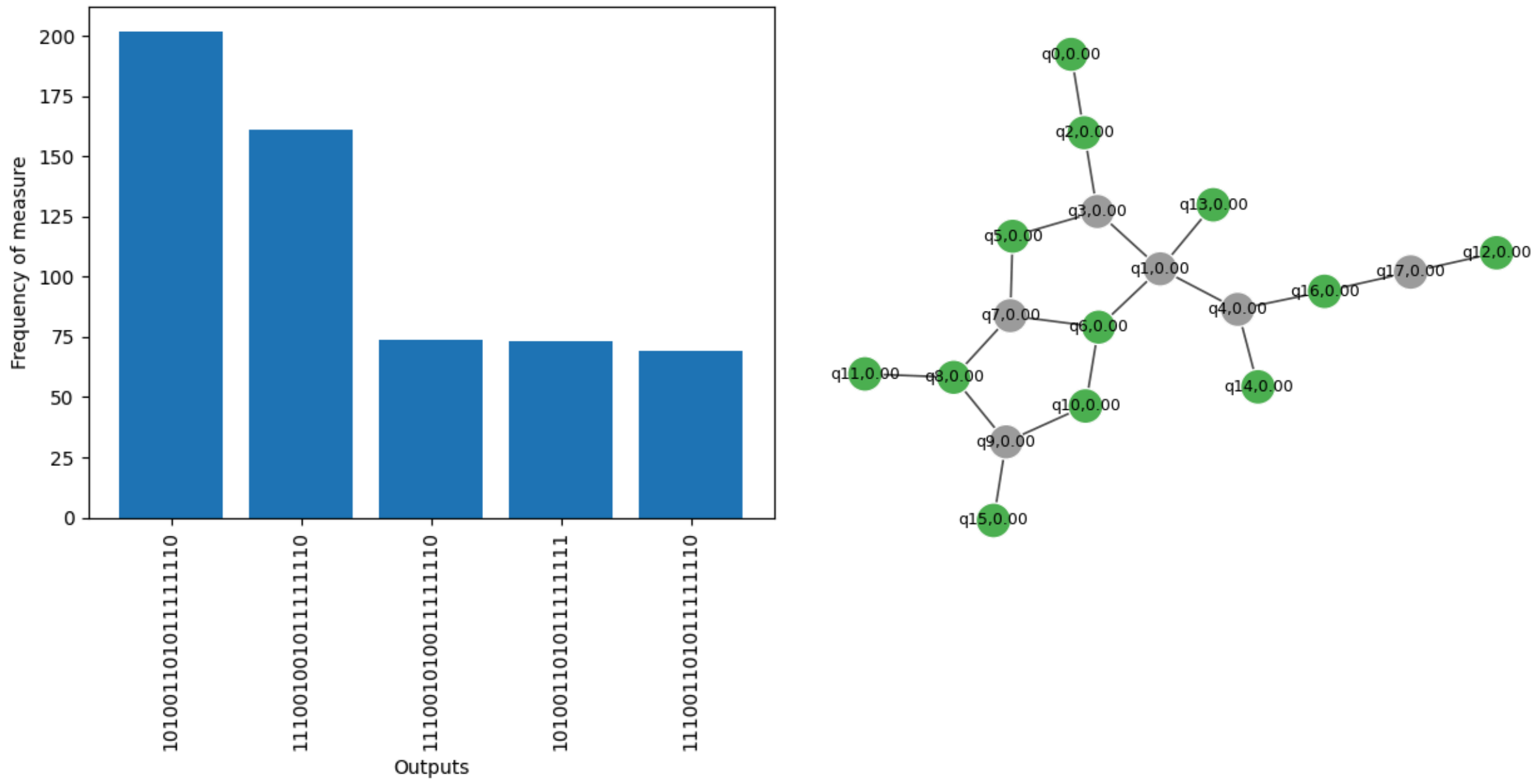

Figure (13): Left: Histogram of measured final states over 1000 repetitions. Right: Graph representation of the most frequently measured activation pattern, which fails to reproduce the correct MIS.

### 13.0.2. New Method with Local Detuning

Applying the proposed method with locally adapted detunings significantly improves the results. As shown in Figure 14, the correct MIS emerges as the dominant outcome, with over 600 occurrences out of 1000 measurements.

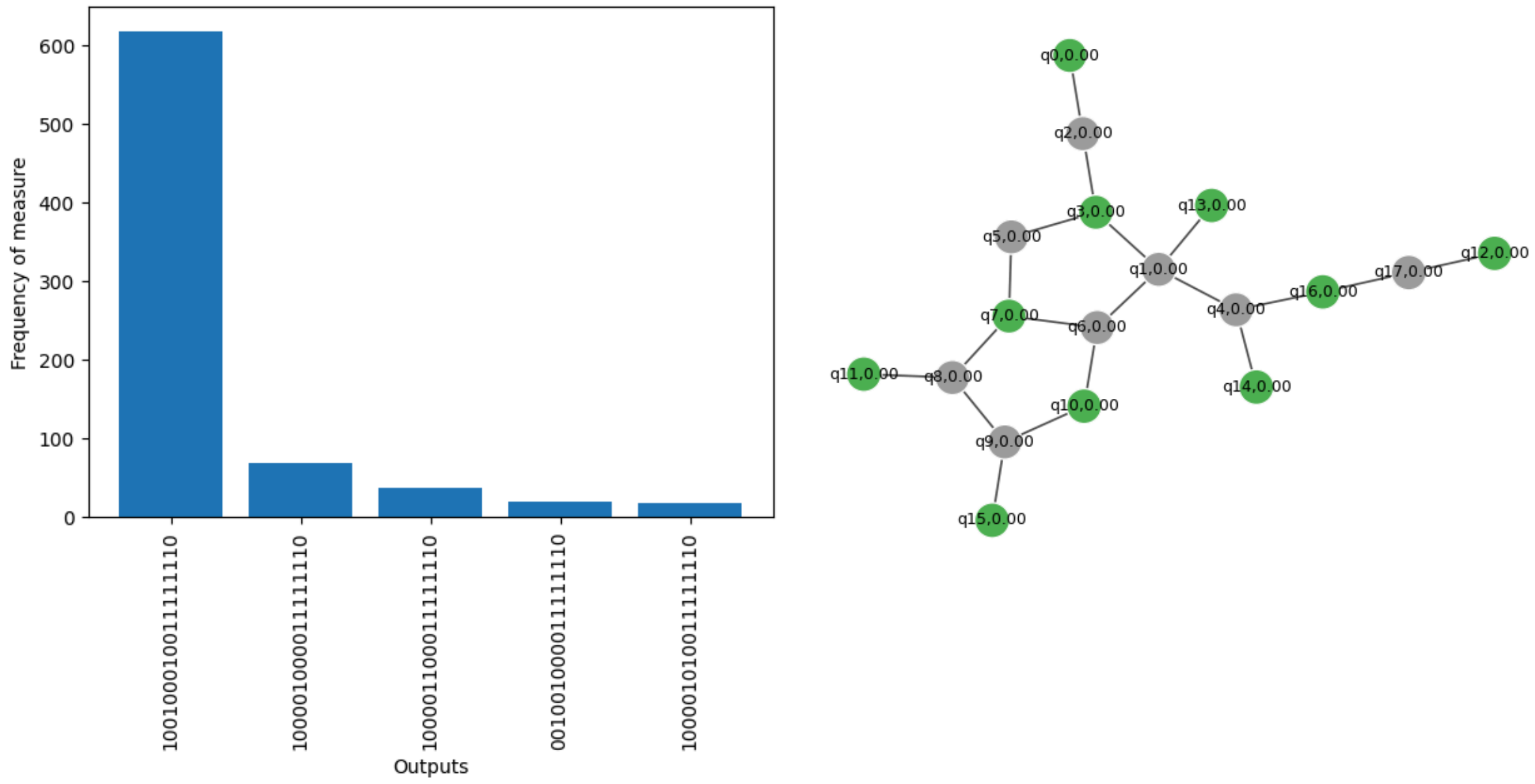

Figure (14): Left: Histogram of measurement outcomes under local detuning. Right: Graph representation of the most frequently measured configuration, corresponding to the correct MIS.

### 13.0.3. New Method with DMM

Using the Detuning Map Modulator (DMM) approach, we obtain results close to the ideal local detuning scheme. The correct MIS remains the most frequently measured outcome, appearing just under 600 times out of 1000, as shown in Figure 15. This confirms that the DMM provides a realistic approximation of the local detuning strategy under current hardware constraints.

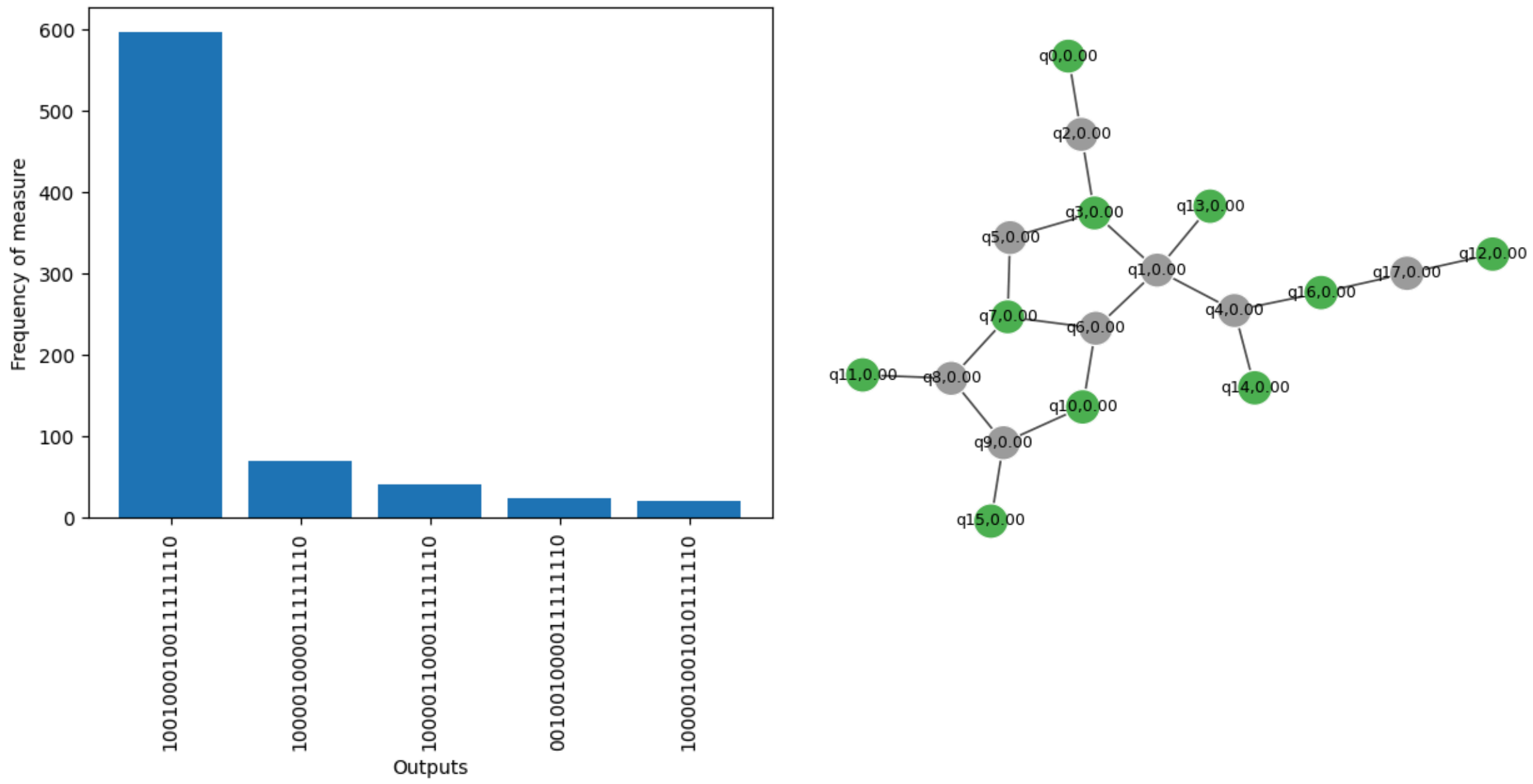

Figure (15): Left: Histogram of measurement outcomes under the DMM scheme. Right: Graph representation of the most frequently measured configuration, which reproduces the correct MIS, albeit with slightly reduced confidence compared to ideal local detuning.

#### 13.0.4. New Method with Global Detuning

Finally, we test the global detuning strategy by setting $\Delta = \text{mean}(\Delta_i)$. In this case, due to the relatively small variance among $\Delta_i$ values for the chosen graph topology, the method performs well: the correct MIS appears as the most measured outcome over 400 times (Figure 16).

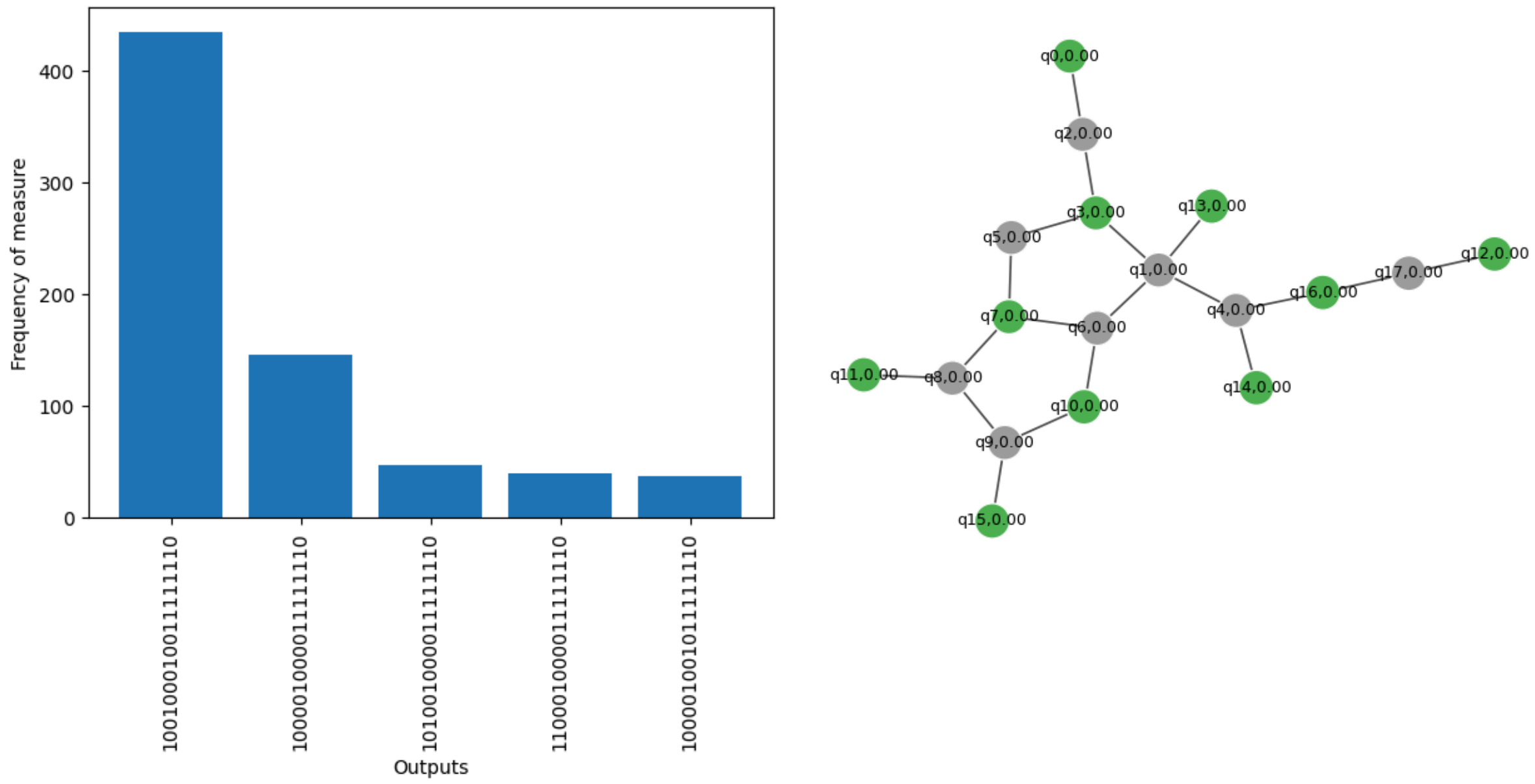

Figure (16): Left: Histogram of measurement outcomes under global detuning. Right: Graph representation of the most frequently measured configuration, corresponding to the correct MIS, albeit with lower probability than local detuning and DMM.

### 13.1. Code Example for local detuning

This is a standalone ready to run code for getting MIS

First make sure you have the necessary libraries which are `pulser`, `numpy`, `matplotlib`, `networkx`, `sklearn`

In [1]:
```python
import numpy as np
import networkx
import matplotlib.pyplot as plt
import pulser
import scipy as sp
```

Now for the graph we use an adjacency matrix of course you can adapt it to any format you want as pulser only needs the positions and the waveforms

In [2]:
```python
Q = np.array(
    [
        [0.0, 1.0, 1.0, 0.0, 0.0, 0.0],
        [1.0, 0.0, 1.0, 0.0, 0.0, 0.0],
        [1.0, 1.0, 0.0, 1.0, 0.0, 0.0],
        [0.0, 0.0, 1.0, 0.0, 1.0, 1.0],
        [0.0, 0.0, 0.0, 1.0, 0.0, 1.0],
        [0.0, 0.0, 0.0, 1.0, 1.0, 0.0],
    ]
)
Q
```

Out[2]:
```
array([[0., 1., 1., 0., 0., 0.],
       [1., 0., 1., 0., 0., 0.],
       [1., 1., 0., 1., 0., 0.],
       [0., 0., 1., 0., 1., 1.],
       [0., 0., 0., 1., 0., 1.],
       [0., 0., 0., 1., 1., 0.]])
```

This is a helper function to get nice graphical representations

In [48]:
```python
def draw(bitstring, positions: np.ndarray, ax=None):
    G = networkx.from_numpy_array(Q)
    pos = positions

    # Define a color map for states
    state_color_map = {
        "1": "#4CAF50",  # green
        "0": "#9E9E9E",  # gray
    }

    # Draw nodes grouped by activation state
    for k, color in state_color_map.items():
        nodes = {i: pos[i] for i in range(len(bitstring)) if bitstring[i] == k}

        # Scale node size based on label length (width grows with text length)
        node_sizes = [300 + 20 * len(f"q{i}") for i in nodes.keys()]

        networkx.draw_networkx_nodes(
            G,
            pos,
            nodelist=nodes.keys(),
            node_color=color,
            node_size=node_sizes,
            edgecolors="white",
            linewidths=1,
            ax=ax,
        )

    # Labels
    labels = {i: f"q{i}" for i in range(len(bitstring))}
    networkx.draw_networkx_labels(
        G, pos, labels=labels, font_size=9, font_color="black", ax=ax
    )

    # Draw edges
    networkx.draw_networkx_edges(
        G, pos, edge_color="black", width=1.2, alpha=0.7, ax=ax
    )

    plt.axis("off")
    plt.tight_layout(pad=2.0)
```

Now the important part which is the positions keep in mind that this depends on the value of omega as it defines the rydberg blockade and thus which atoms are connected

In [4]:
```
positions = np.array(
    [
        [-8.14402479, -9.56823364],
        [-1.74386237, -12.41500245],
        [-1.87085513, -4.08263706],
        [1.52776929, 3.55825435],
        [7.80092456, 9.04386765],
        [1.40075438, 11.89061848],
    ]
)
```

In [5]:
```
draw("".join(["0" for _ in Q]), positions)
```

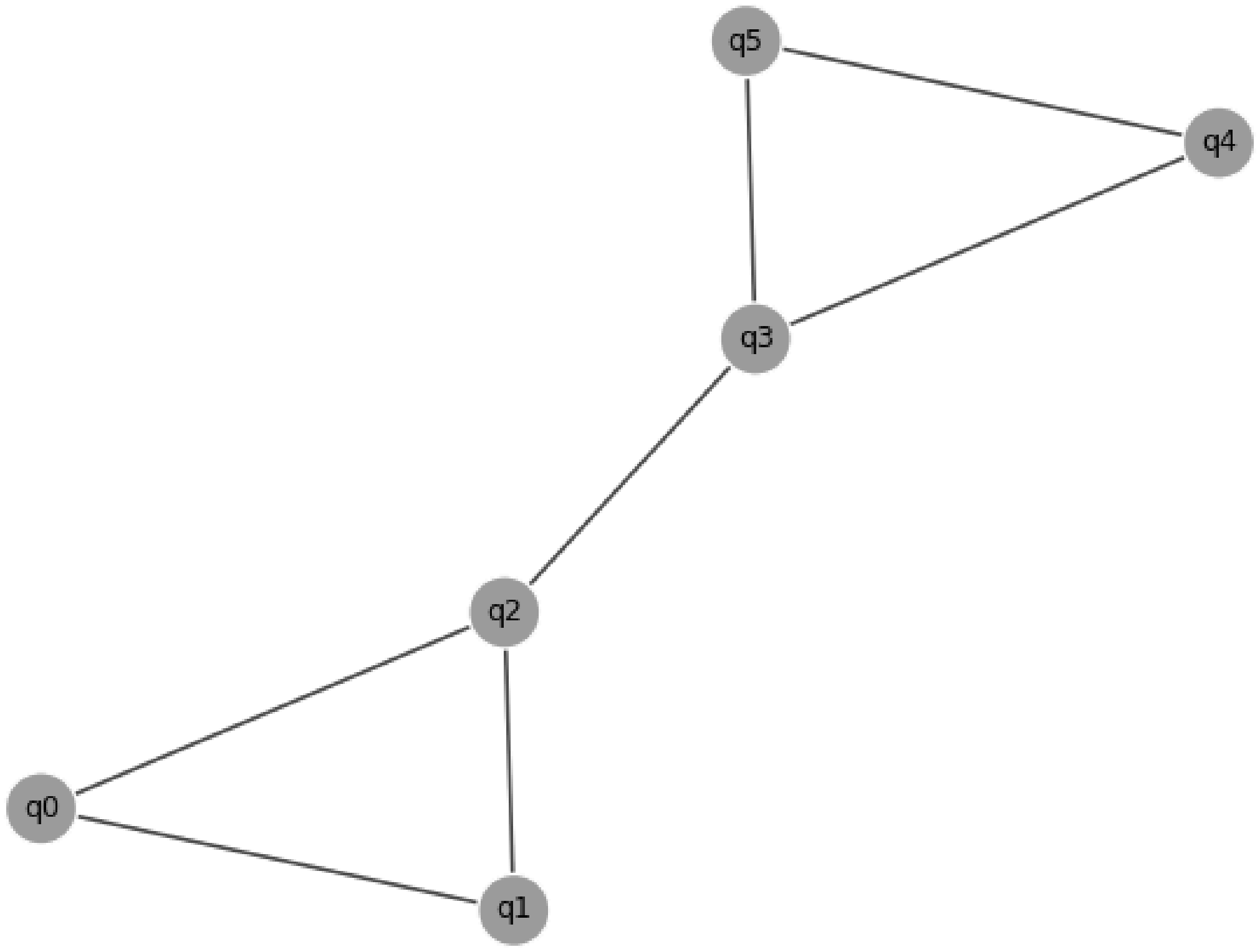

This part is the Pulser initialization you can play arround here with values to suit your use case

In [6]:
```
device = pulser.MockDevice
```

In [7]:
```
omega_max = 12.0
duration = 6000
```

In [8]:
```
register = pulser.Register.from_coordinates(positions, prefix="q")
register.draw(
    draw_half_radius=True,
```

```
blockade_radius=device.rydberg_blockade_radius(omega_max)
)
```

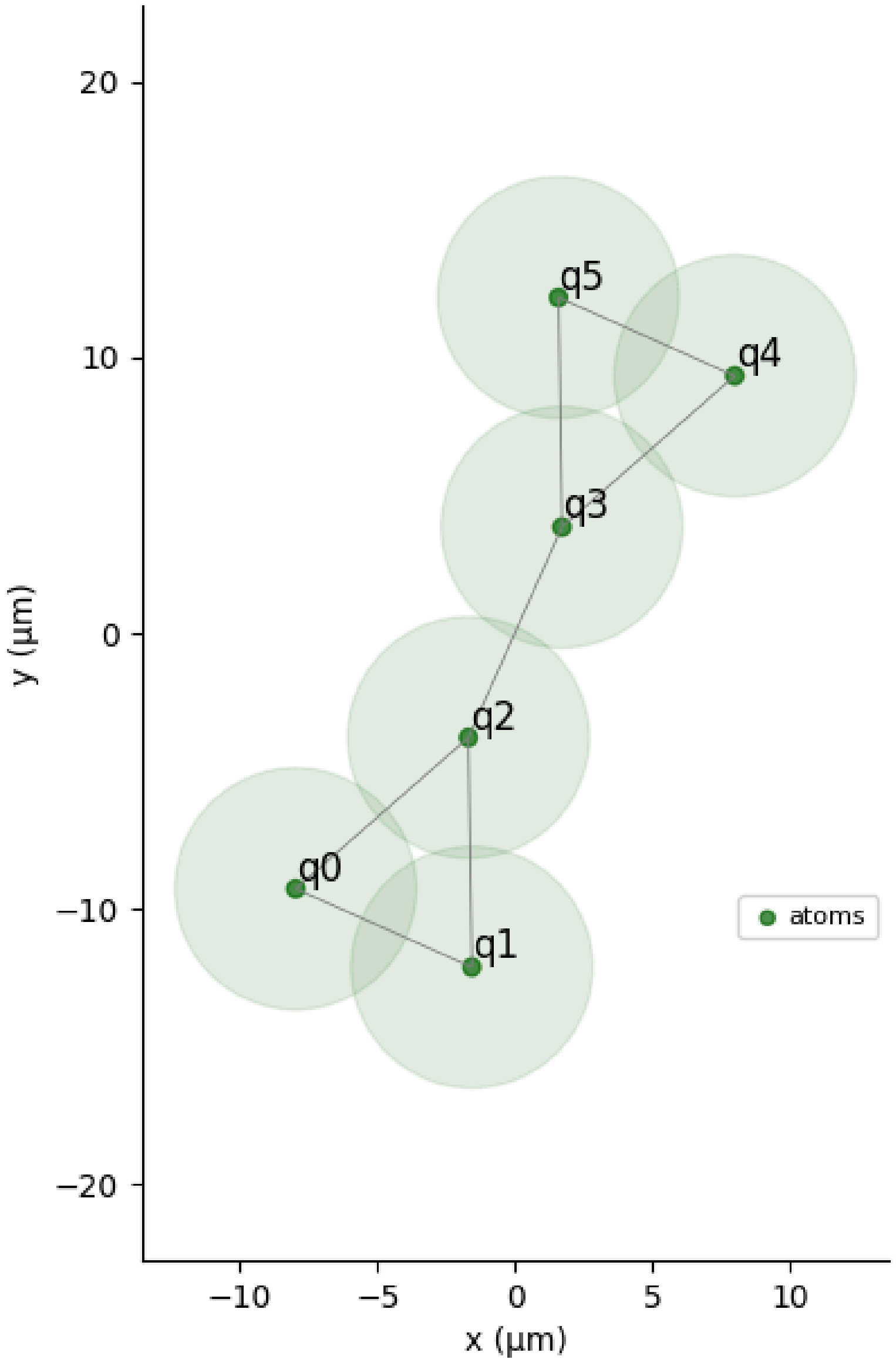

In [9]:
```
def filter_zeros(x):
    if x == 0:
        return float("inf")
    else:
        return x


vect_fill_zeros = np.vectorize(filter_zeros)
```

This is the method to get the $\delta s$ it corresponds to the equations shown earlier

In [10]:
```
connected_mask = np.array(Q, dtype=bool)
dists = sp.spatial.distance.squareform(sp.spatial.distance.pdist(positions))
np.fill_diagonal(dists, float("inf"))
```

```
C6 = device.interaction_coeff
V = C6 / (dists**6)
total_interaction = V.sum(axis=1)
unconnected_interaction = (V * (~connected_mask)).sum(axis=1)
connected_interaction = vect_fill_zeros(V * connected_mask).min(axis=1)
epsilon = (V * connected_mask).mean(axis=1)
detunings = (
    unconnected_interaction + connected_interaction * 0.9
)  # for the mwis modify the 0.9 to correspond to the weights instead remember to
interpolate the weights between 0.1 and 0.9 as 0 and 1 should be reserved for
forcing activations/deactivations
epsilons = detunings / (detunings.max())
detuning = detunings.max()
```

Now we need to create the sequence which is the "source code" of our process here we use the theoretical method with real local detunings

In [11]:

```
channel = "rydberg_local"
sequence = pulser.Sequence(register, device)
for i in range(len(positions)):
    detuning = detunings[i]
    pulse = pulser.Pulse(
        pulser.InterpolatedWaveform(6000, [0, omega_max, 0]),
        pulser.InterpolatedWaveform(6000, [-detuning, detuning]),
        0,
    )
    sequence.declare_channel(f"q{i}", channel, initial_target=f"q{i}")
    sequence.add(pulse, f"q{i}")
sequence.draw()
```

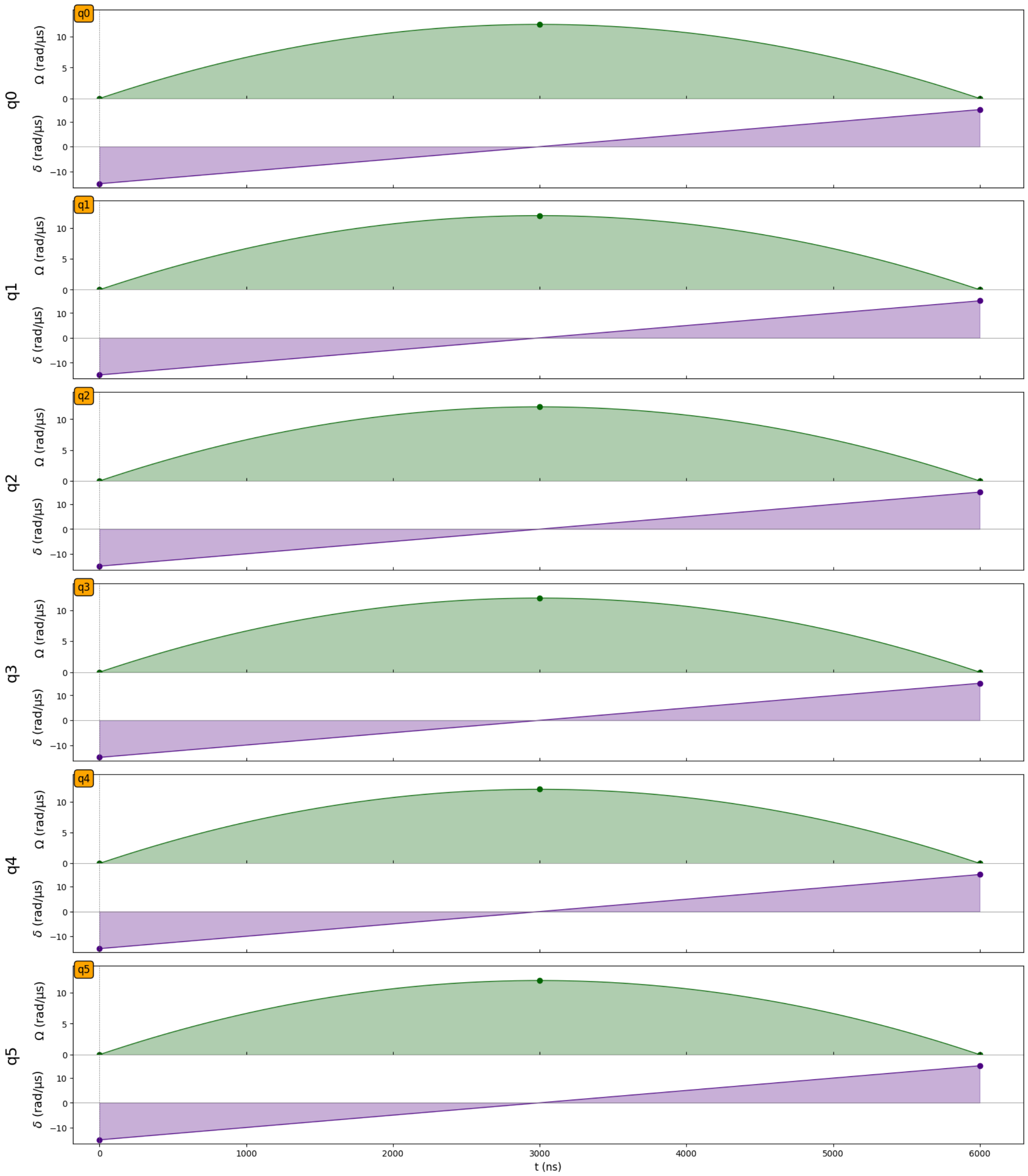

And here is the execution keep in mind that this works for graphs under 12 nodes as anymore nodes requires using another backend and will probably be slow on your device for our tests we used the pasqal_cloud library to outsource the execution of the sequence

In [12]:
```
backend = pulser.backends.QutipBackend(sequence)
results = backend.run(progress_bar=True)
data = results.sample_final_state()

sorted_data = sorted(data.items(), key=lambda x: x[1], reverse=True)
normalized_data = list(map(lambda x: (x[0], x[1] / 1000), sorted_data))
normalized_data_dict = dict(normalized_data[:10])
```

```
plt.bar(normalized_data_dict.keys(), normalized_data_dict.values())
plt.xlabel("Measured States")
plt.xticks(rotation=90)
plt.ylabel("Probability")
```

```
10.0%. Run time:   0.01s. Est. time left: 00:00:00:00
20.0%. Run time:   0.03s. Est. time left: 00:00:00:00
30.0%. Run time:   0.05s. Est. time left: 00:00:00:00
40.0%. Run time:   0.06s. Est. time left: 00:00:00:00
50.0%. Run time:   0.08s. Est. time left: 00:00:00:00
60.0%. Run time:   0.09s. Est. time left: 00:00:00:00
70.0%. Run time:   0.11s. Est. time left: 00:00:00:00
80.0%. Run time:   0.12s. Est. time left: 00:00:00:00
90.0%. Run time:   0.14s. Est. time left: 00:00:00:00
100.0%. Run time:   0.15s. Est. time left: 00:00:00:00
Total run time:   0.16s
```

```
Out[12]:  Text(0, 0.5, 'Probability')
```

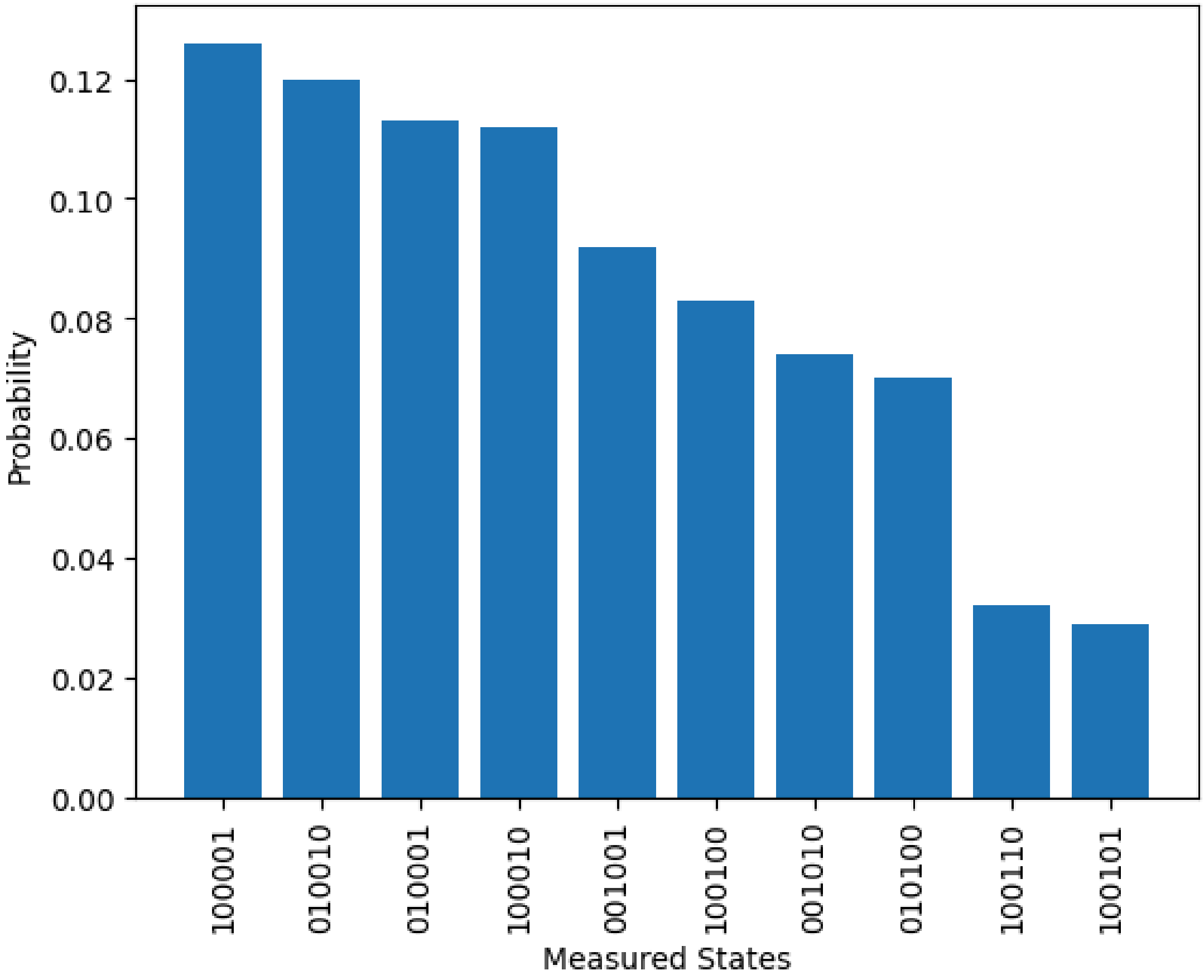

Now we can see the results which are all our possible MIS you can change the range to plot more top values the probabilities or very low this is due to the graph having multiple possible solutions and thus are endstate is likely a superposition of these solutions

```
In [50]:  fig, ax = plt.subplots(2, 4)
          top_mesurments = list(normalized_data_dict.keys())[:8]
```

```python
for i in range(2):
    for j in range(4):
        b = top_mesurments[i * 4 + j]
        ax[i][j].set_title(b)
        ax[i][j].axis("off")
        draw(b, positions, ax=ax[i][j])
```

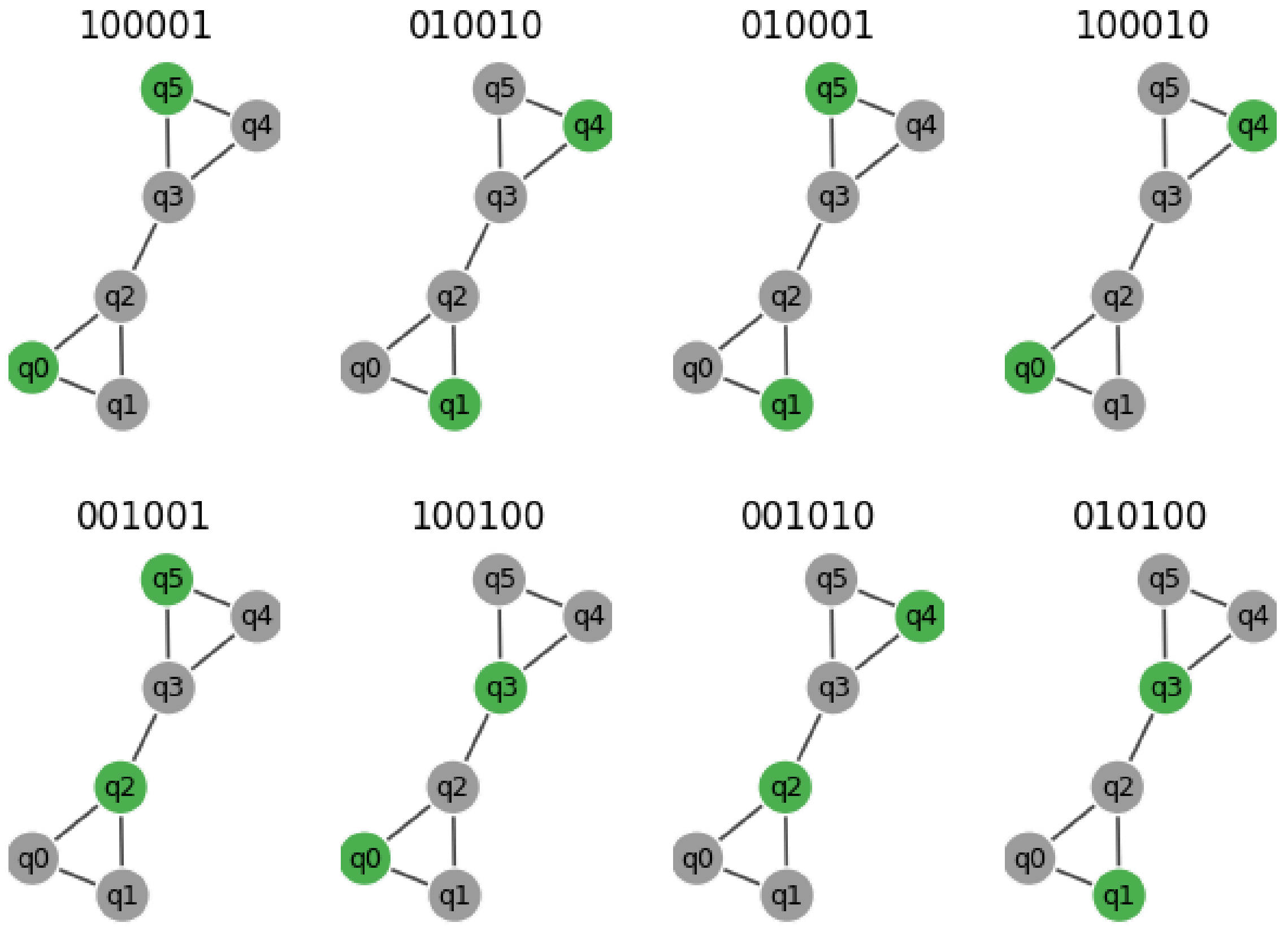

### 13.2. Instances of solving MIS

This section presents illustrative examples of results obtained when solving the Maximum Independent Set (MIS) problem using different methods. For each instance, we show the graph embedding and the probability of measuring an MIS solution — that is, an independent set equivalent to the MIS, not necessarily the unique brute-force MIS.

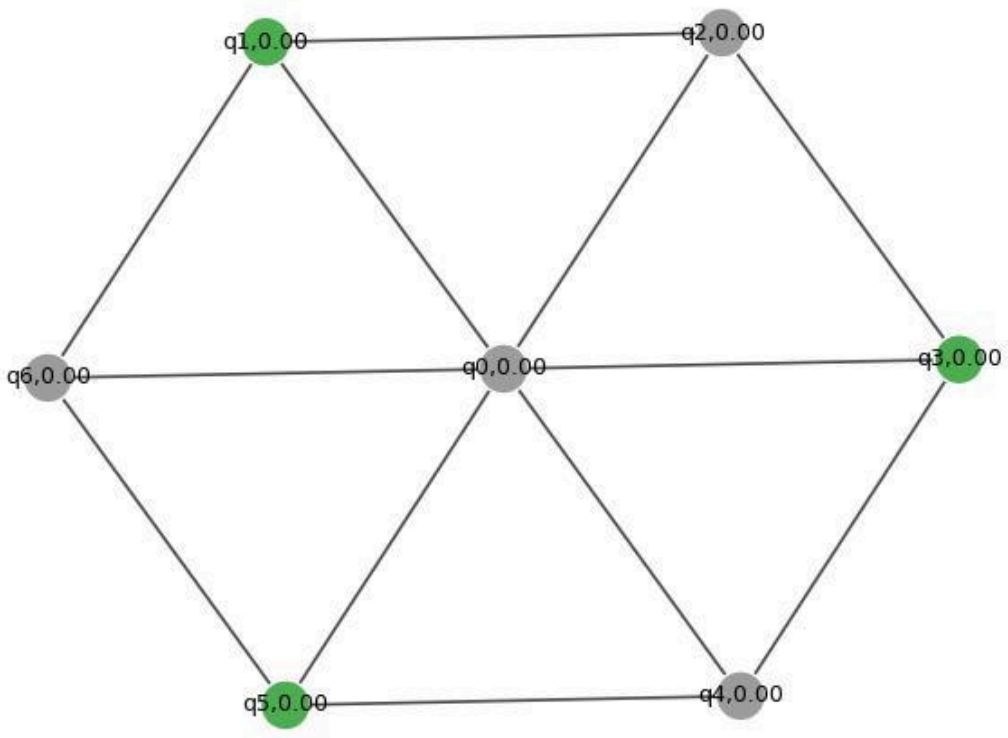

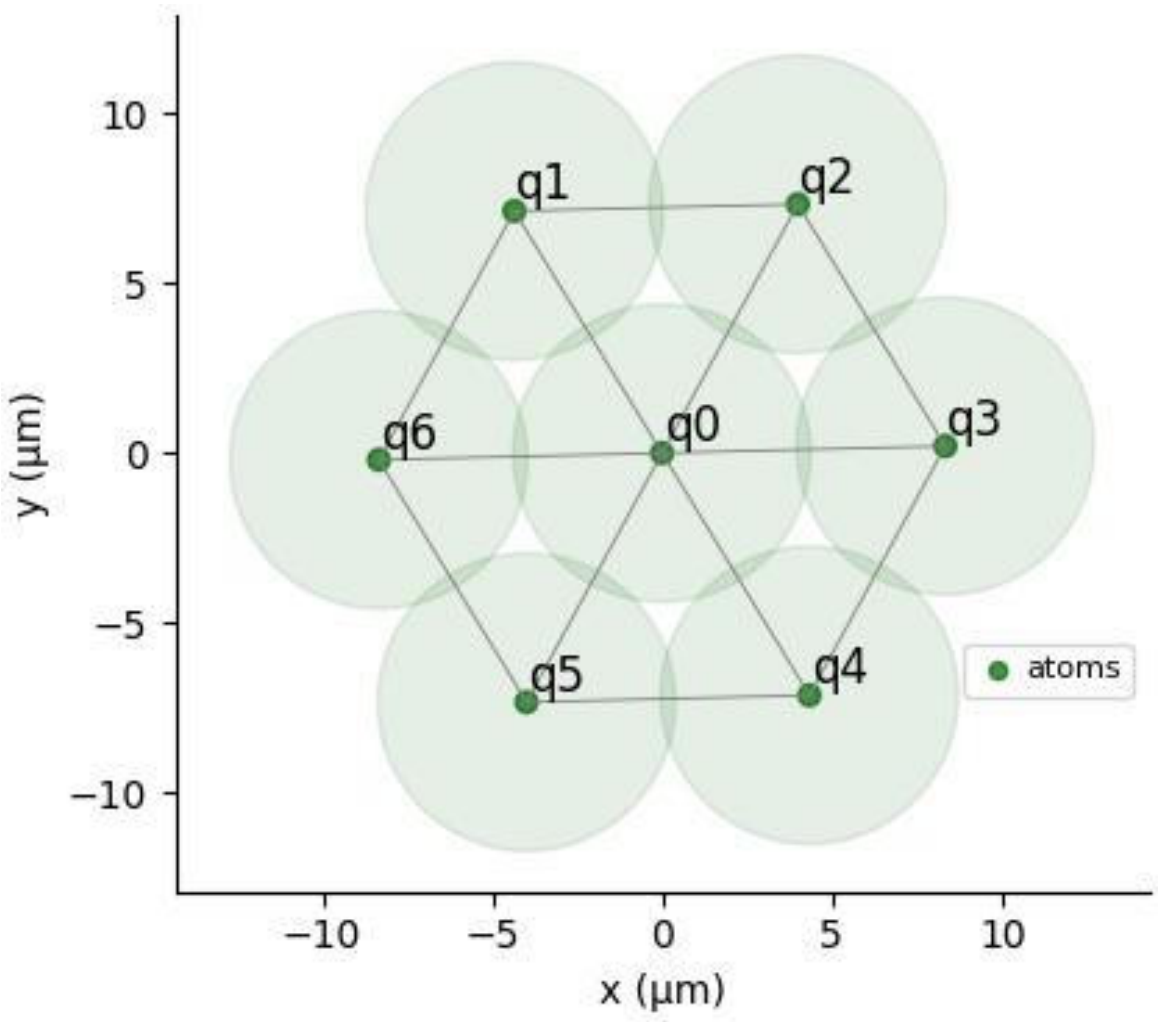

| Local Detuning | DMM | Global Detuning | Current Method |
|---|---|---|---|
| 0.972 | 0.9720000000000001 | 0.904 | 0 |

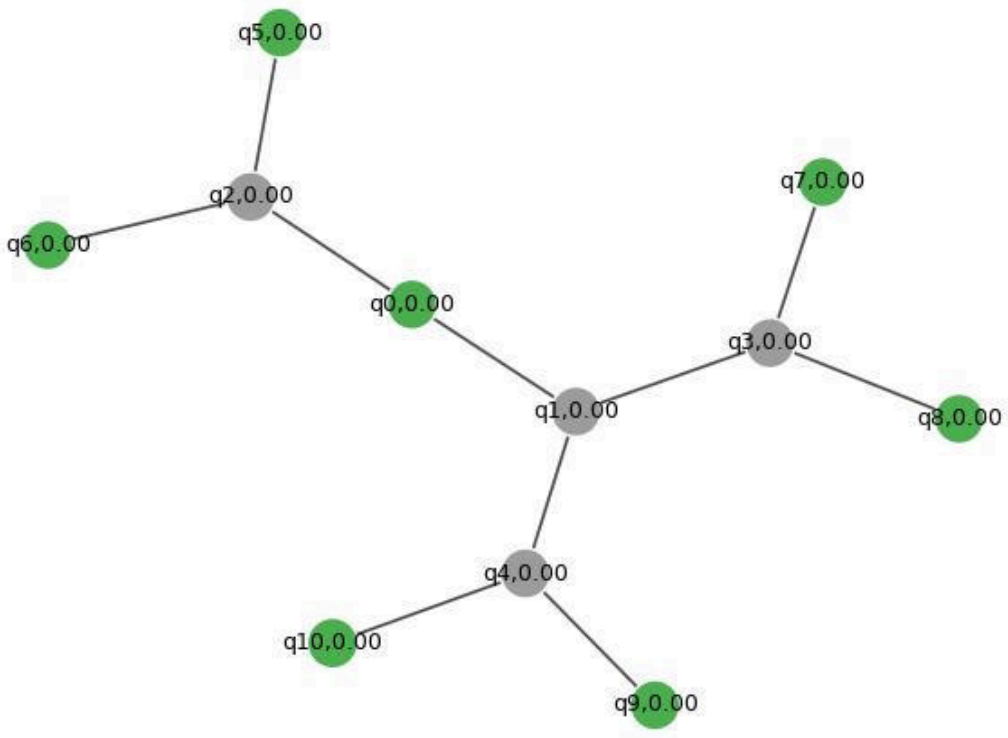

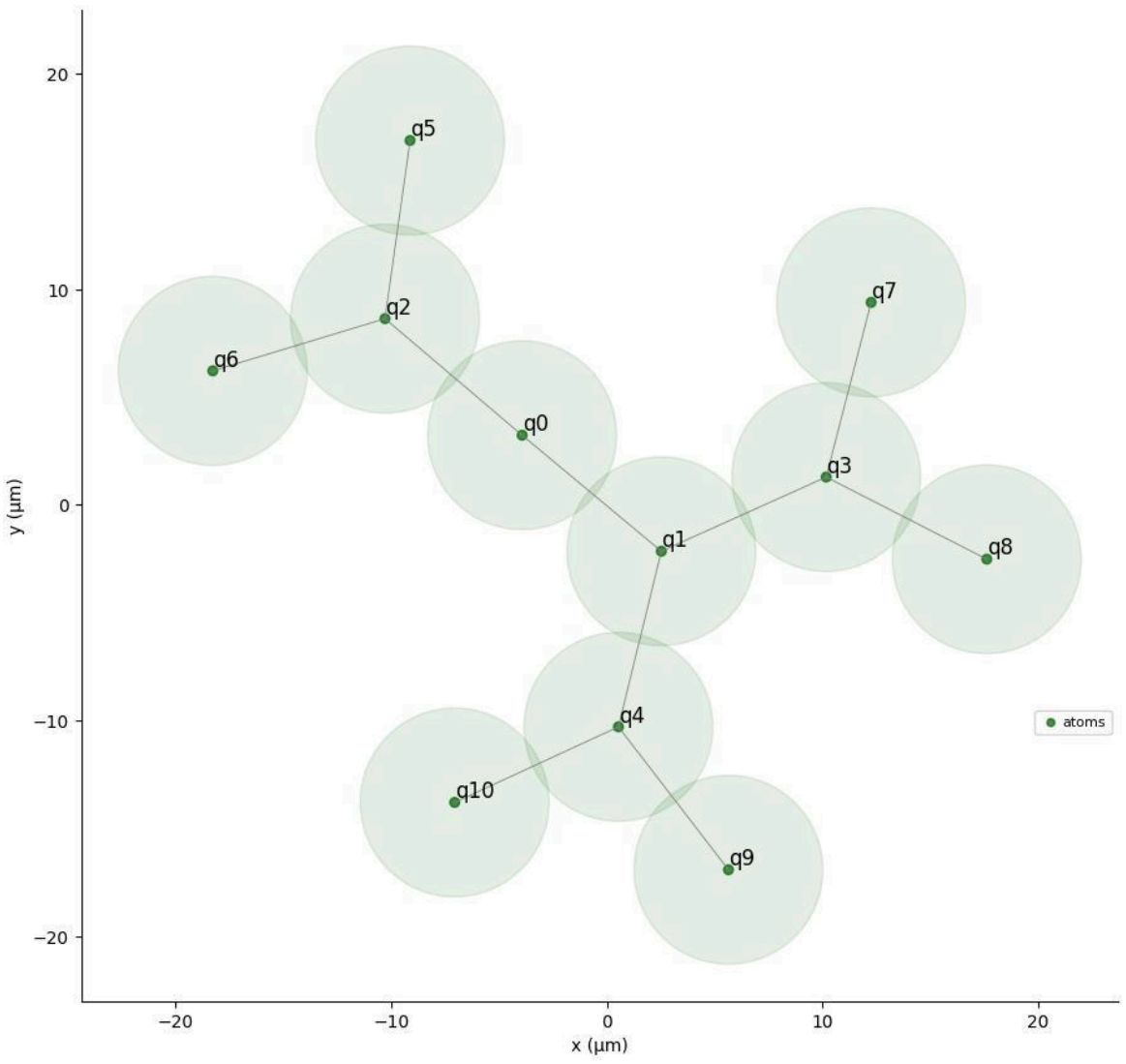

| Local Detuning | DMM | Global Detuning | Current Method |
|---|---|---|---|
| 0.976 | 0.9610000000000001 | 0.965 | 0.96 |

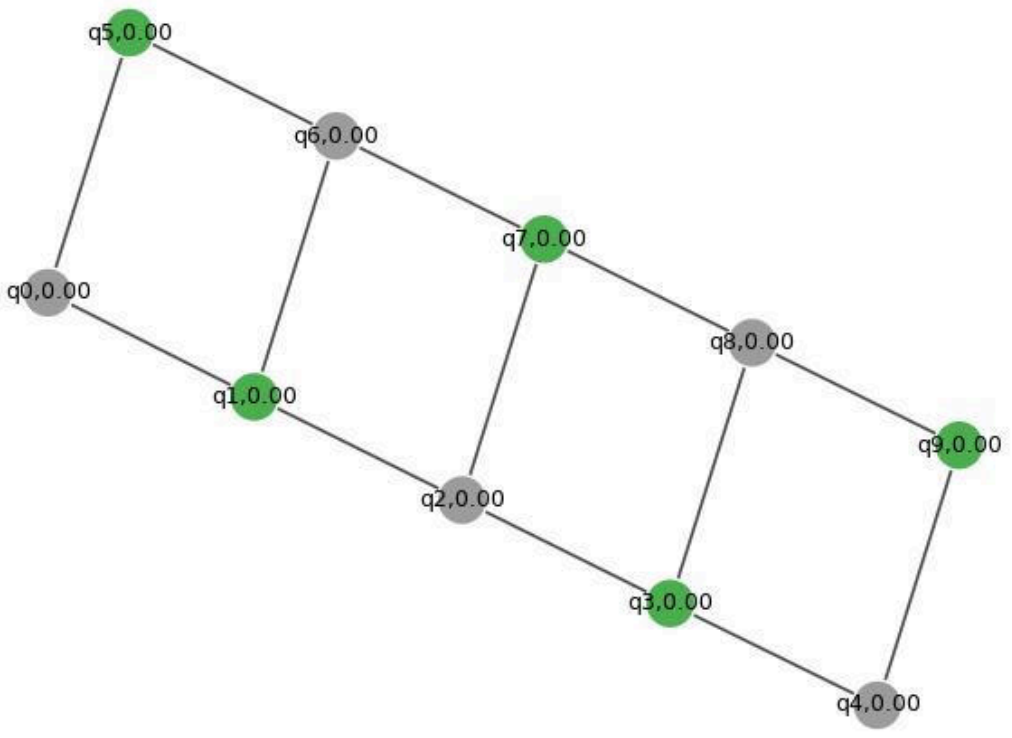

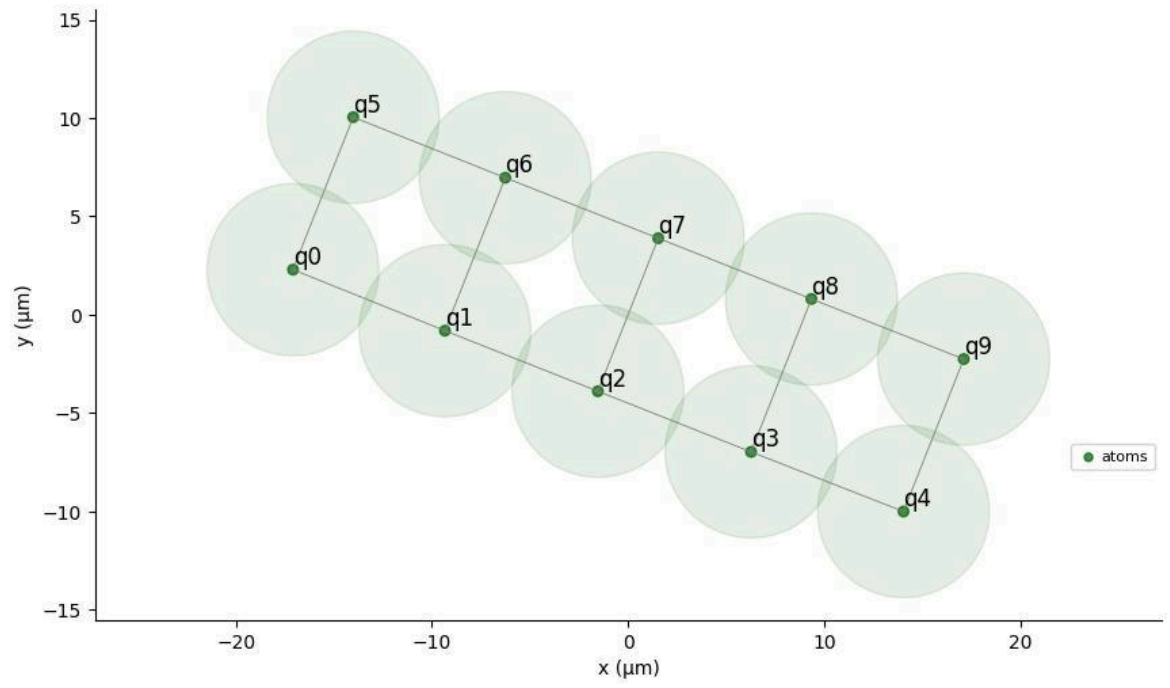

| Local Detuning | DMM | Global Detuning | Current Method |
|---|---|---|---|
| 0.9359999999999999 | 0.944 | 0.917 | 0.952 |

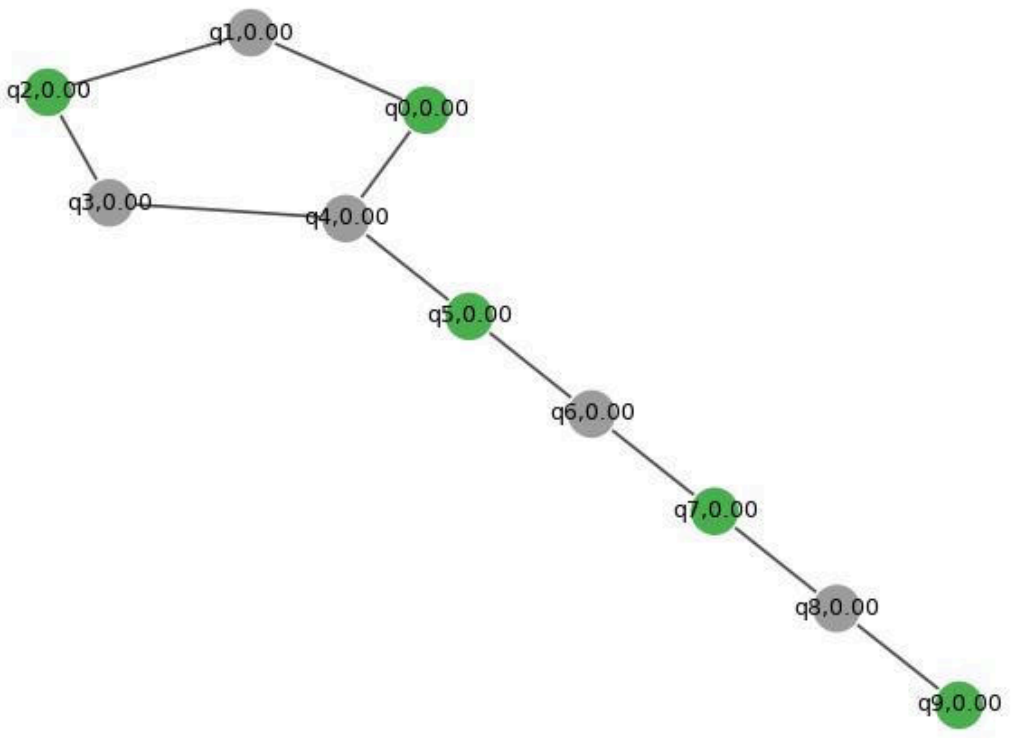

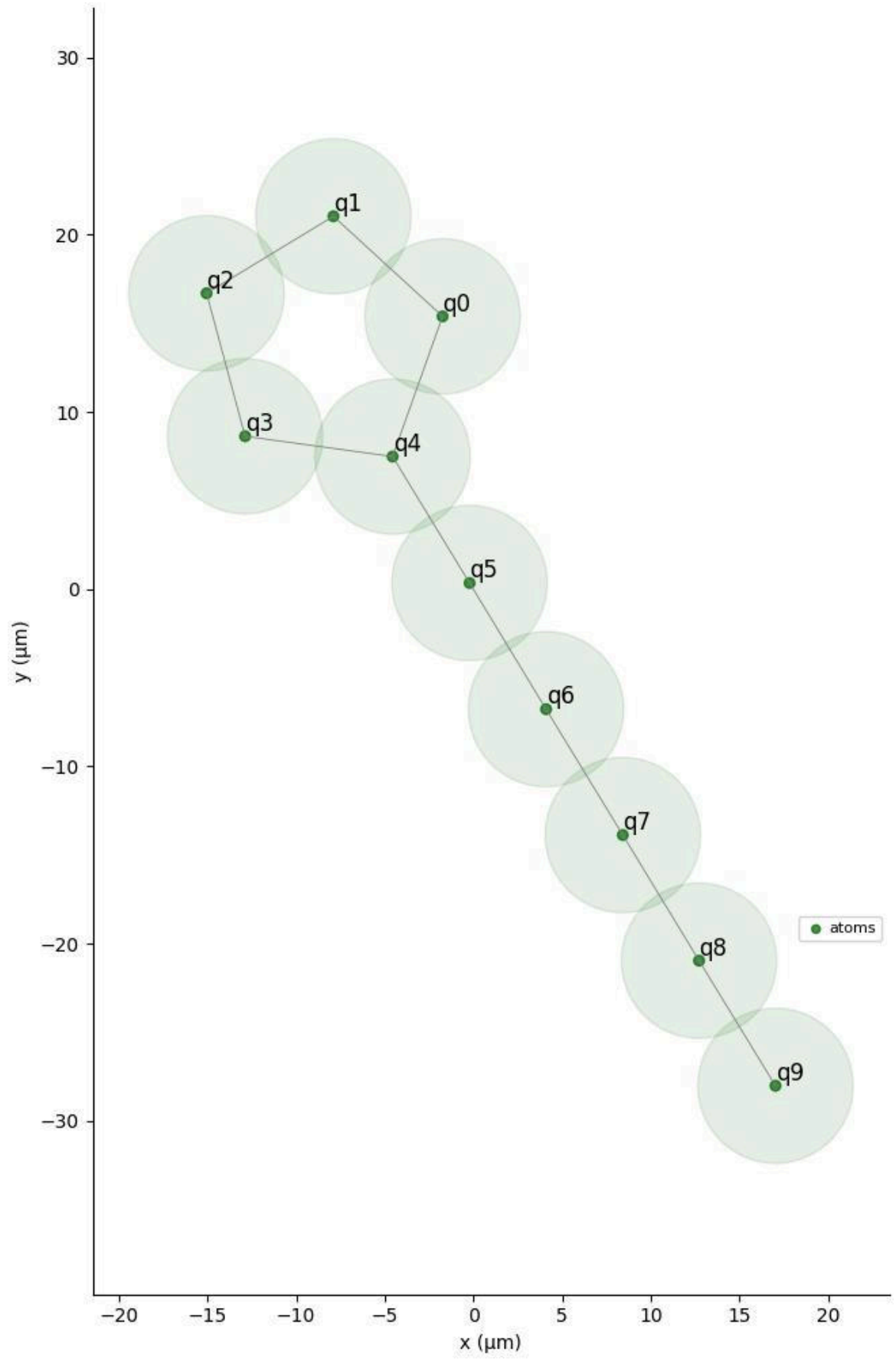

| Local Detuning | DMM | Global Detuning | Current Method |
|---|---|---|---|
| 0.9159999999999999 | 0.903 | 0.912 | 0.9089999999999999 |

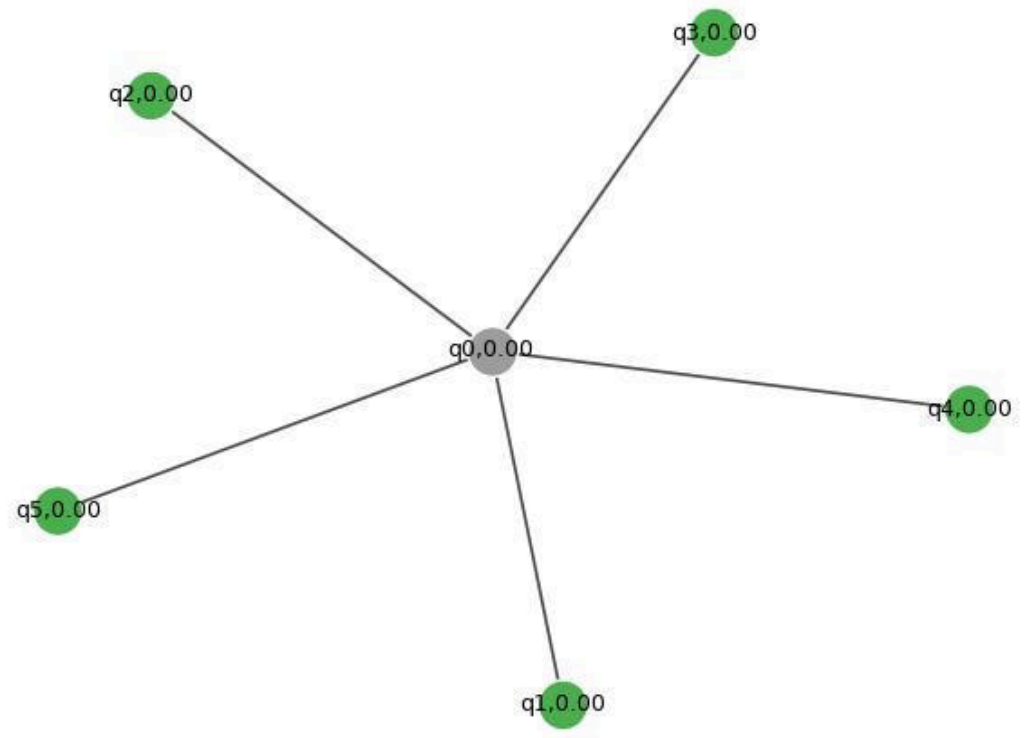

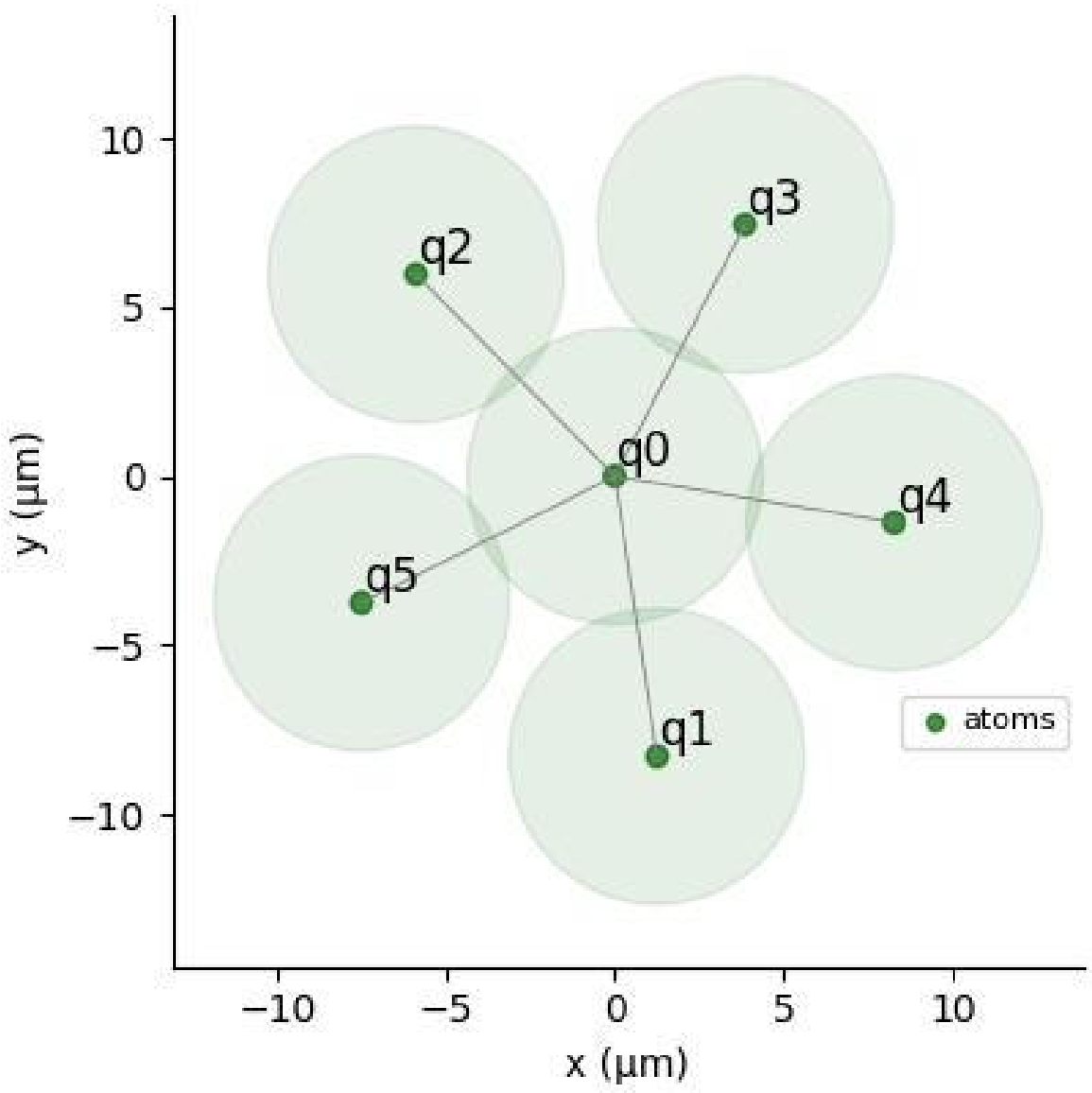

| Local Detuning | DMM | Global Detuning | Current Method |
|---|---|---|---|
| 0.993 | 0.998 | 0.993 | 1 |

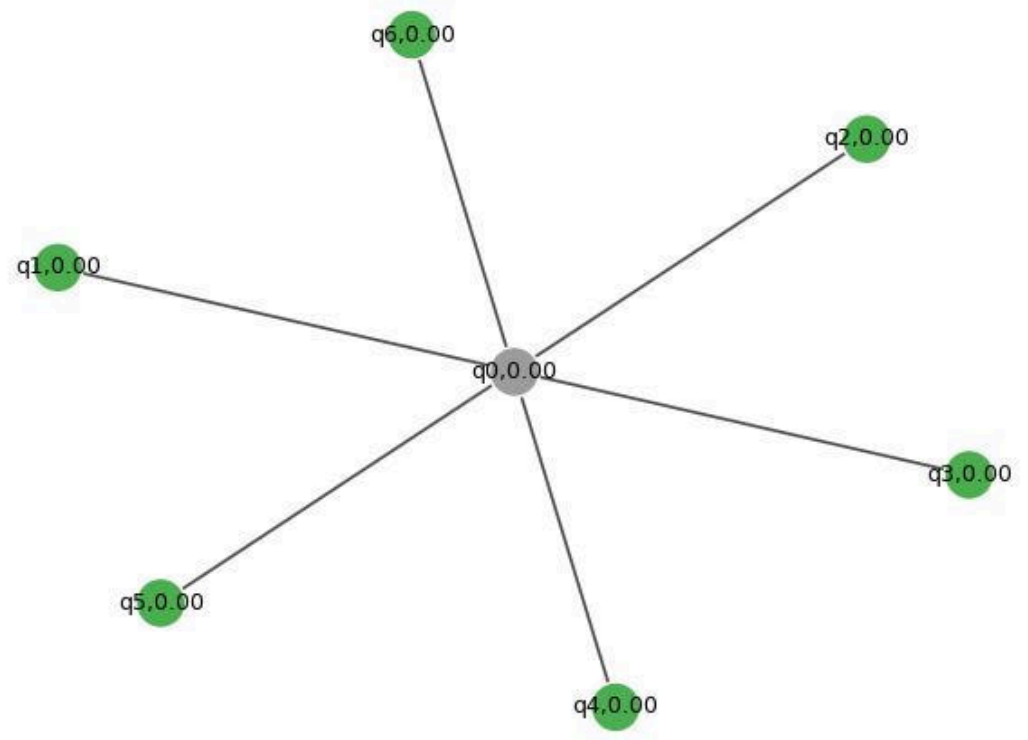

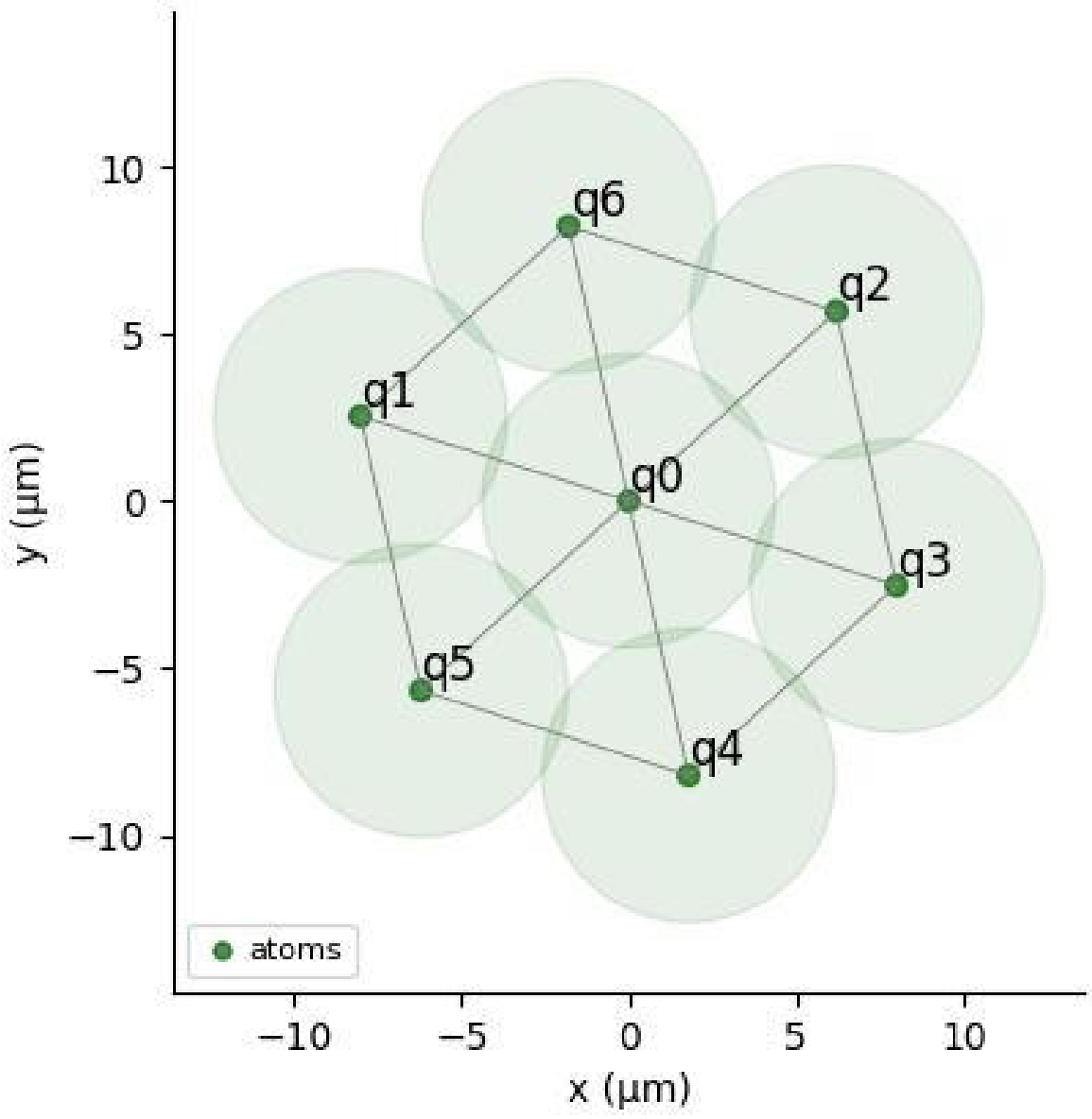

| Local Detuning | DMM | Global Detuning | Current Method |
|---|---|---|---|
| 0.992 | 0.998 | 0.906 | 0.172 |

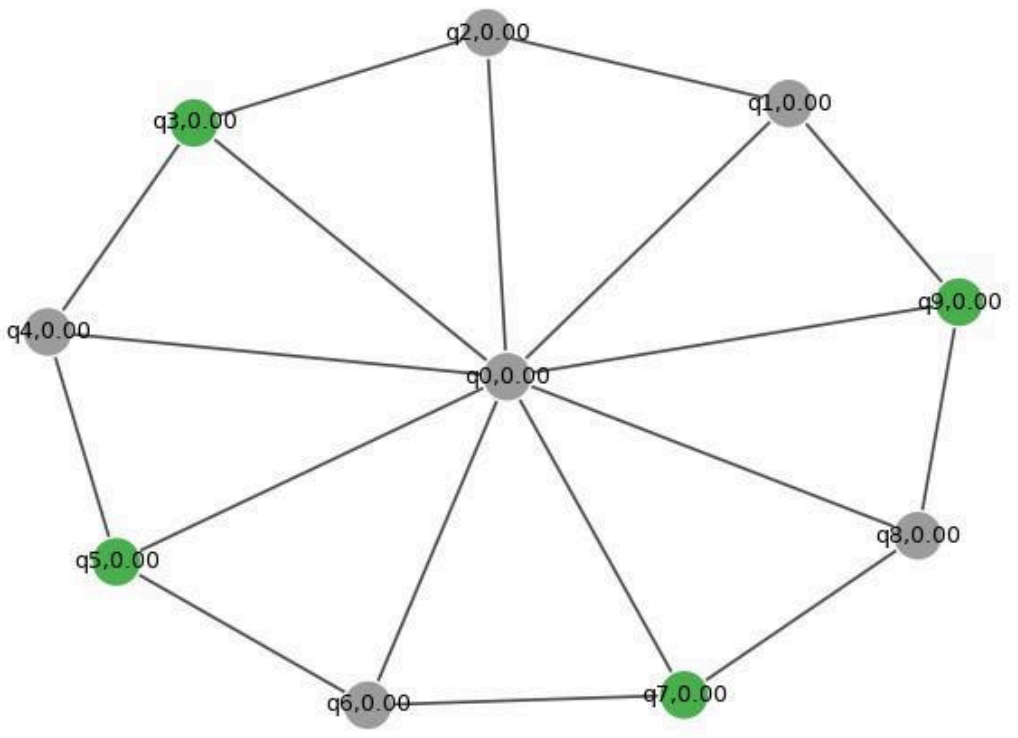

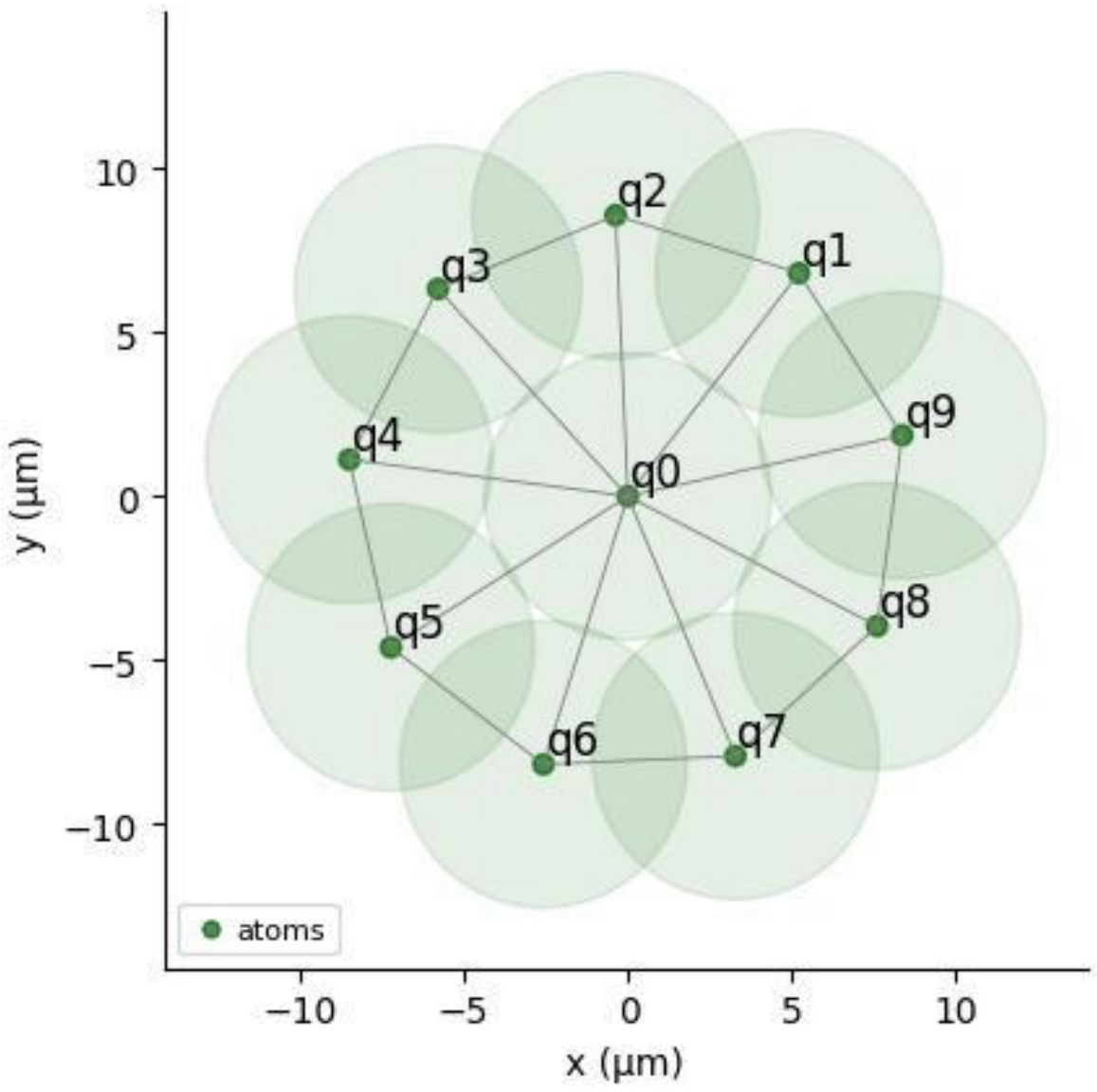

| Local Detuning | DMM | Global Detuning | Current Method |
| --- | --- | --- | --- |
| 0.9949999999999999 | 0.9969999999999999 | 0.9969999999999999 | 0.9999999999999999 |

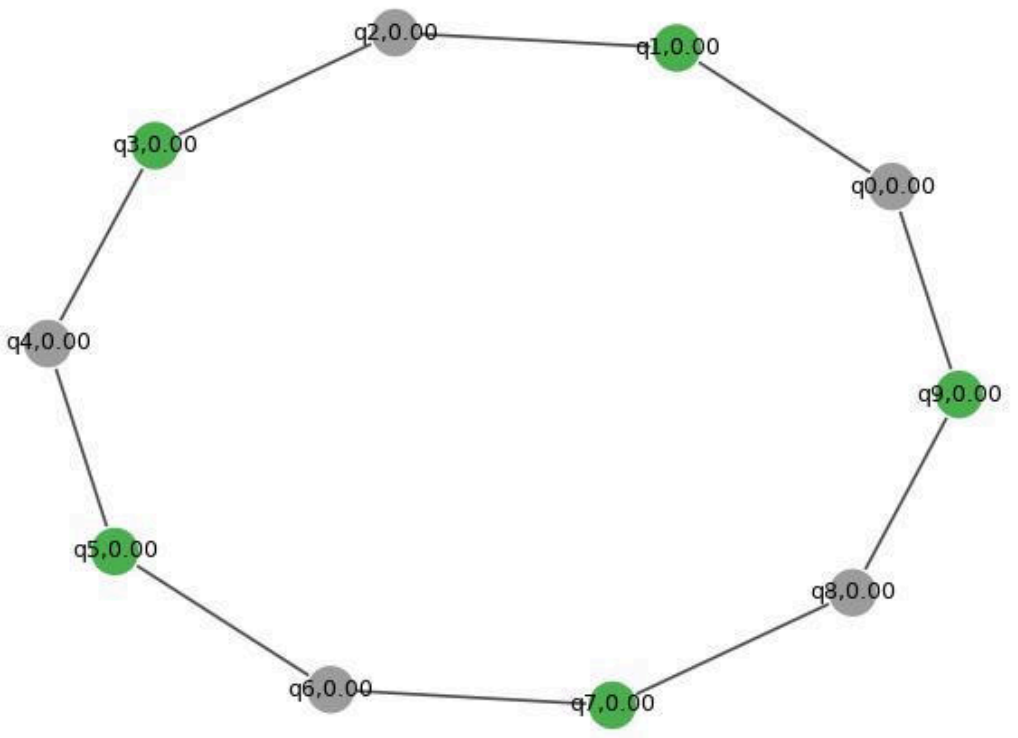

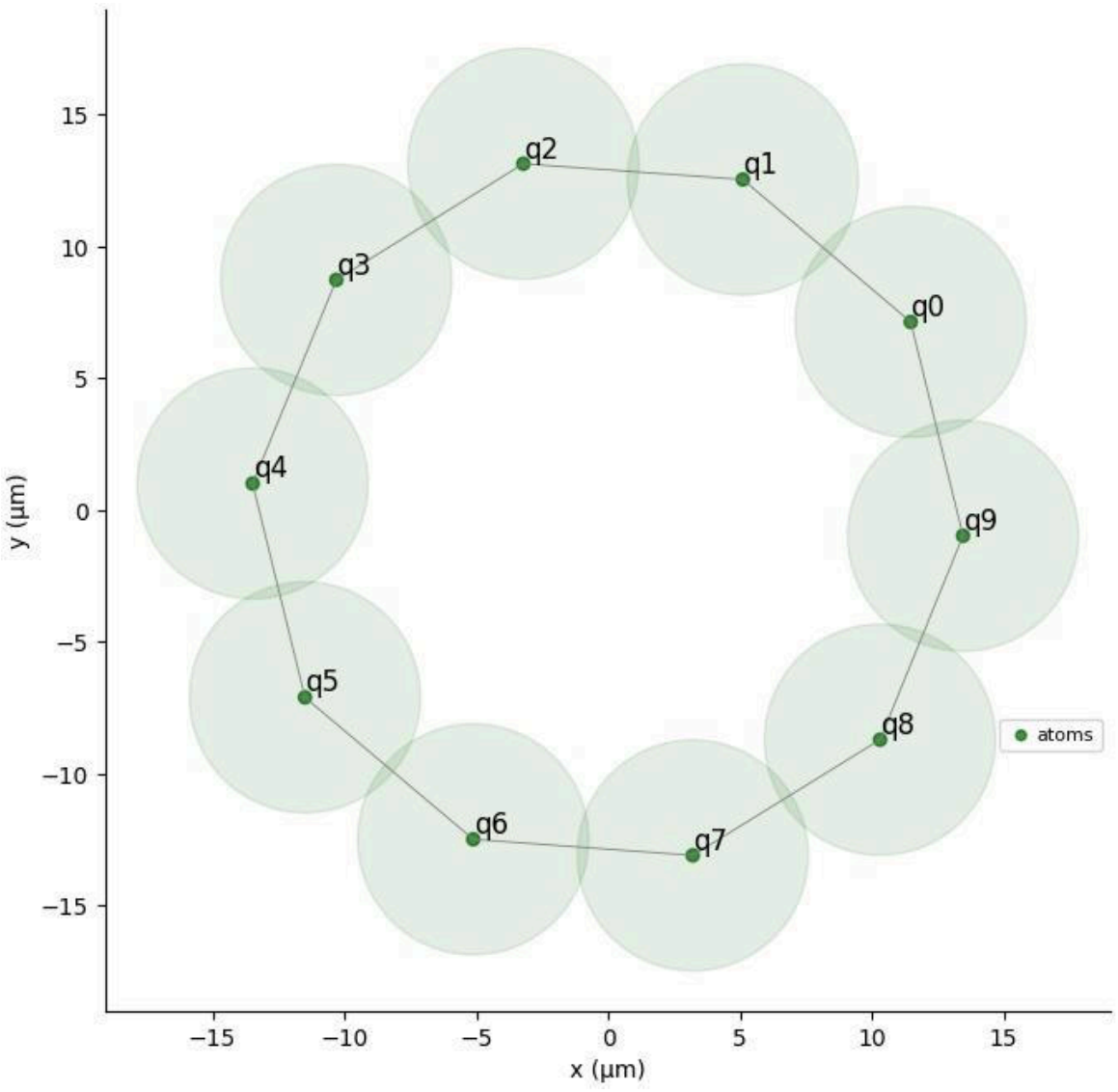

| Local Detuning | DMM | Global Detuning | Current Method |
| --- | --- | --- | --- |
| 0.892 | 0.893 | 0.874 | 0.385 |

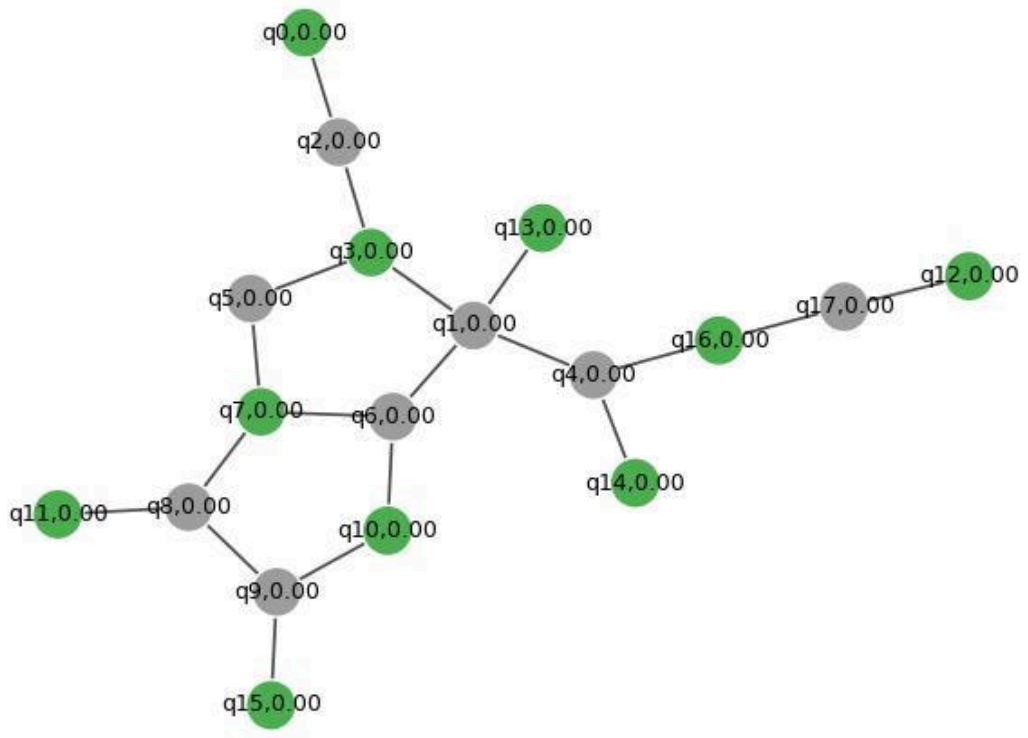

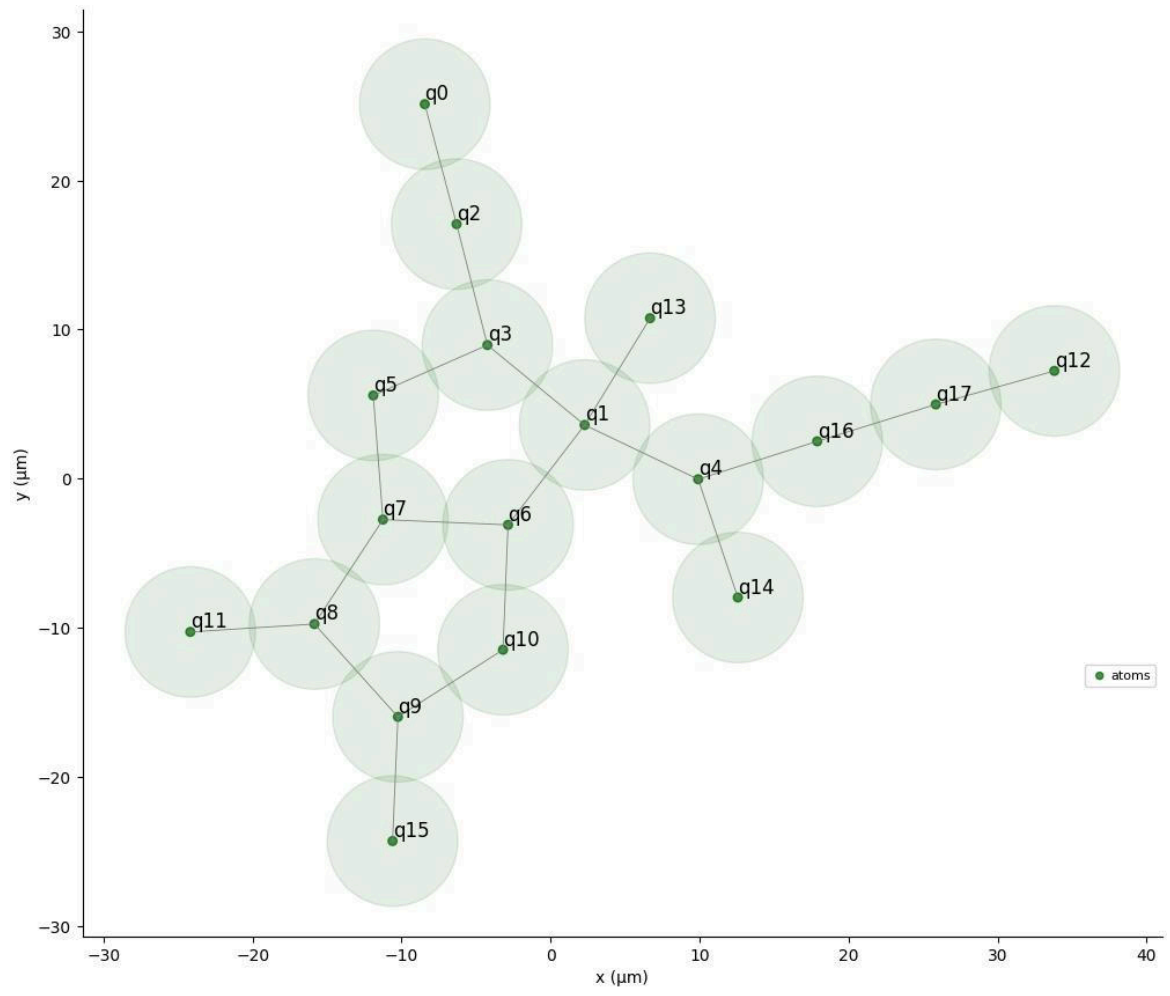

| Local Detuning | DMM | Global Detuning | Current Method |
|---|---|---|---|
| 0.579 | 0.603 | 0.448 | 0.305 |

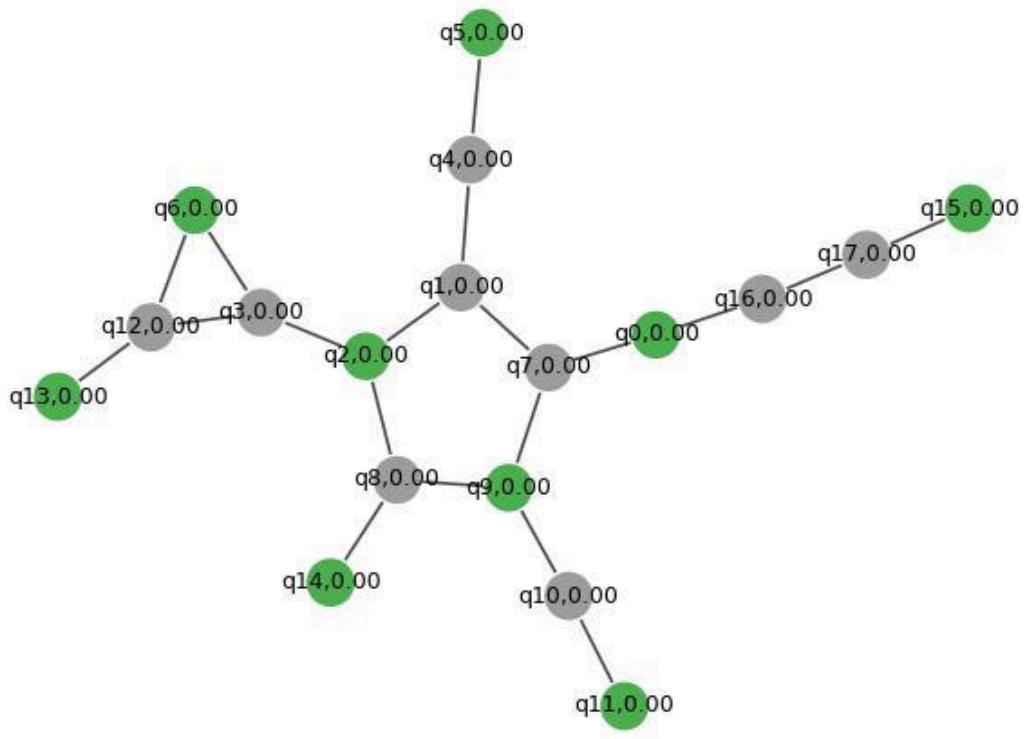

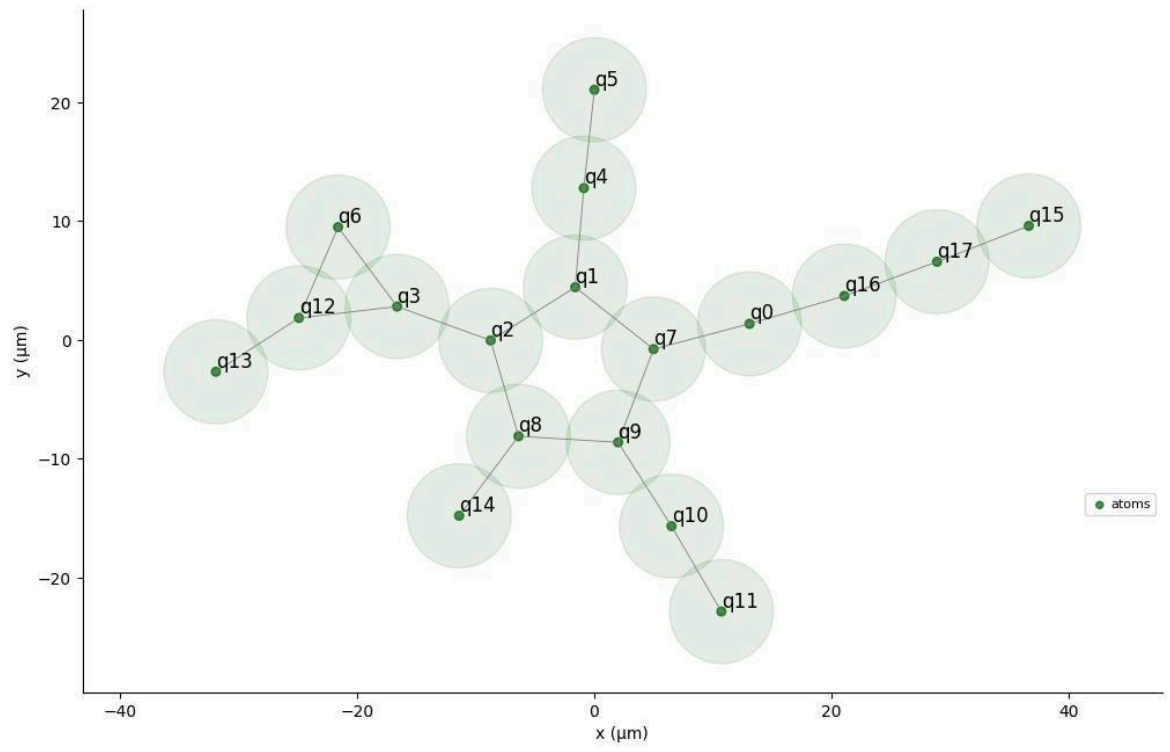

| Local Detuning | DMM | Global Detuning | Current Method |
|---|---|---|---|
| 0.8680000000000003 | 0.8880000000000003 | 0.8880000000000002 | 0.8970000000000002 |

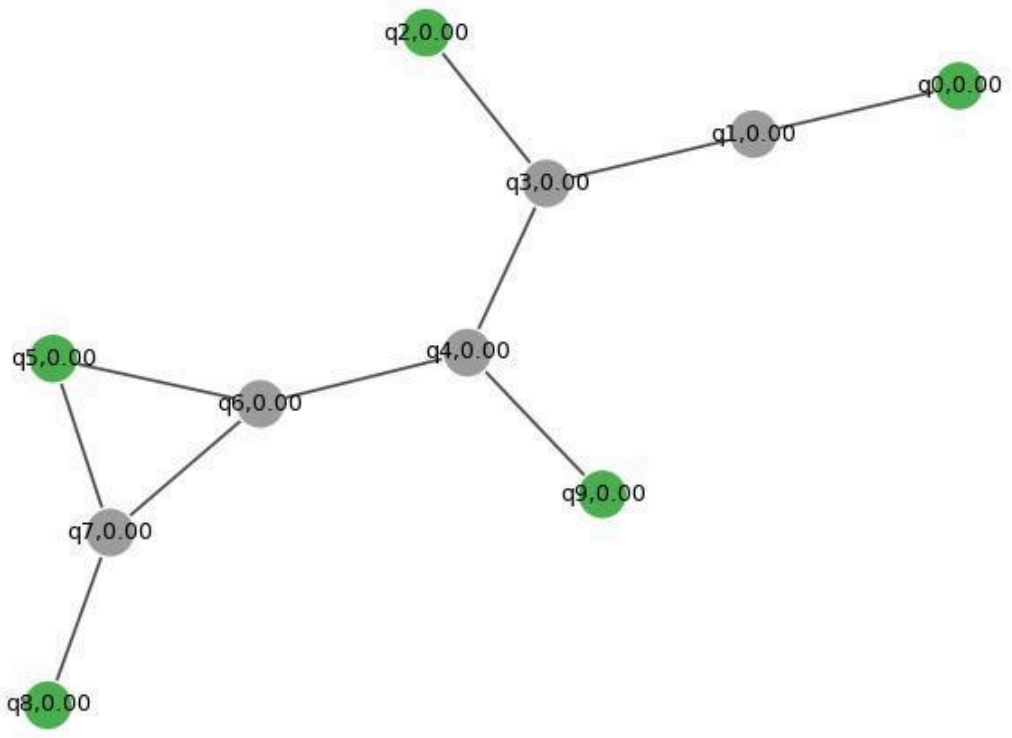

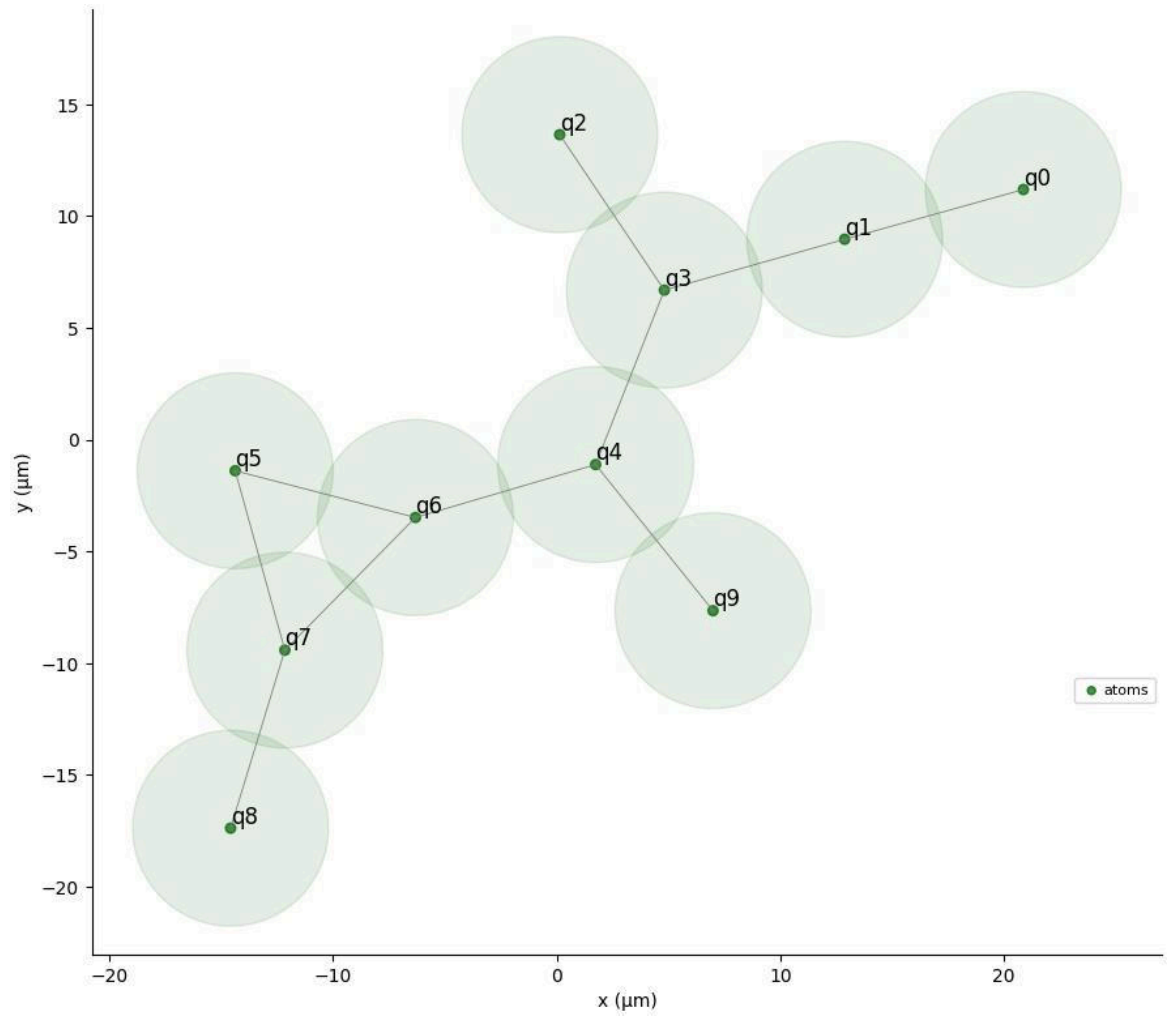

| Local Detuning | DMM | Global Detuning | Current Method |
| --- | --- | --- | --- |
| 0.8930000000000001 | 0.9210000000000002 | 0.901 | 0.91 |

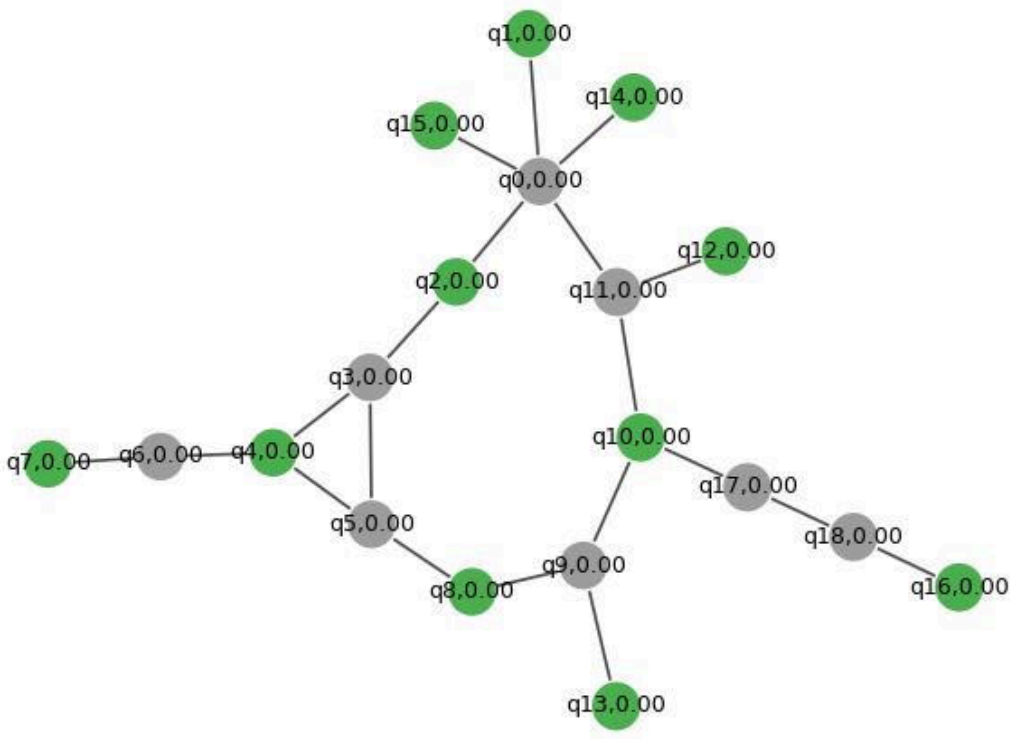

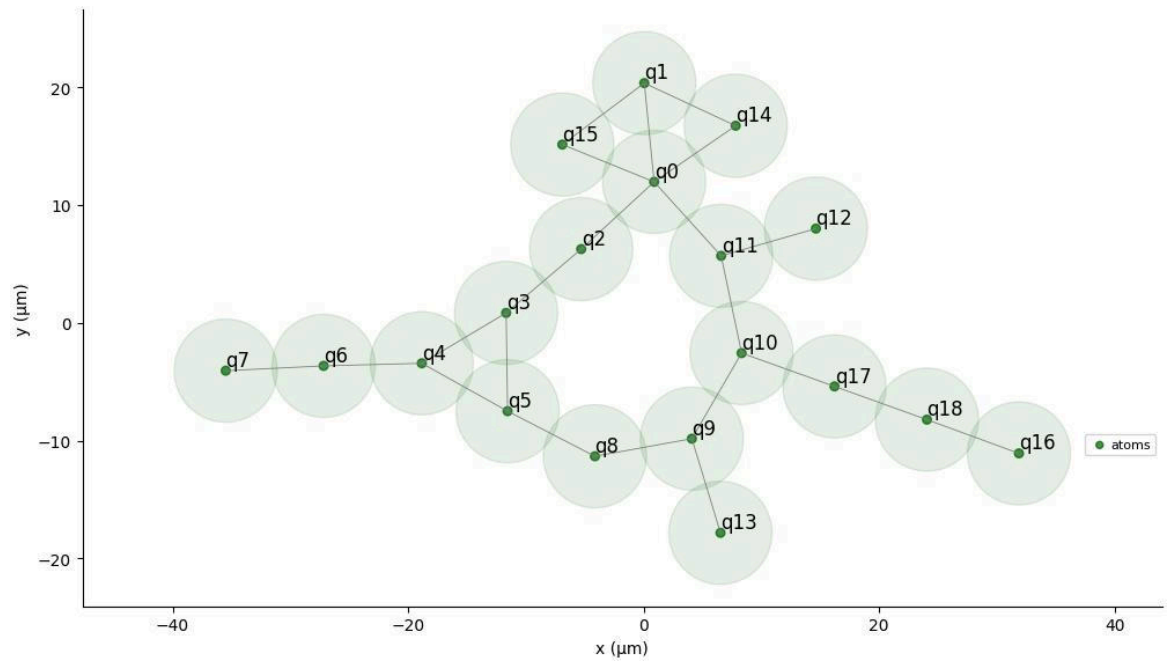

| Local Detuning | DMM | Global Detuning | Current Method |
| --- | --- | --- | --- |
| 0.8700000000000001 | 0.748 | 0.003 | 0 |

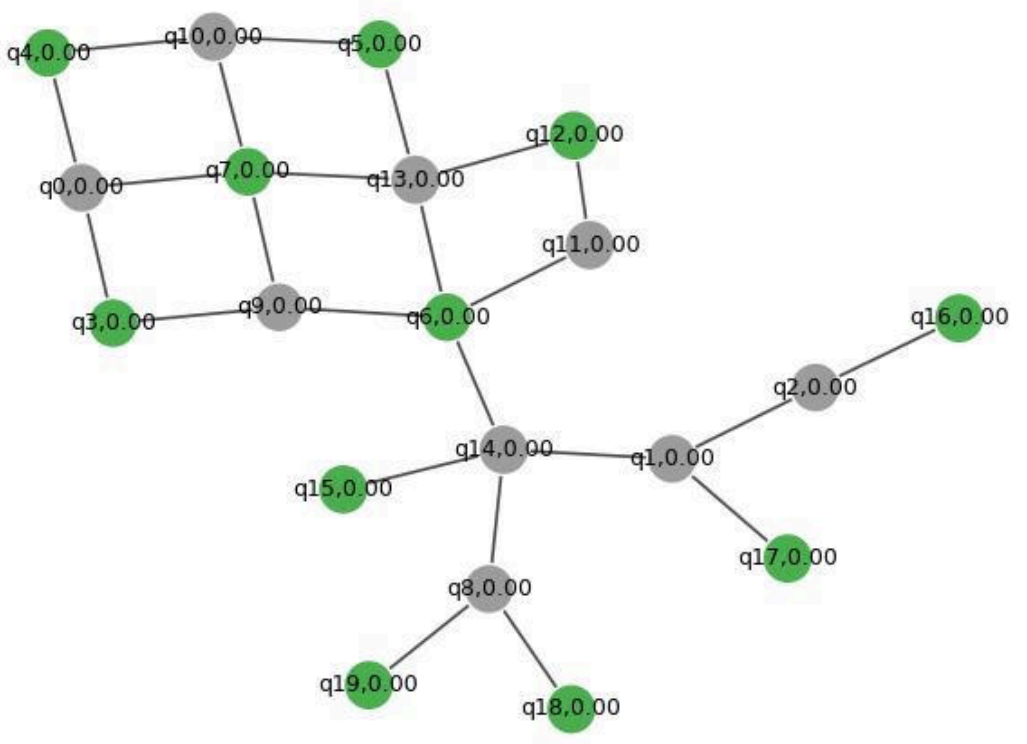

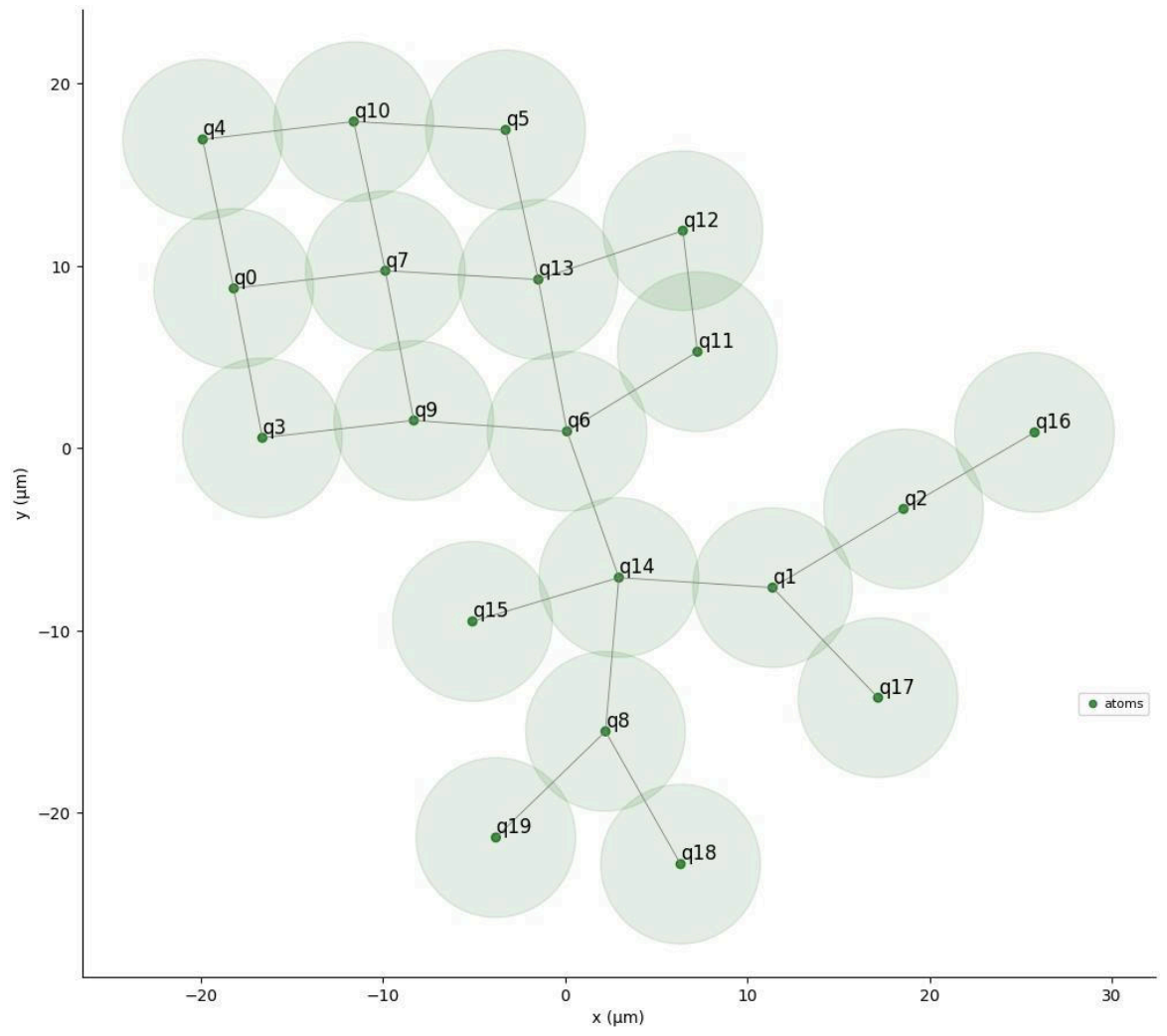

| Local Detuning | DMM | Global Detuning | Current Method |
| --- | --- | --- | --- |
| 0.41000000000000003 | 0.102 | 0.6940000000000001 | 0 |

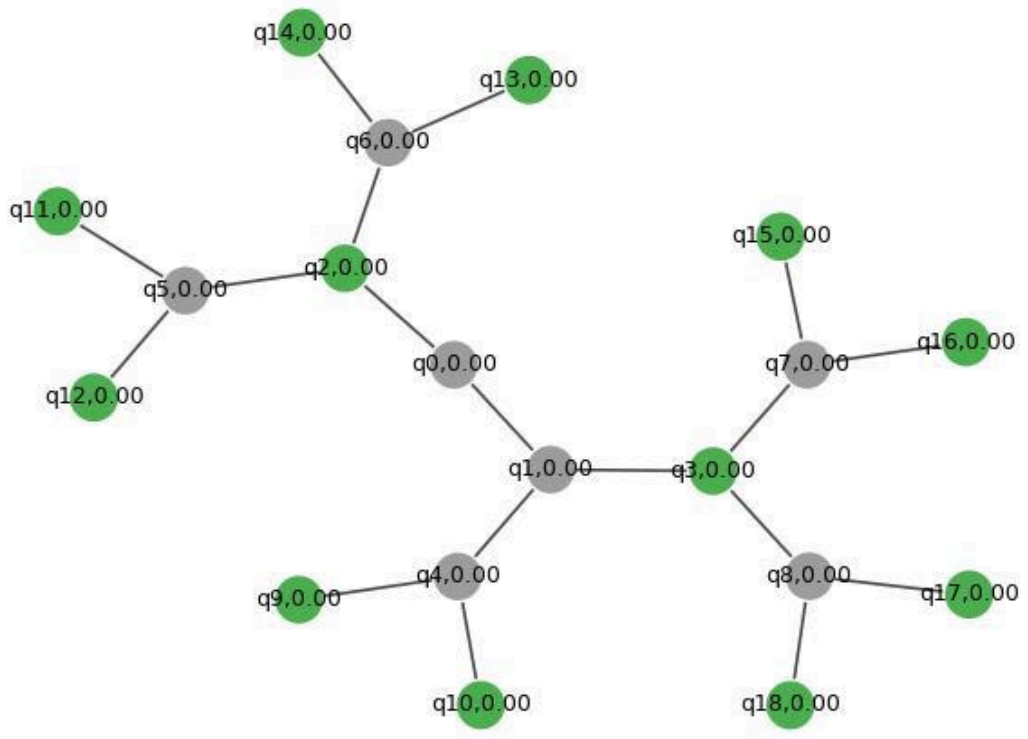

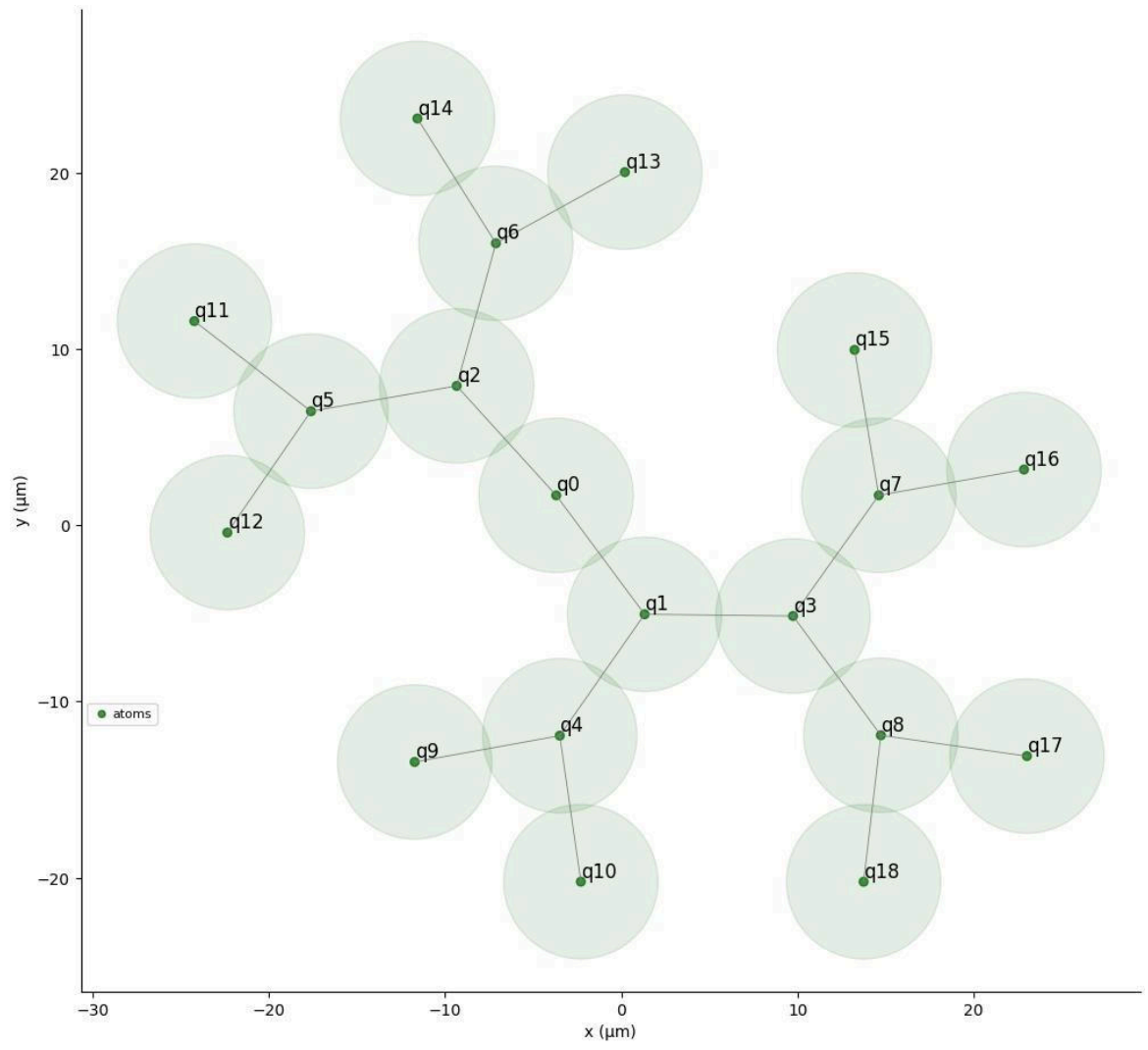

| Local Detuning | DMM | Global Detuning | Current Method |
|---|---|---|---|
| 0.934 | 0.9319999999999999 | 0.937 | 0.198 |

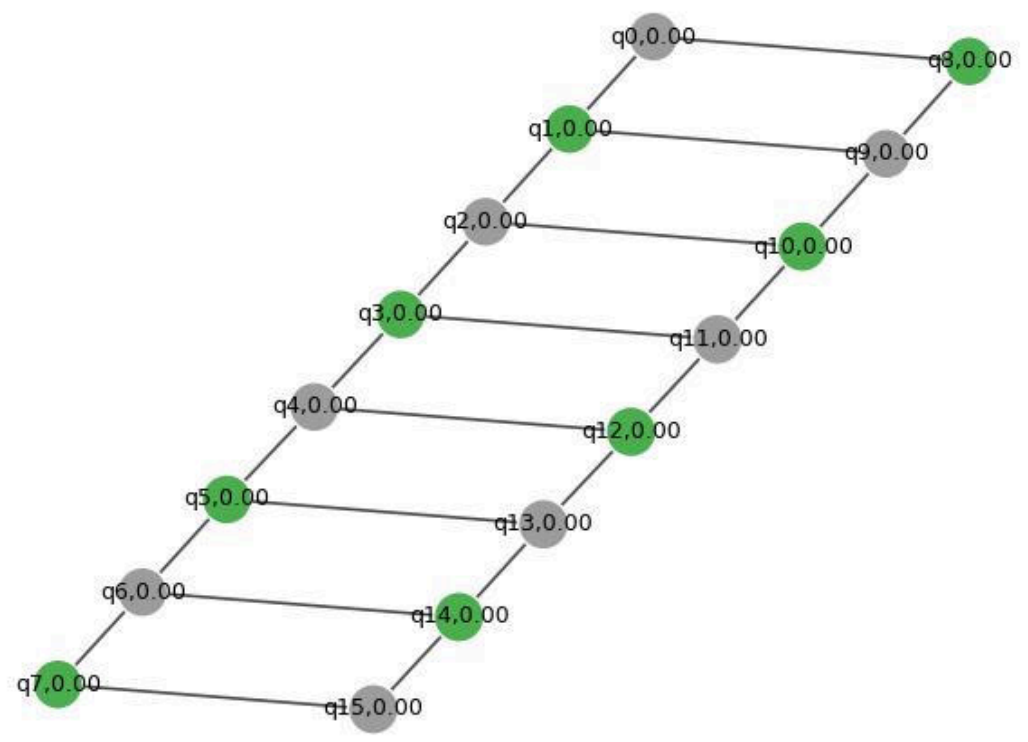

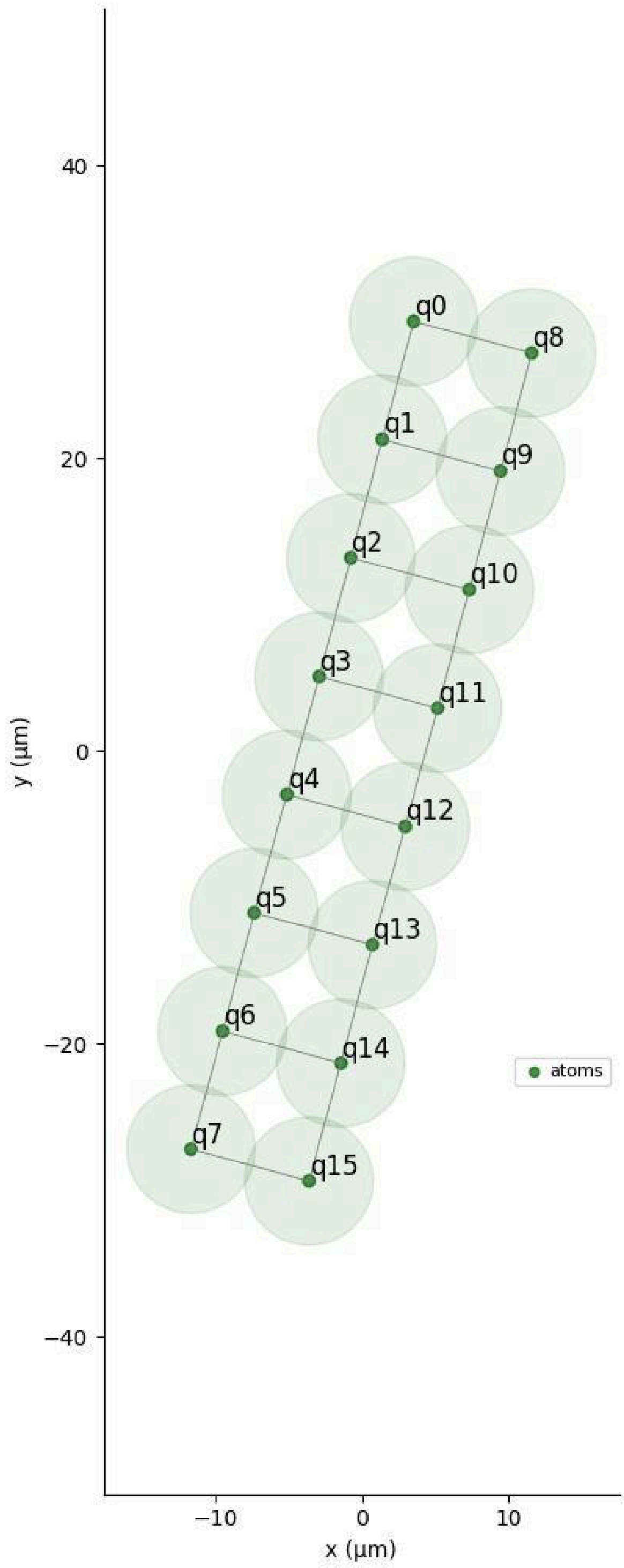

| Local Detuning | DMM | Global Detuning | Current Method |
|---|---|---|---|
| 0.754 | 0.697 | 0.76 | 0.729 |

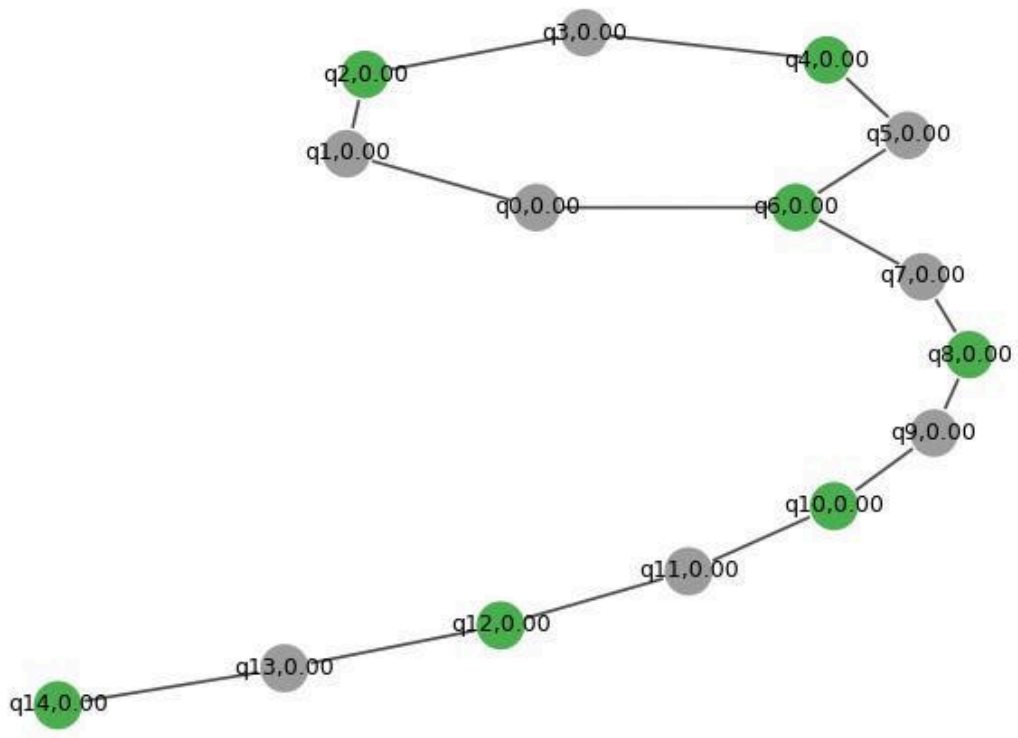

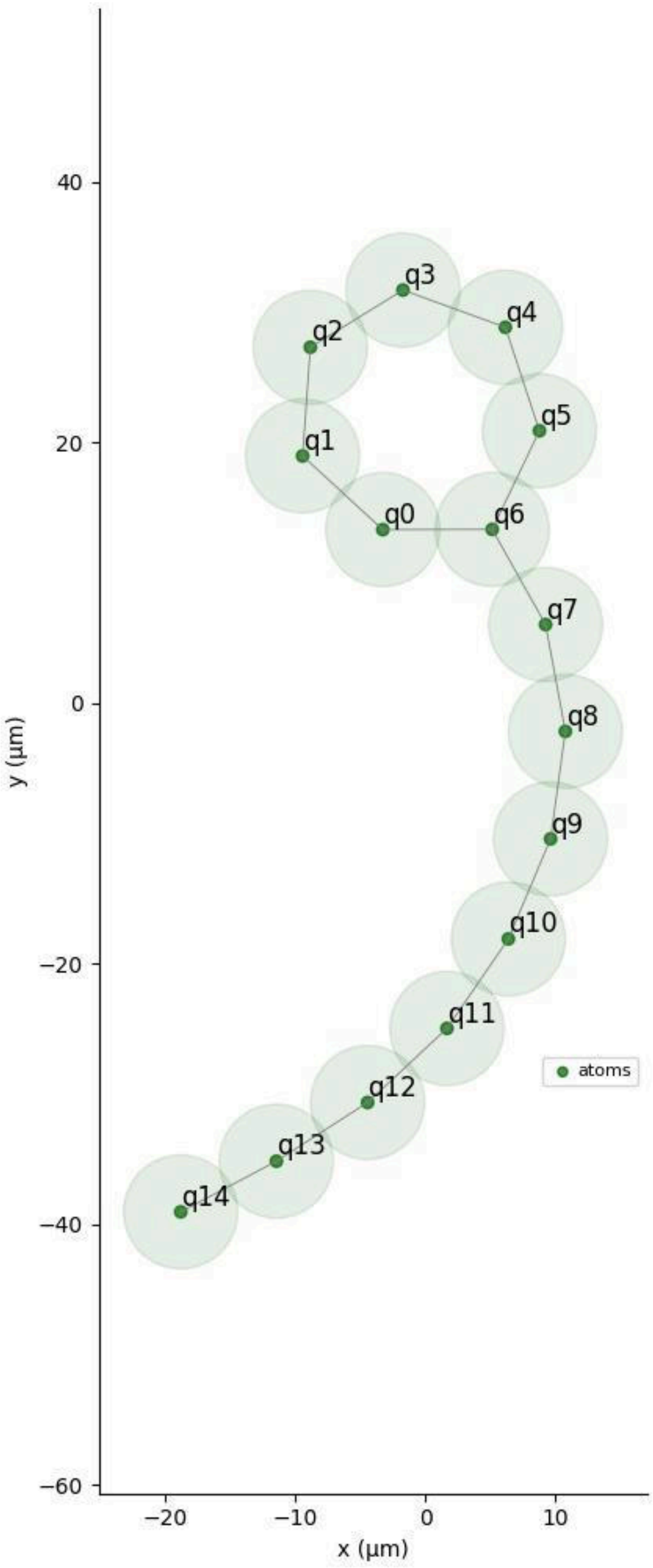

| Local Detuning | DMM | Global Detuning | Current Method |
|---|---|---|---|
| 0.9200000000000004 | 0.9120000000000004 | 0.8970000000000002 | 0.7280000000000003 |

### 13.3. Instances of solving MWIS

This section provides representative examples of results for the Maximum Weighted Independent Set (MWIS) problem. For each instance, it shows the graph embedding, along with each methods's outputs quantified by success probability — the probability of measuring an MWIS with weight exactly equal to that of the classically computed optimum — and optimality ratio — the average closeness of the measured MWIS weight to the classically determined optimum of all the different measured mwis

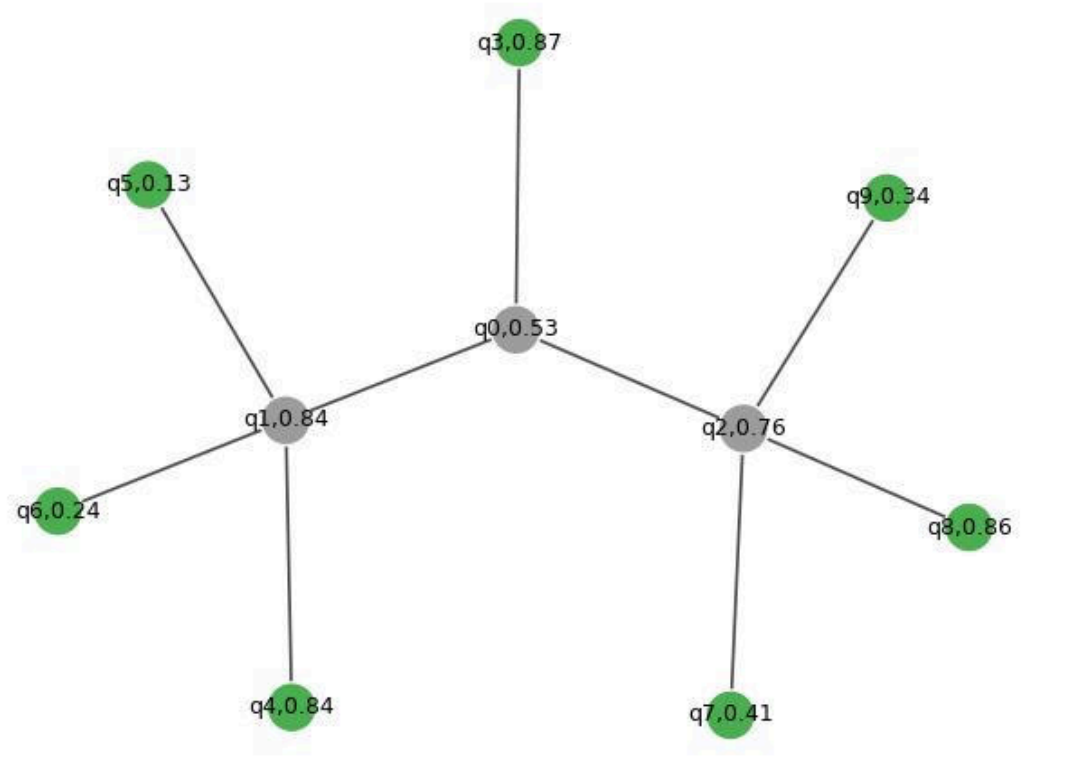

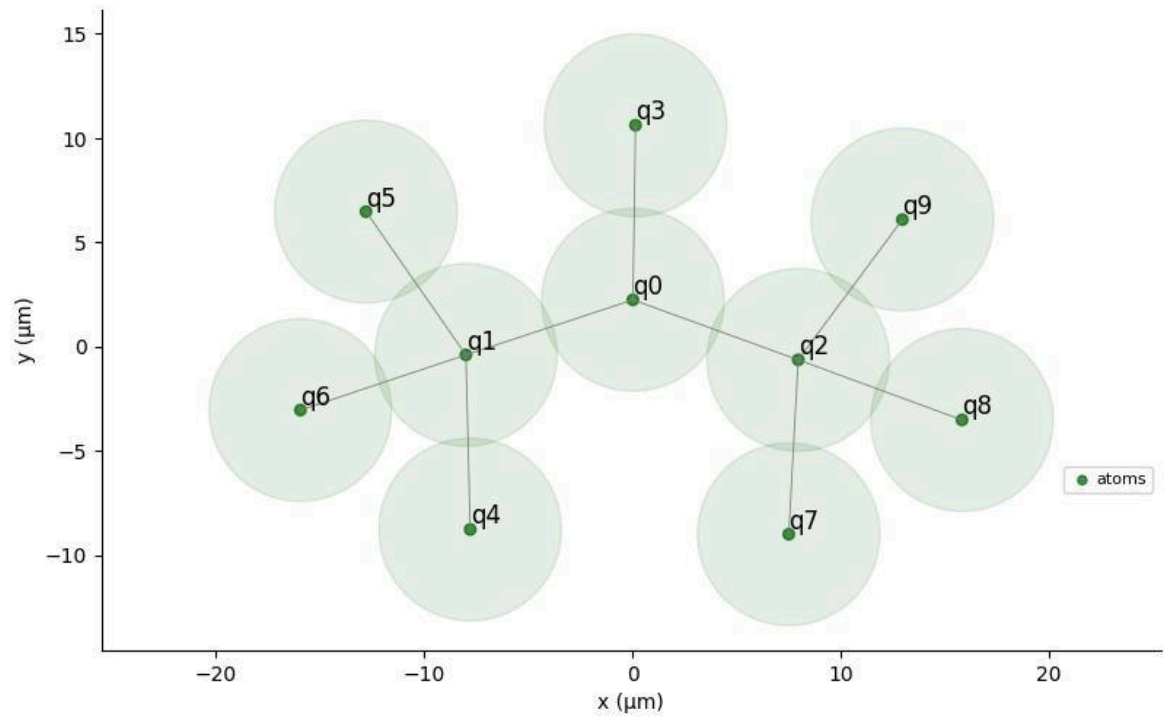

| Local Detuning | | DMM | |
|---|---|---|---|
| **Success probability** | **Optimality Ratio** | **Success probability** | **Optimality Ratio** |
| 0.642 | 0.9794 | 0.559 | 0.9731 |

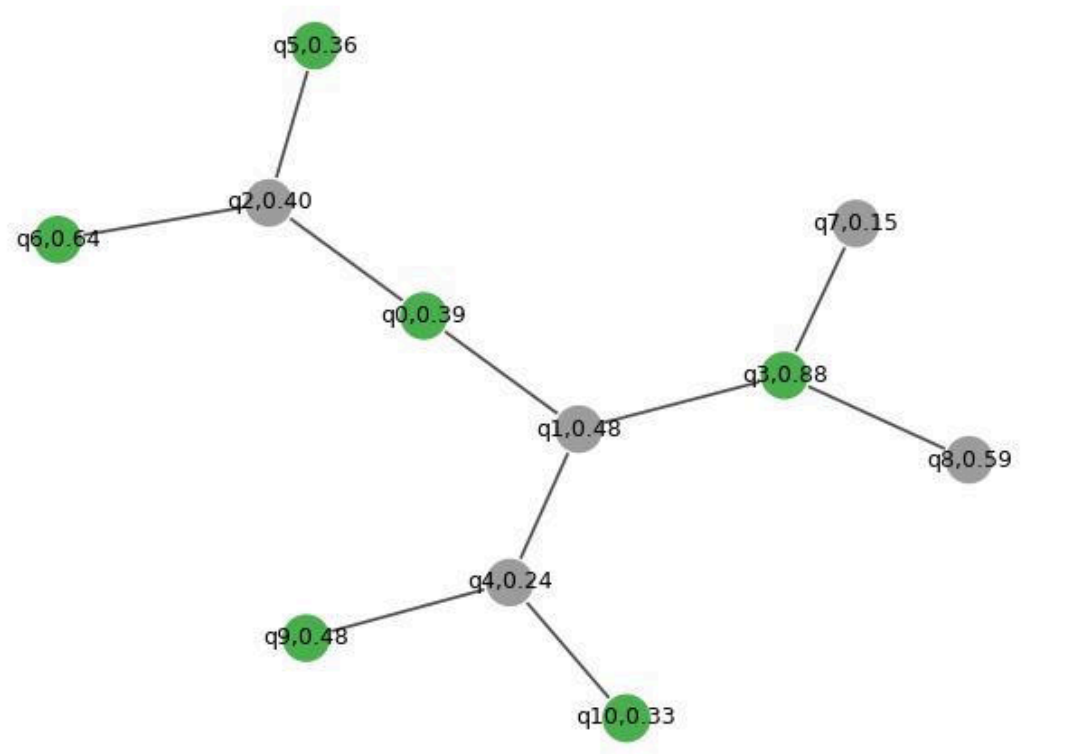

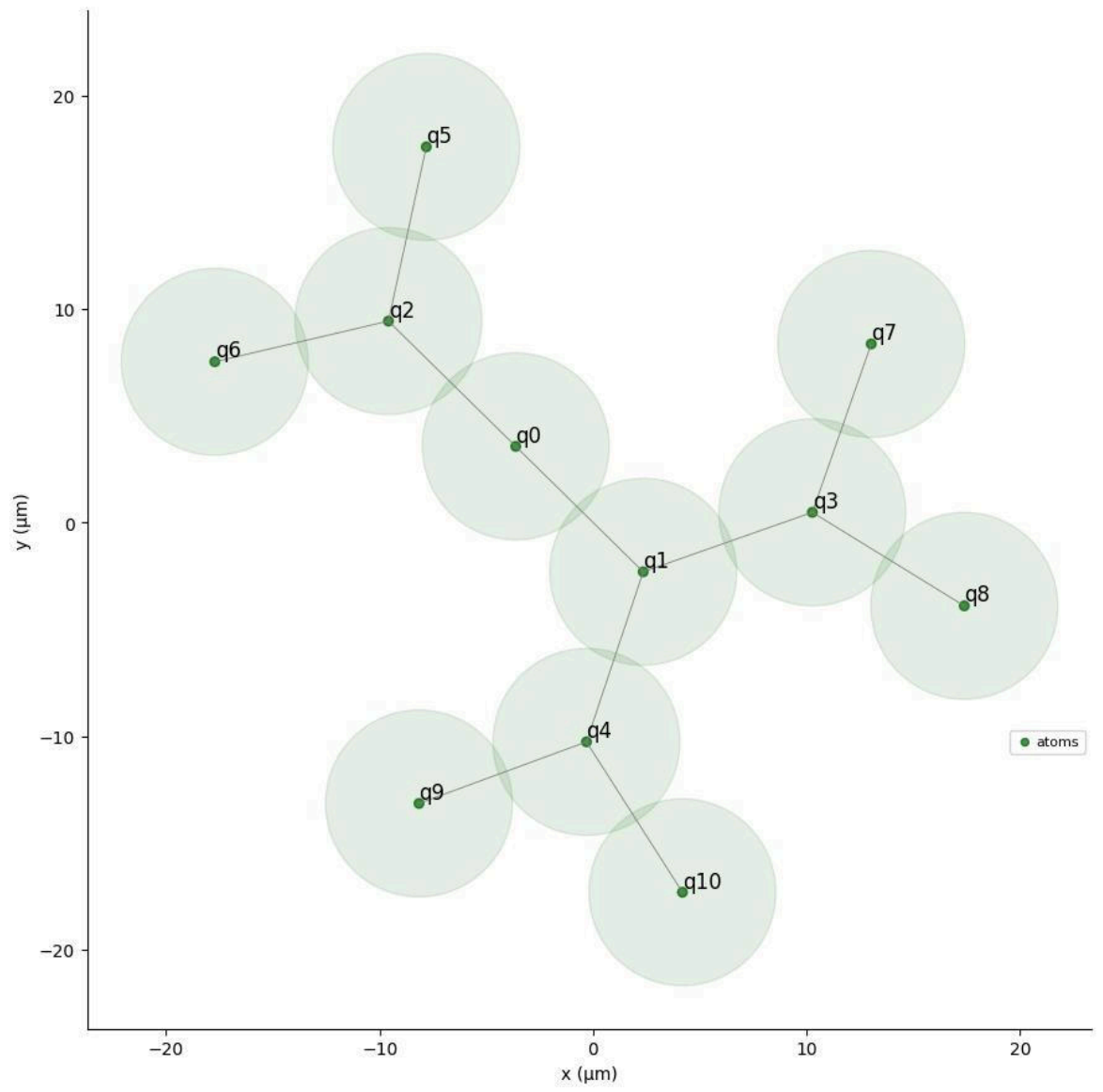

| Local Detuning | | DMM | |
|---|---|---|---|
| **Success probability** | **Optimality Ratio** | **Success probability** | **Optimality Ratio** |
| 0.079 | 0.9303 | 0.212 | 0.9431 |

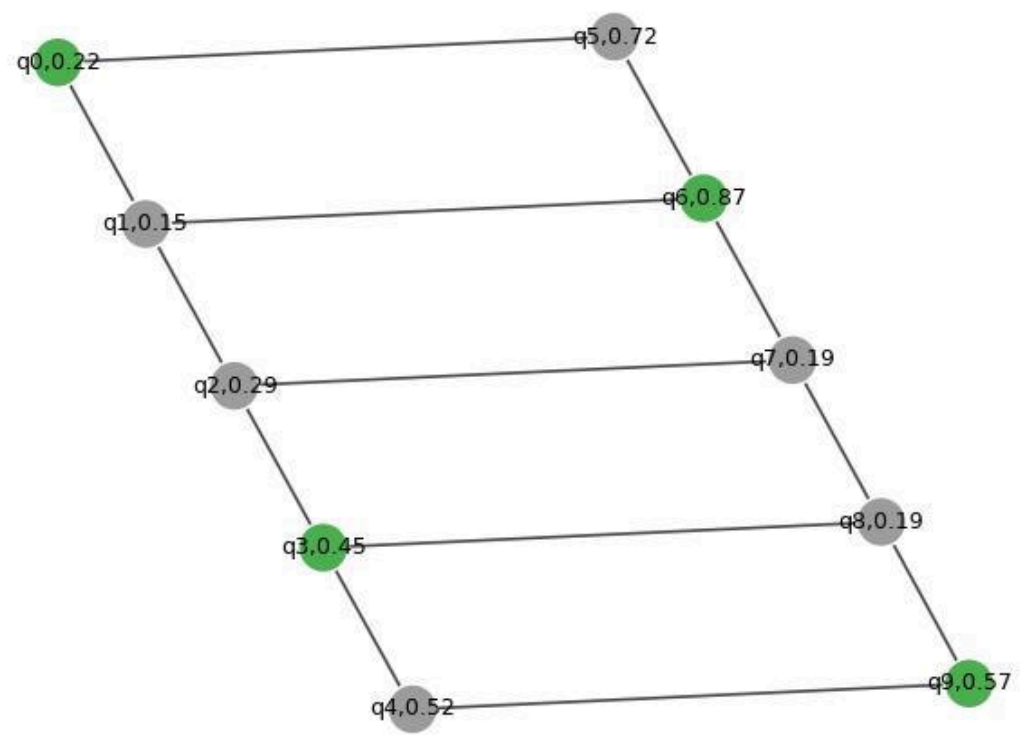

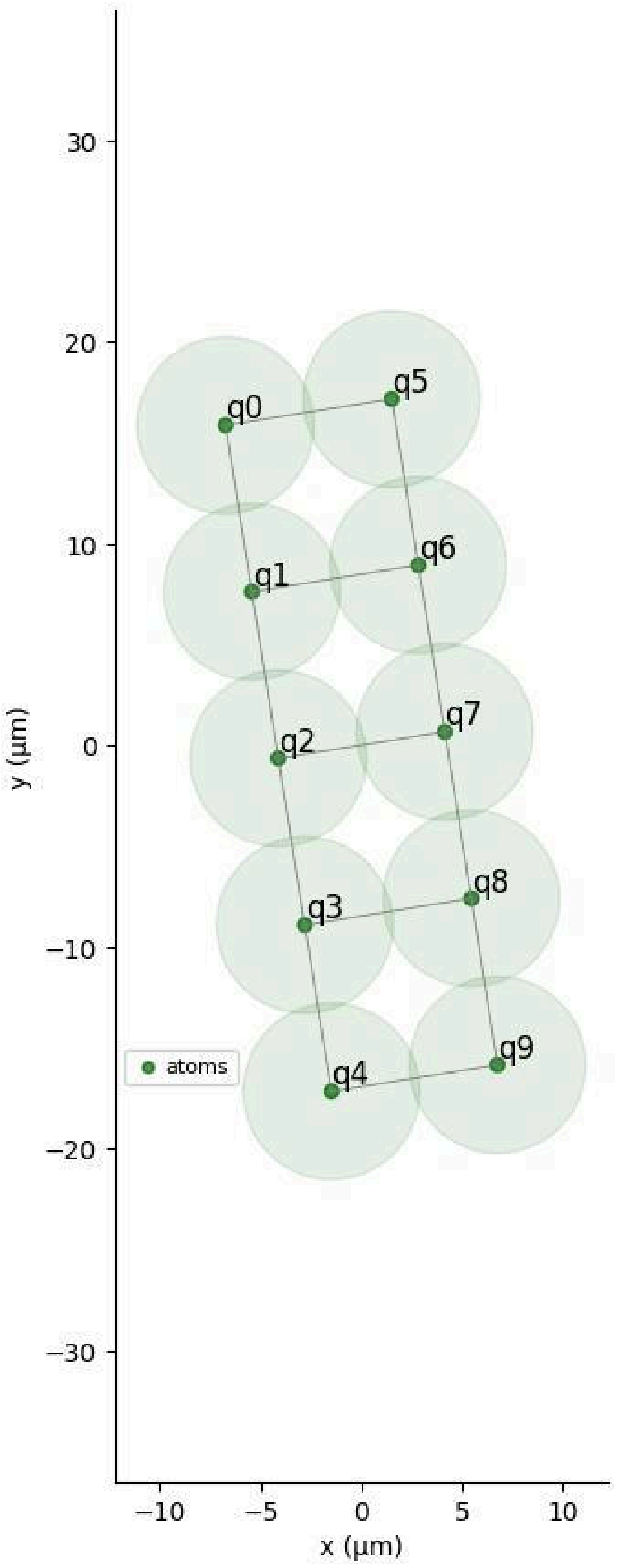

| Local Detuning | | DMM | |
|---|---|---|---|
| **Success probability** | **Optimality Ratio** | **Success probability** | **Optimality Ratio** |
| 0.401 | 0.9222 | 0.395 | 0.9199 |

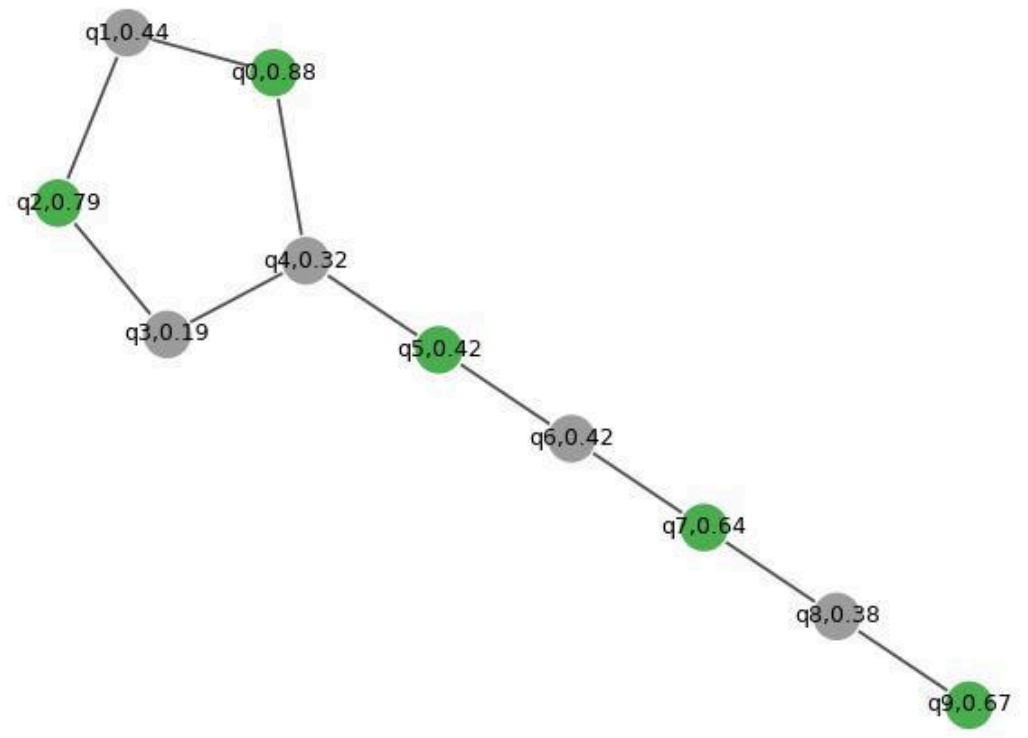

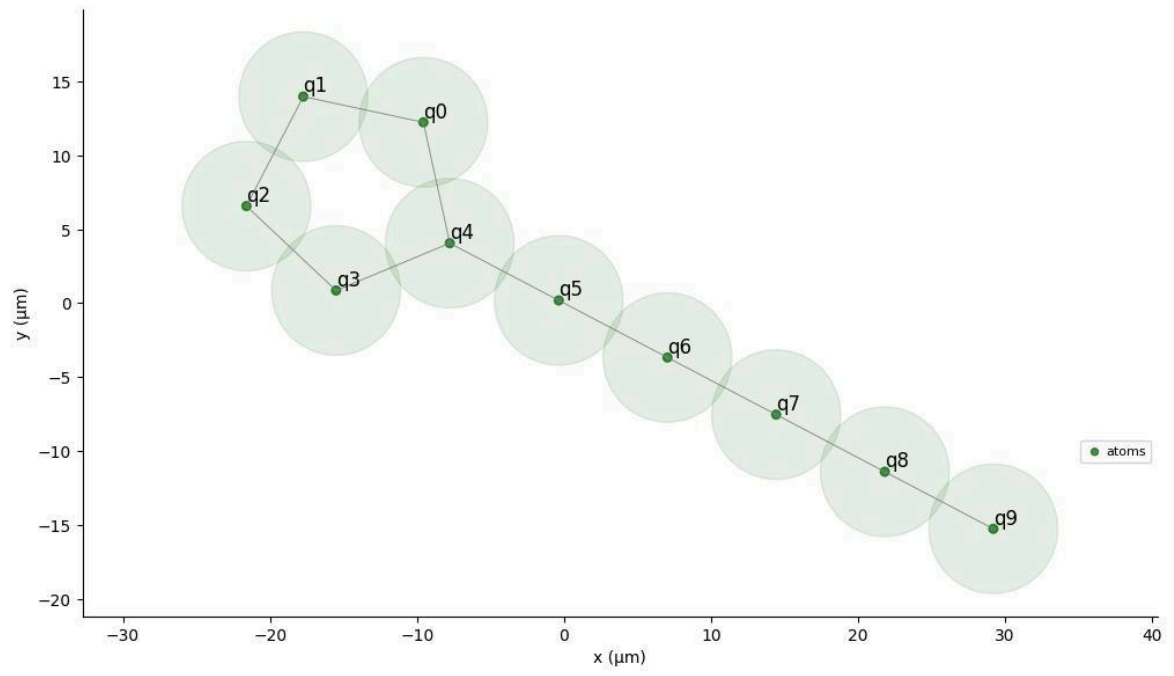

| Local Detuning | | DMM | |
|---|---|---|---|
| **Success probability** | **Optimality Ratio** | **Success probability** | **Optimality Ratio** |
| 0.909 | 0.9806 | 0.979 | 0.9961 |

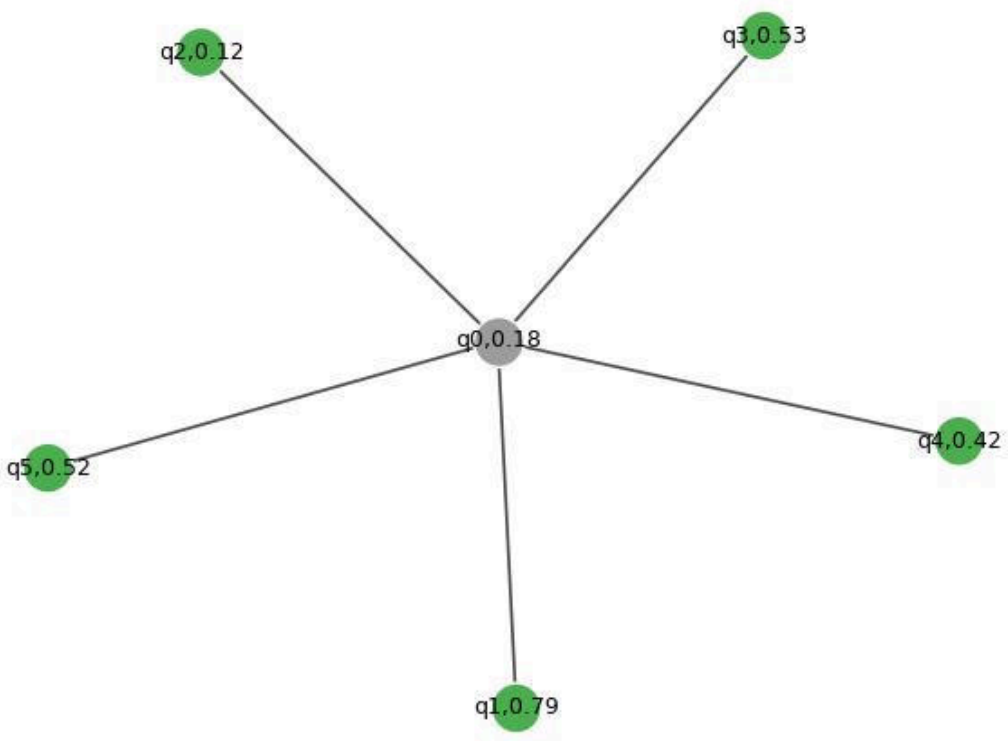

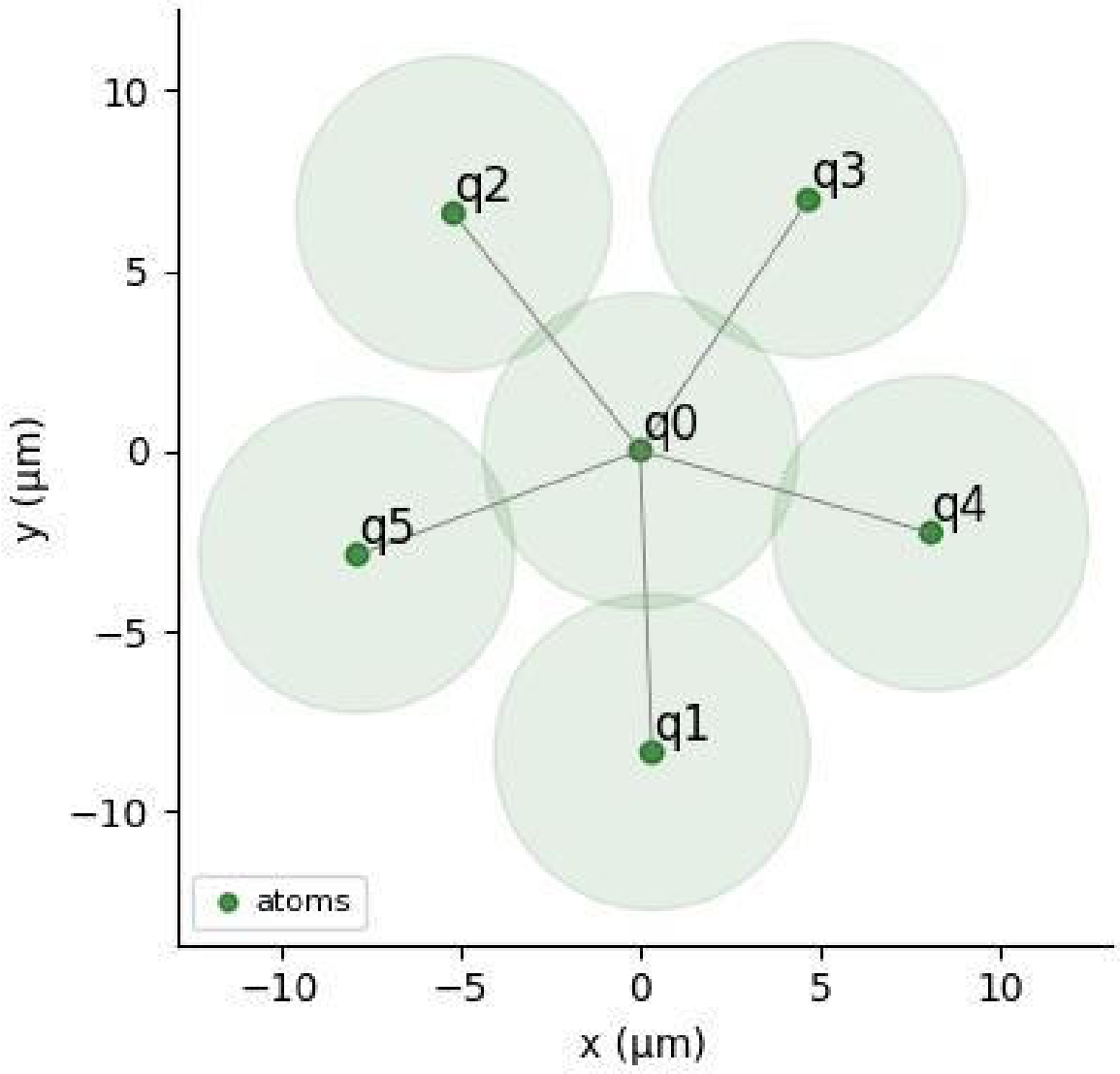

| Local Detuning | | DMM | |
|---|---|---|---|
| **Success probability** | **Optimality Ratio** | **Success probability** | **Optimality Ratio** |
| 0.556 | 0.9387 | 0.421 | 0.9585 |

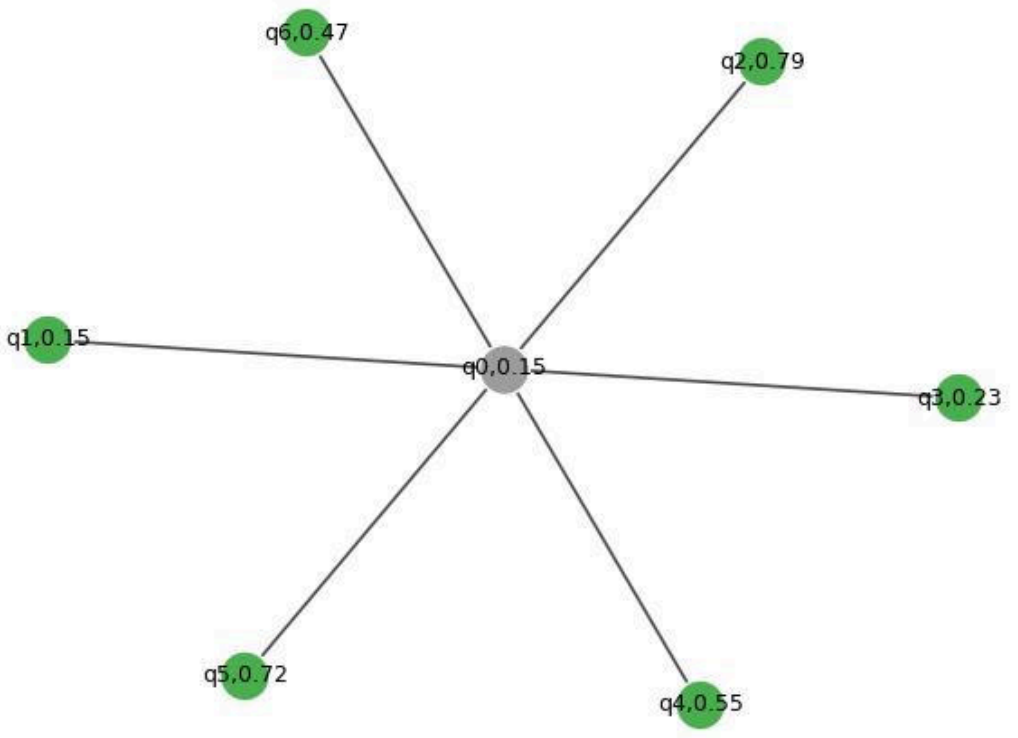

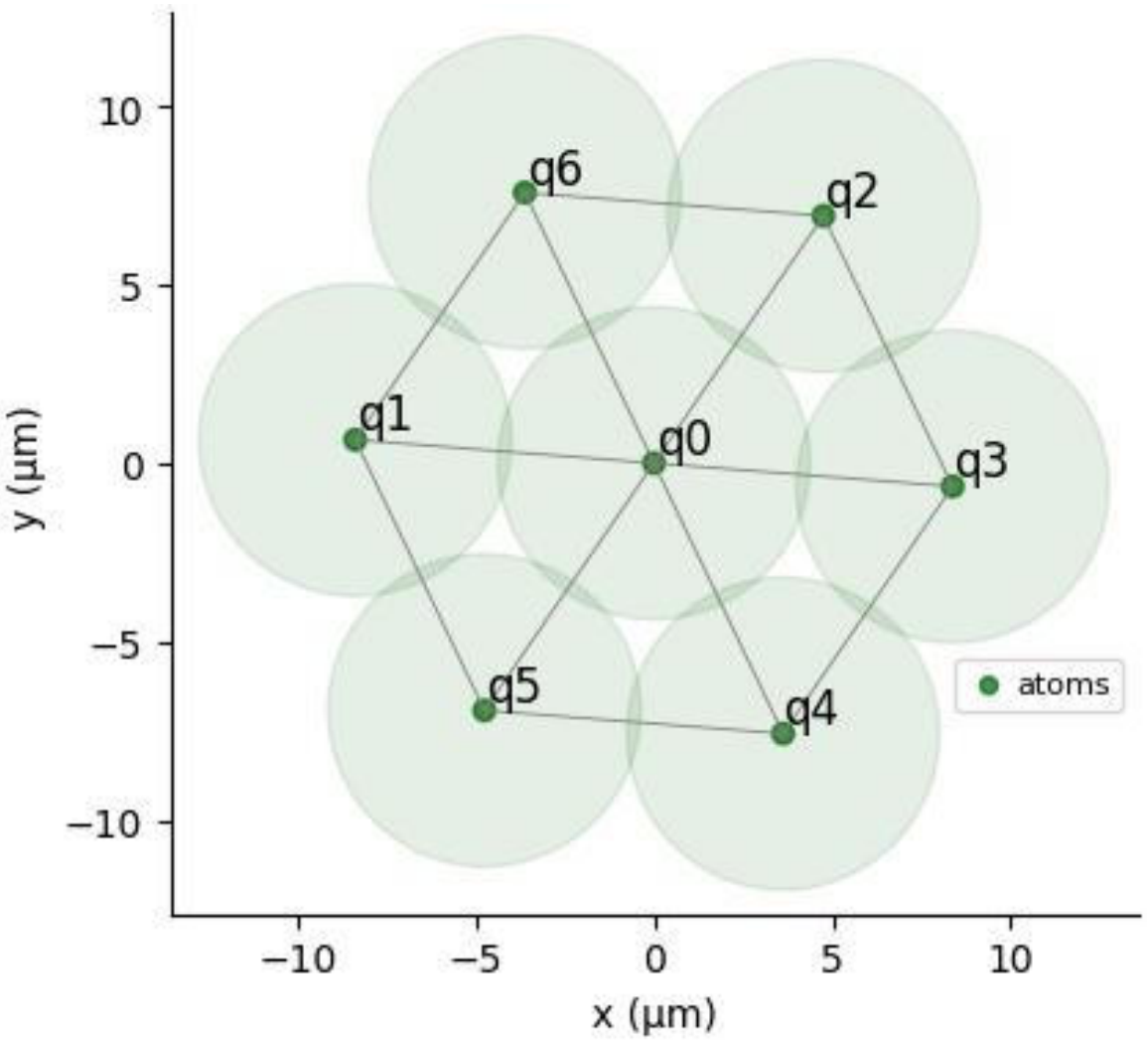

| Local Detuning | | DMM | |
|---|---|---|---|
| **Success probability** | **Optimality Ratio** | **Success probability** | **Optimality Ratio** |
| 0.112 | 0.8791 | 0.07 | 0.8954 |

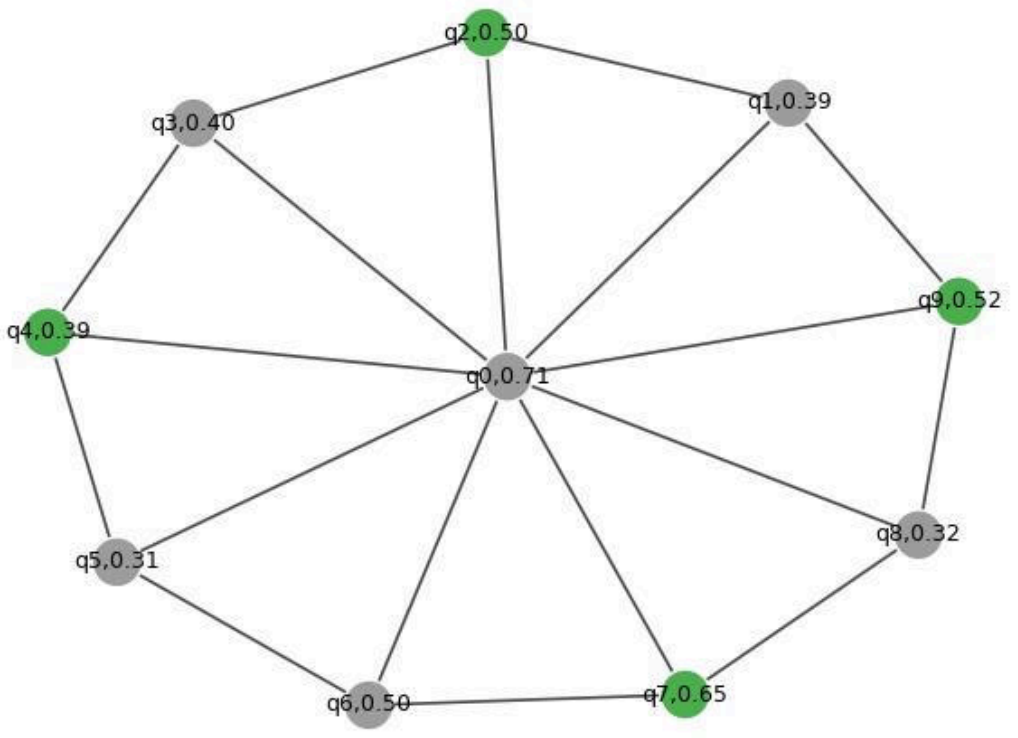

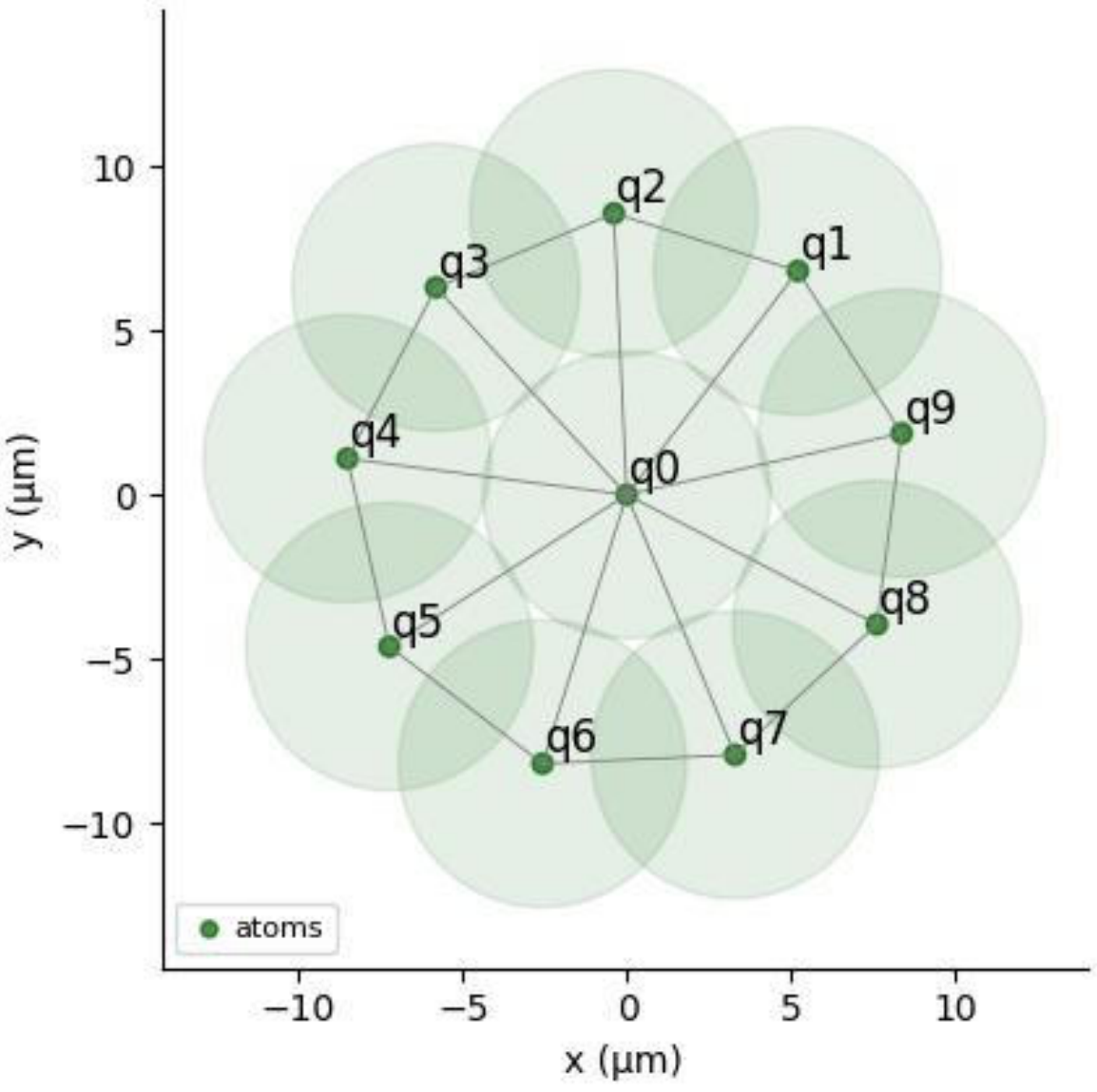

| Local Detuning | | DMM | |
|---|---|---|---|
| **Success probability** | **Optimality Ratio** | **Success probability** | **Optimality Ratio** |
| 0.354 | 0.9296 | 0.417 | 0.9415 |

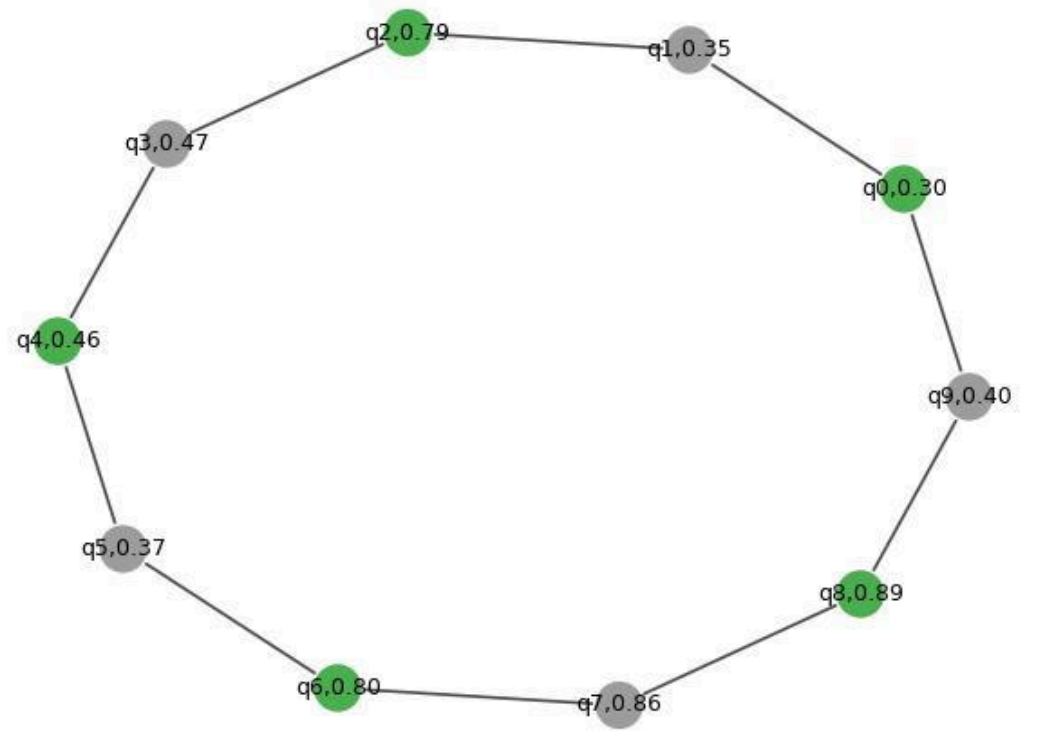

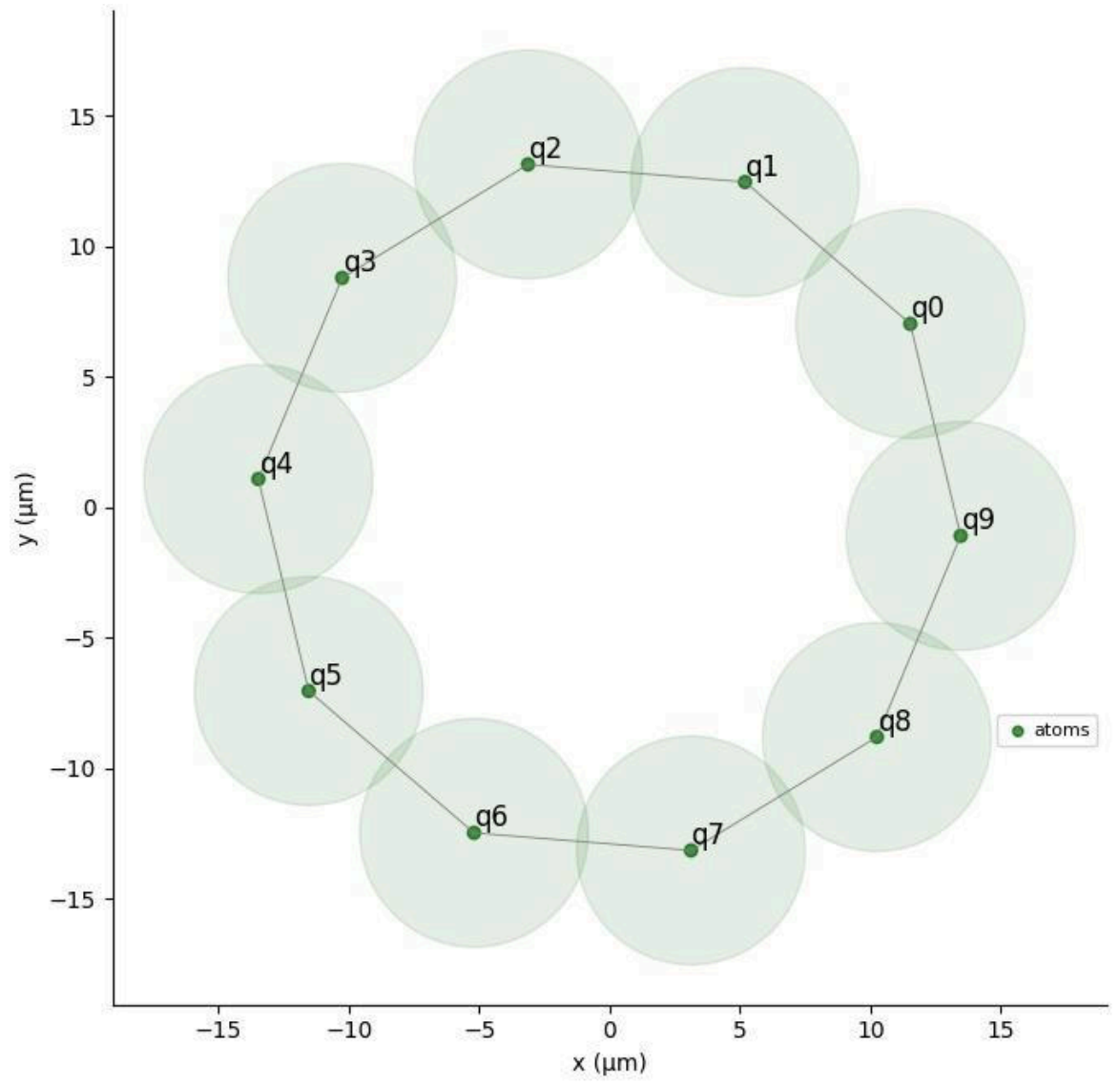

| Local Detuning | | DMM | |
|---|---|---|---|
| **Success probability** | **Optimality Ratio** | **Success probability** | **Optimality Ratio** |
| 0.776 | 0.9473 | 0.874 | 0.9791 |

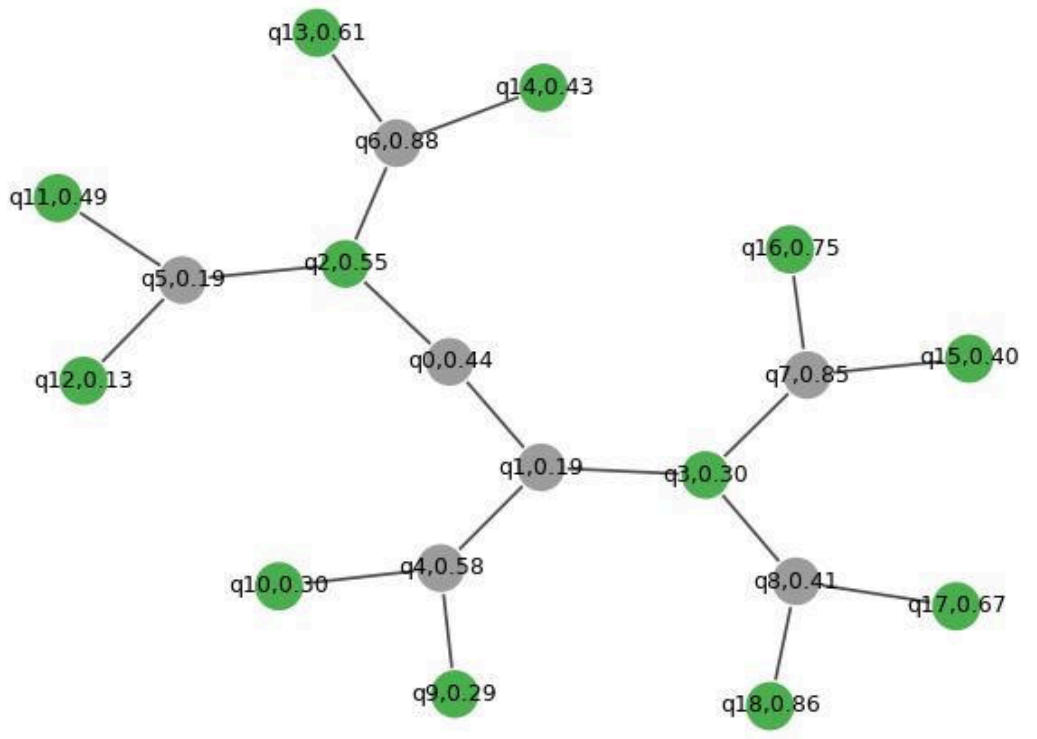

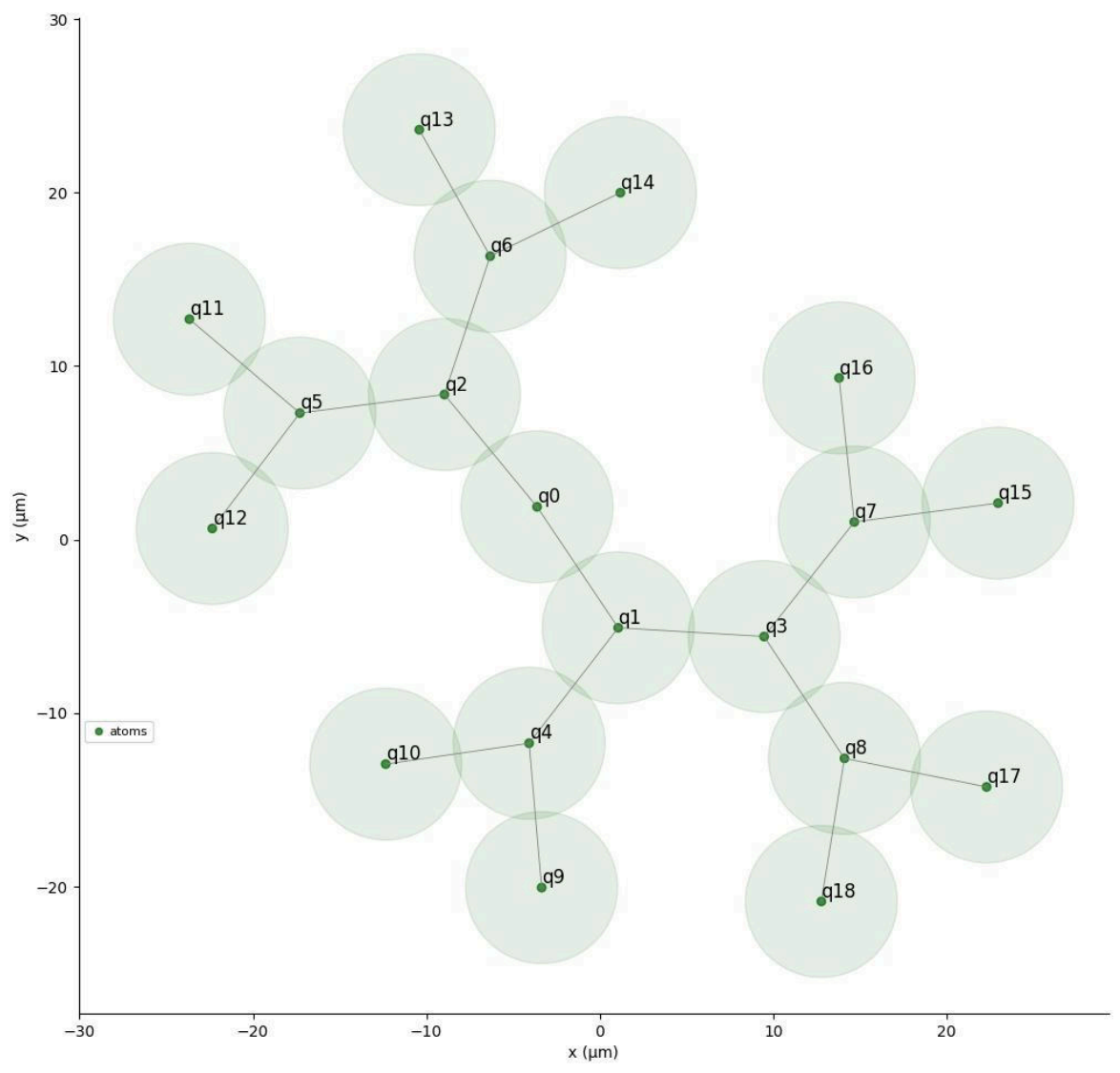

| Local Detuning | | DMM | |
|---|---|---|---|
| **Success probability** | **Optimality Ratio** | **Success probability** | **Optimality Ratio** |
| 0.244 | 0.9622 | 0.207 | 0.9633 |

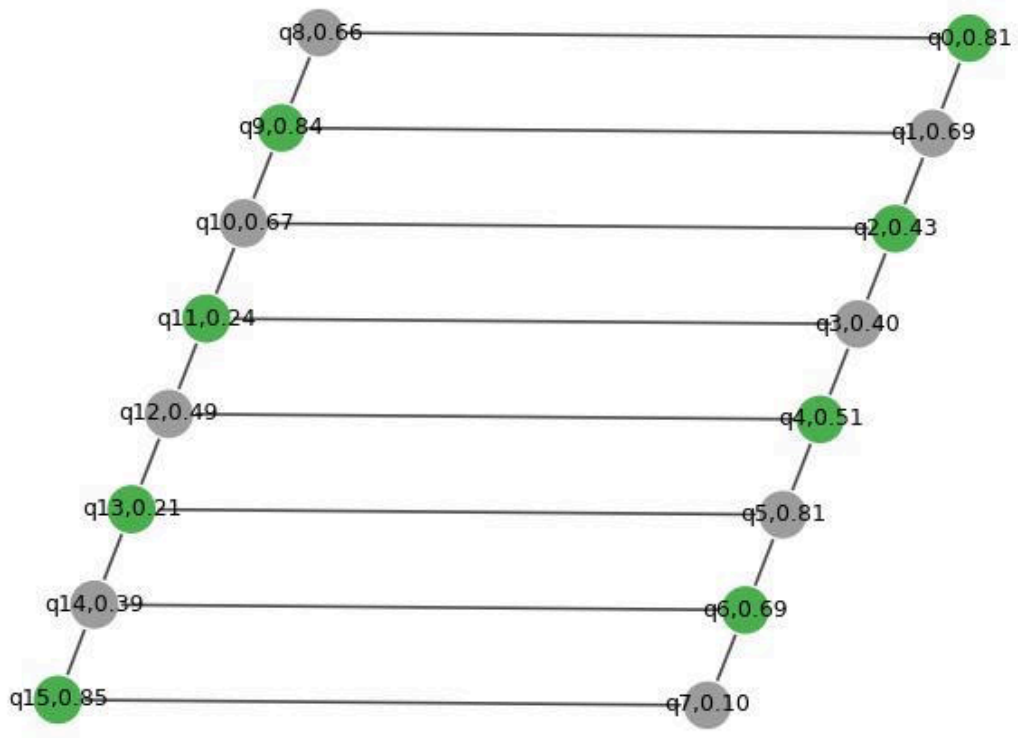

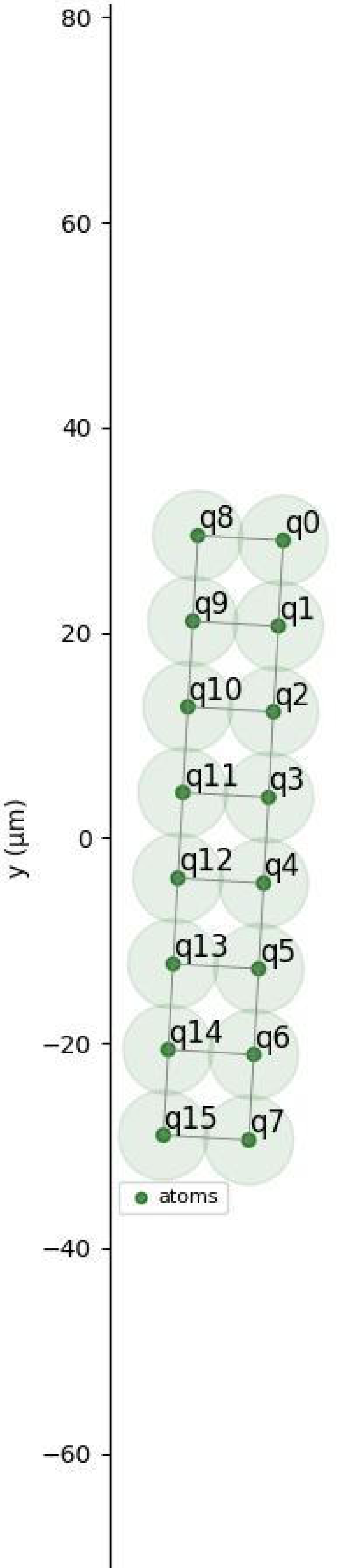

| Local Detuning | | DMM | |
|---|---|---|---|
| **Success probability** | **Optimality Ratio** | **Success probability** | **Optimality Ratio** |
| 0.418 | 0.9473 | 0.4 | 0.9449 |

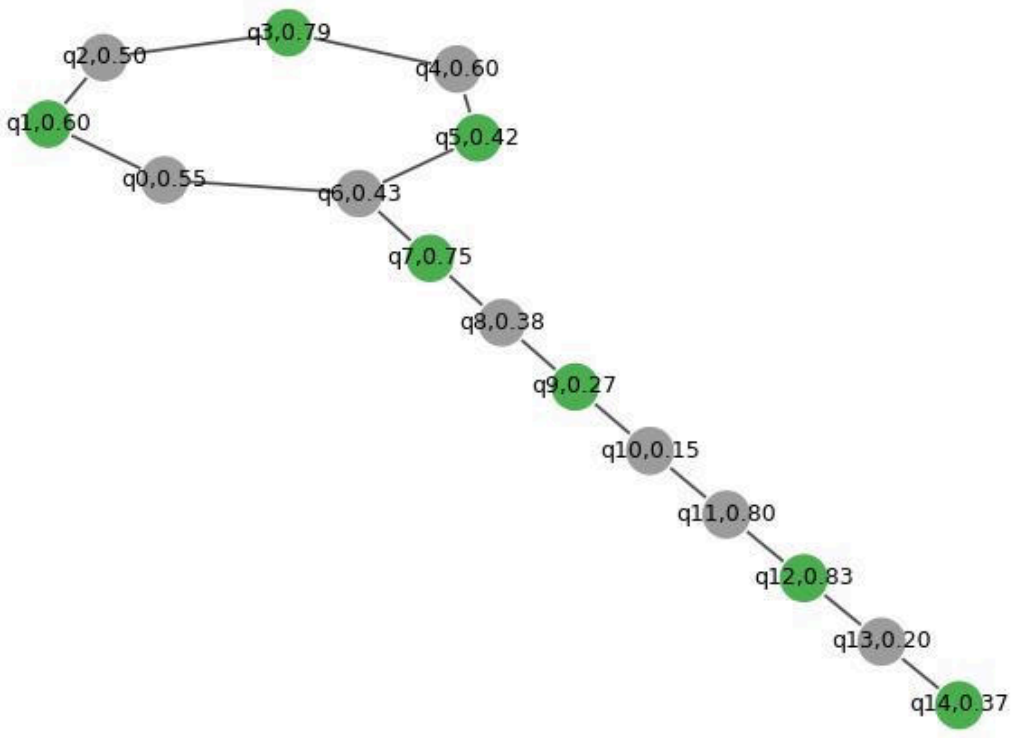

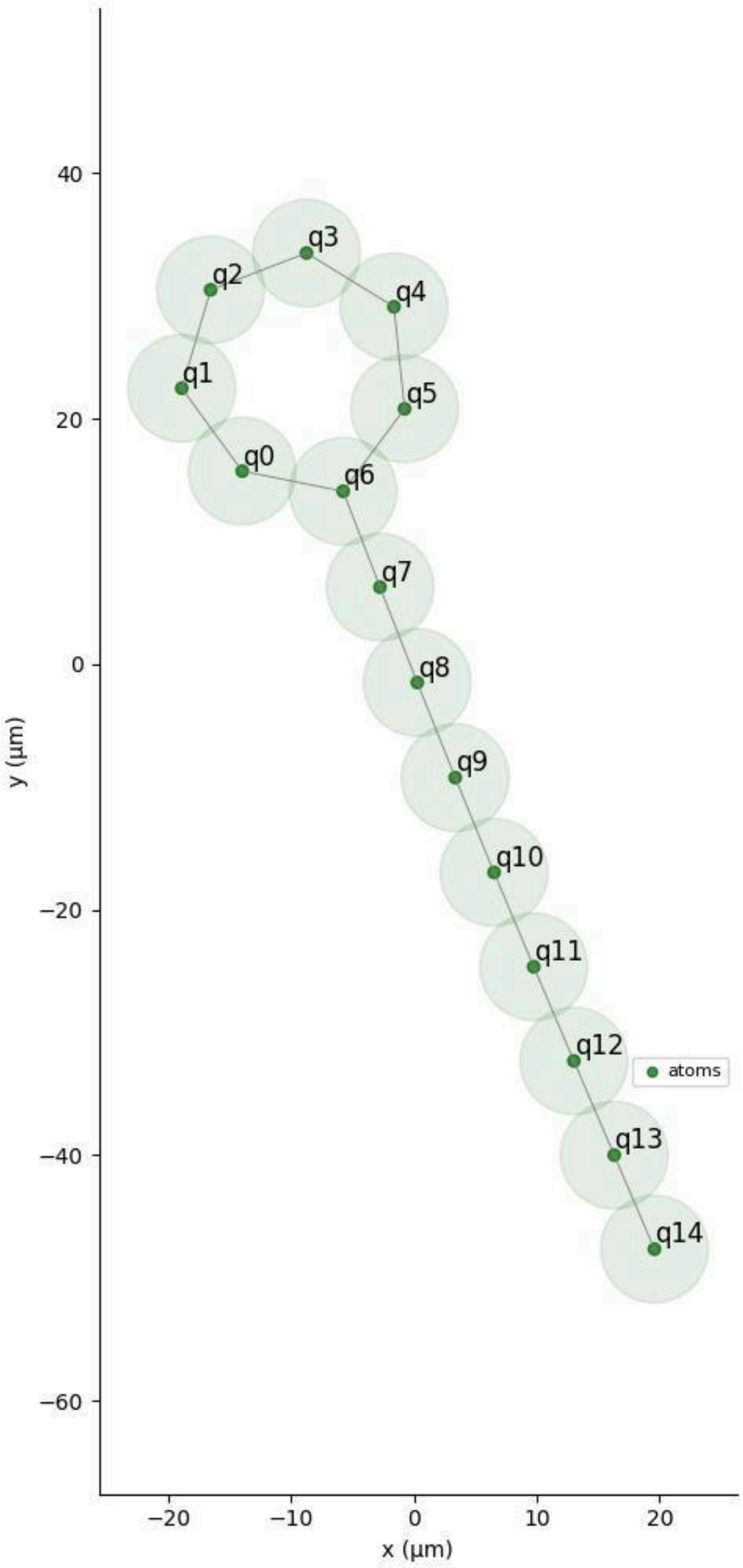

| Local Detuning | | DMM | |
|---|---|---|---|
| **Success probability** | **Optimality Ratio** | **Success probability** | **Optimality Ratio** |
| 0.333 | 0.937 | 0.471 | 0.96 |

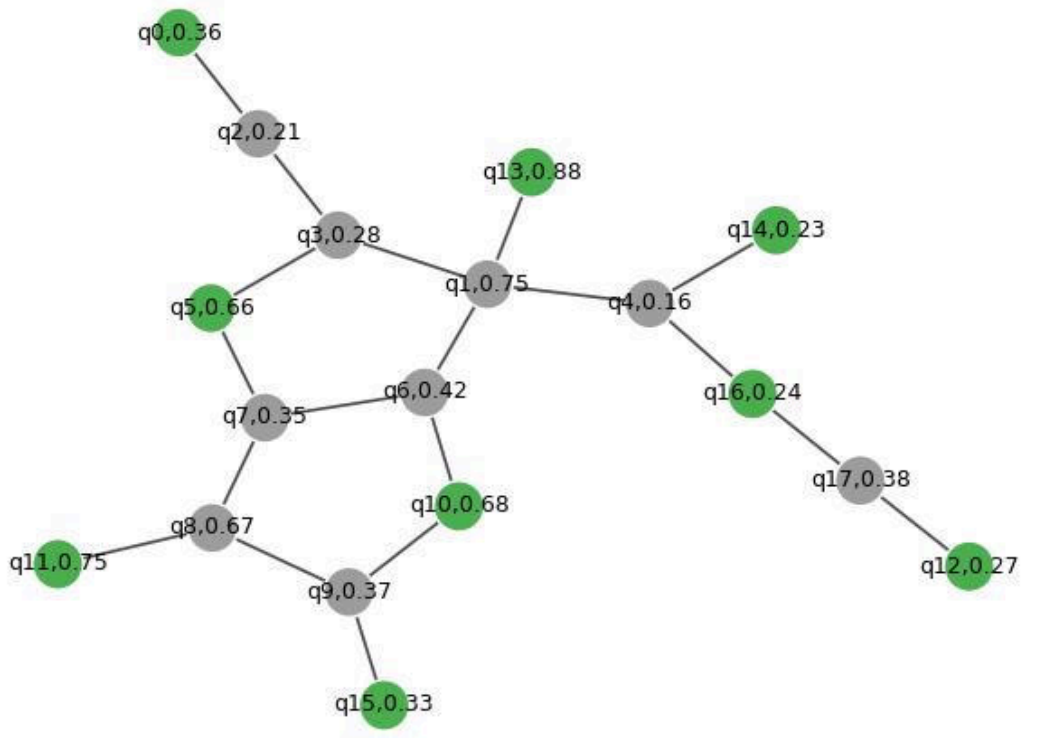

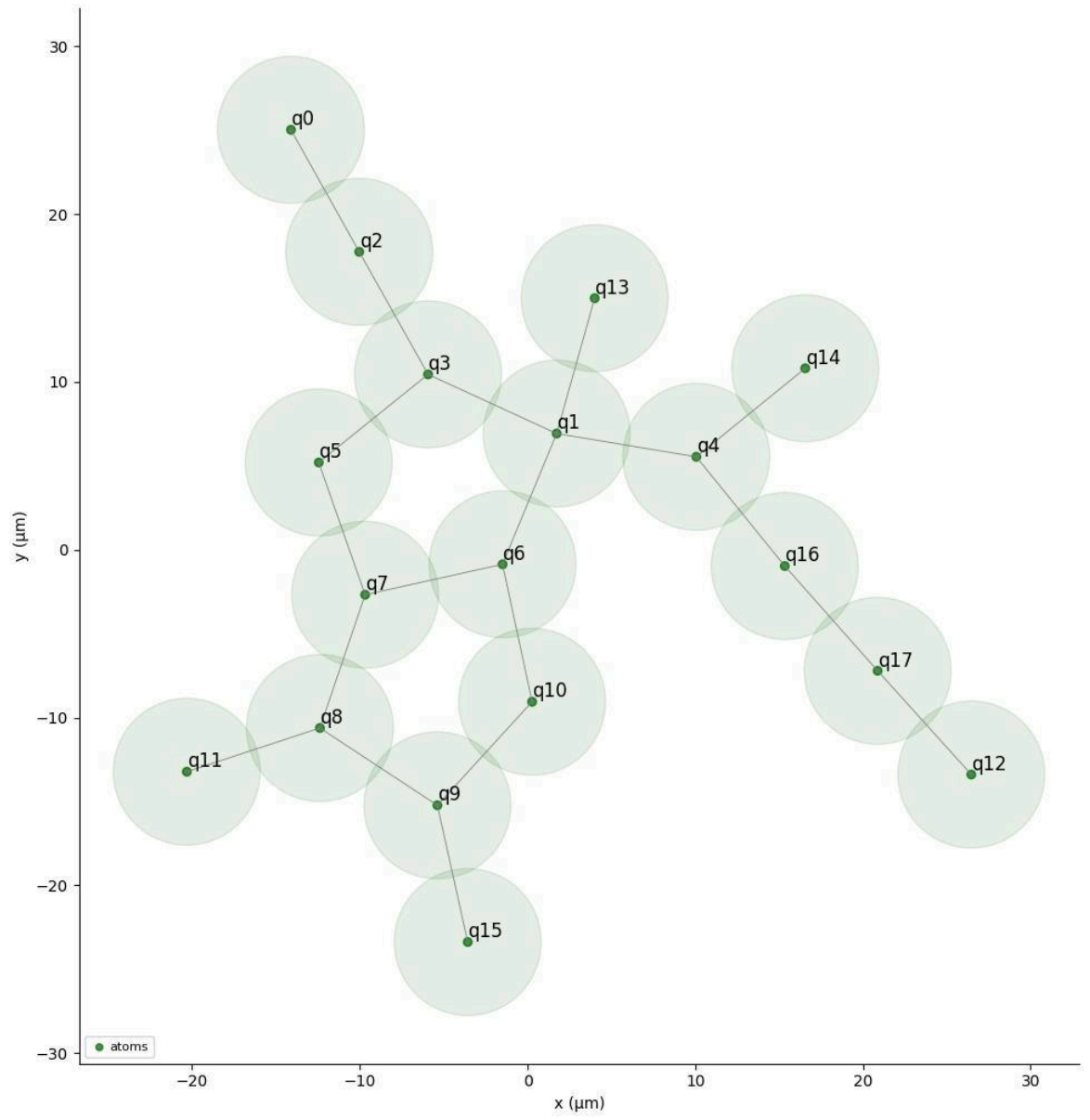

| Local Detuning | | DMM | |
|---|---|---|---|
| **Success probability** | **Optimality Ratio** | **Success probability** | **Optimality Ratio** |
| 0.431 | 0.9586 | 0.194 | 0.9353 |

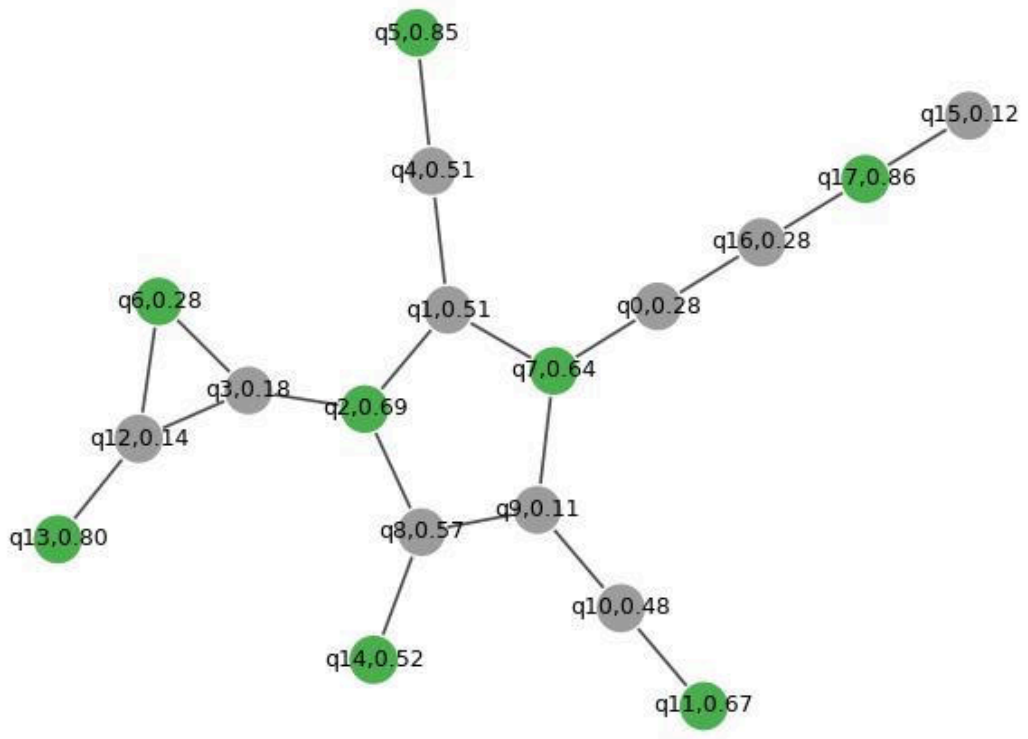

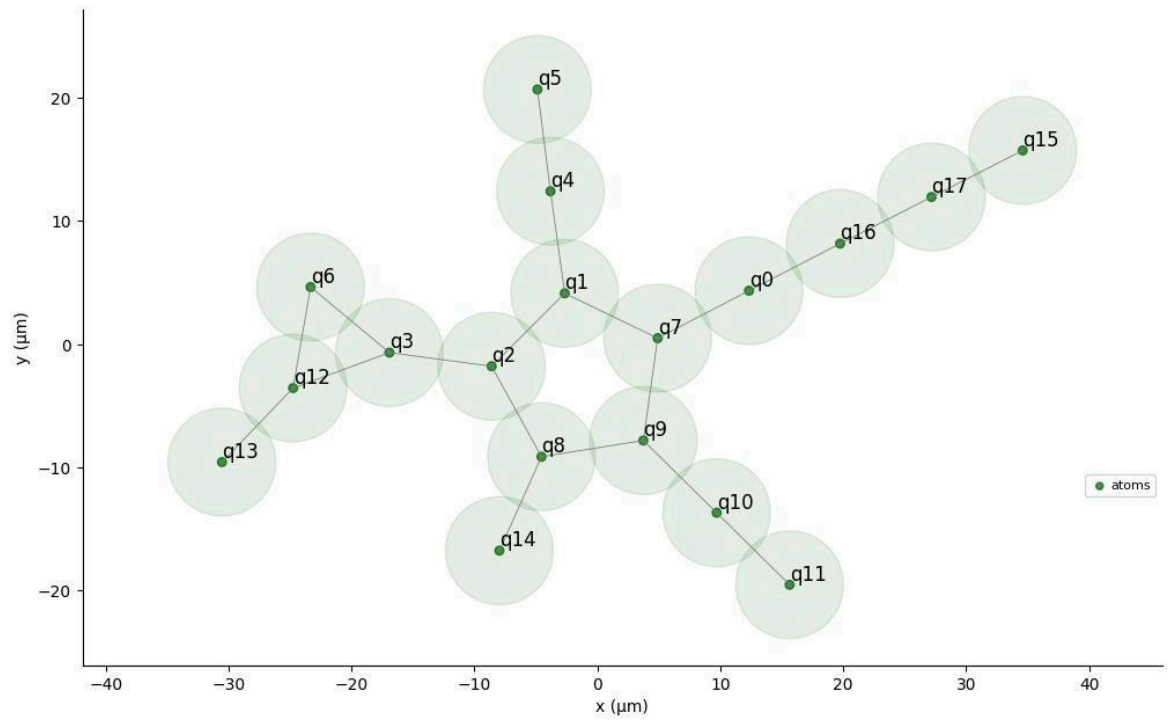

| Local Detuning | | DMM | |
|---|---|---|---|
| **Success probability** | **Optimality Ratio** | **Success probability** | **Optimality Ratio** |
| 0.478 | 0.9459 | 0.642 | 0.9758 |

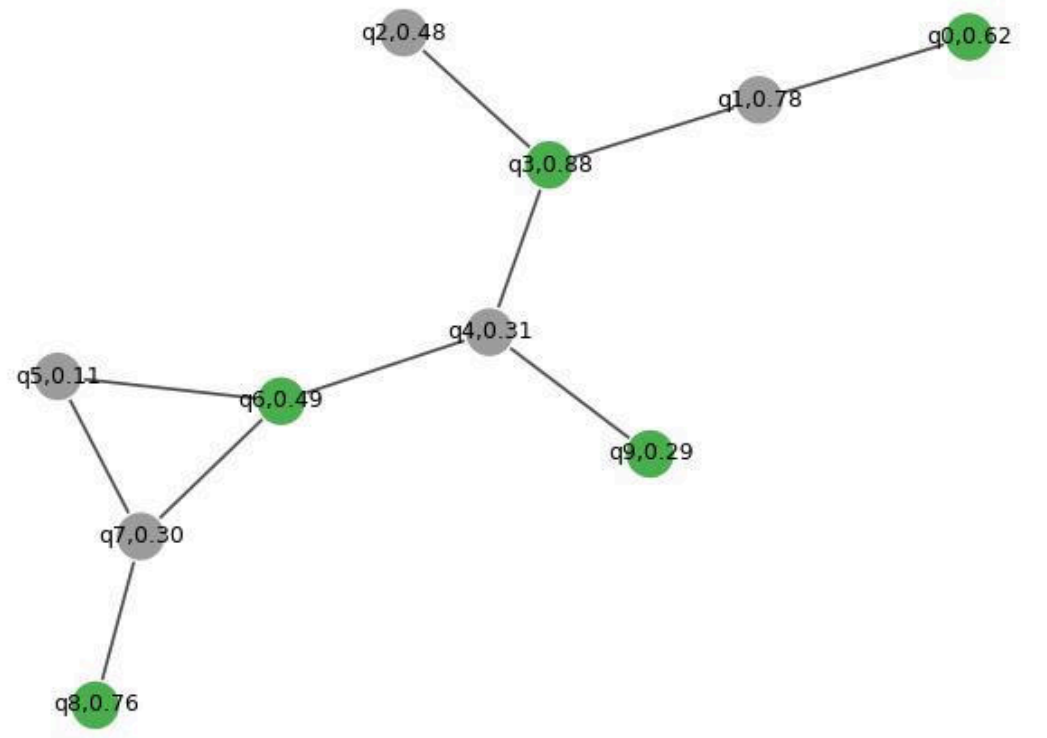

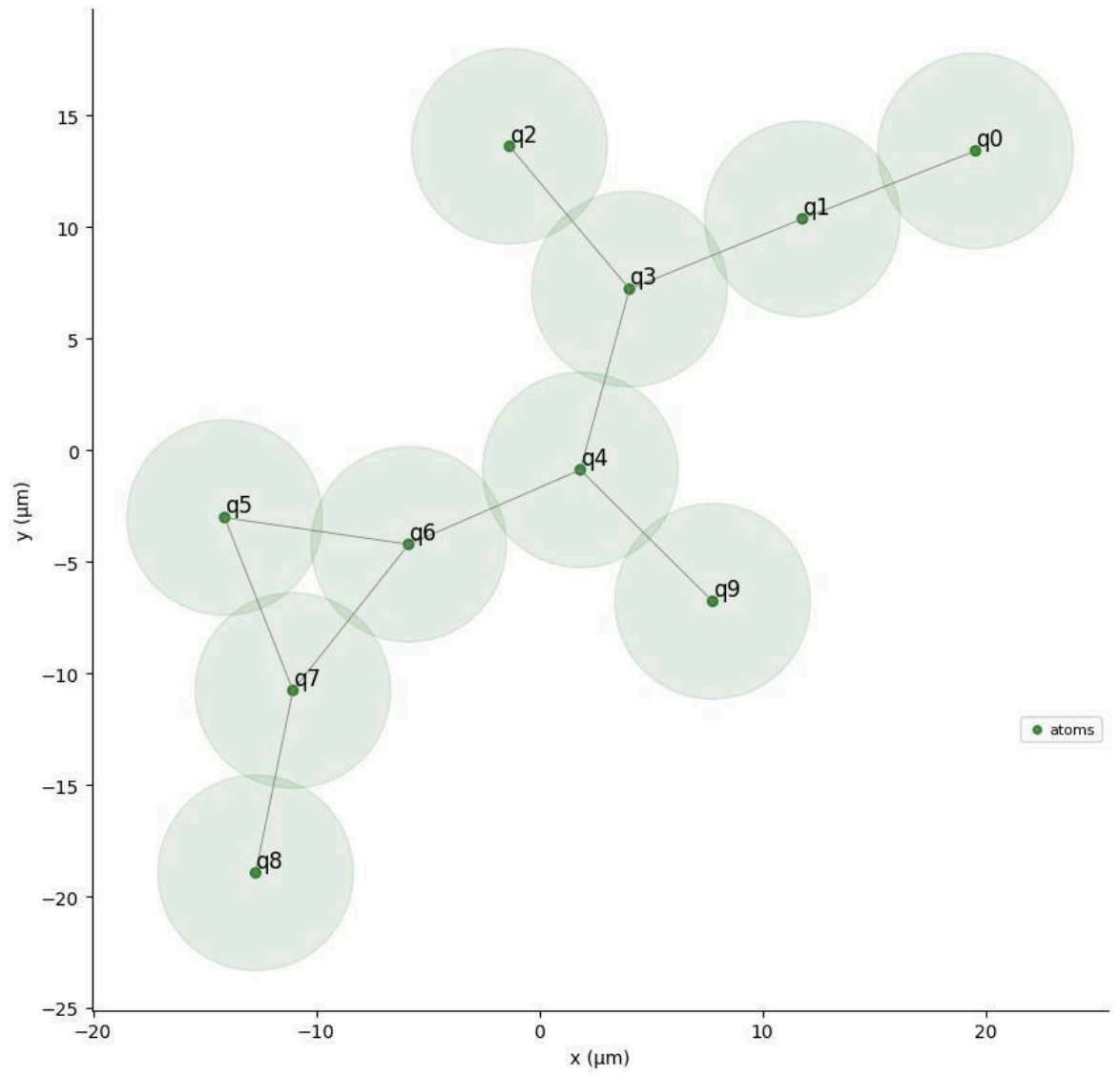

| Local Detuning | | DMM | |
|---|---|---|---|
| **Success probability** | **Optimality Ratio** | **Success probability** | **Optimality Ratio** |
| 0.332 | 0.8852 | 0.496 | 0.9416 |

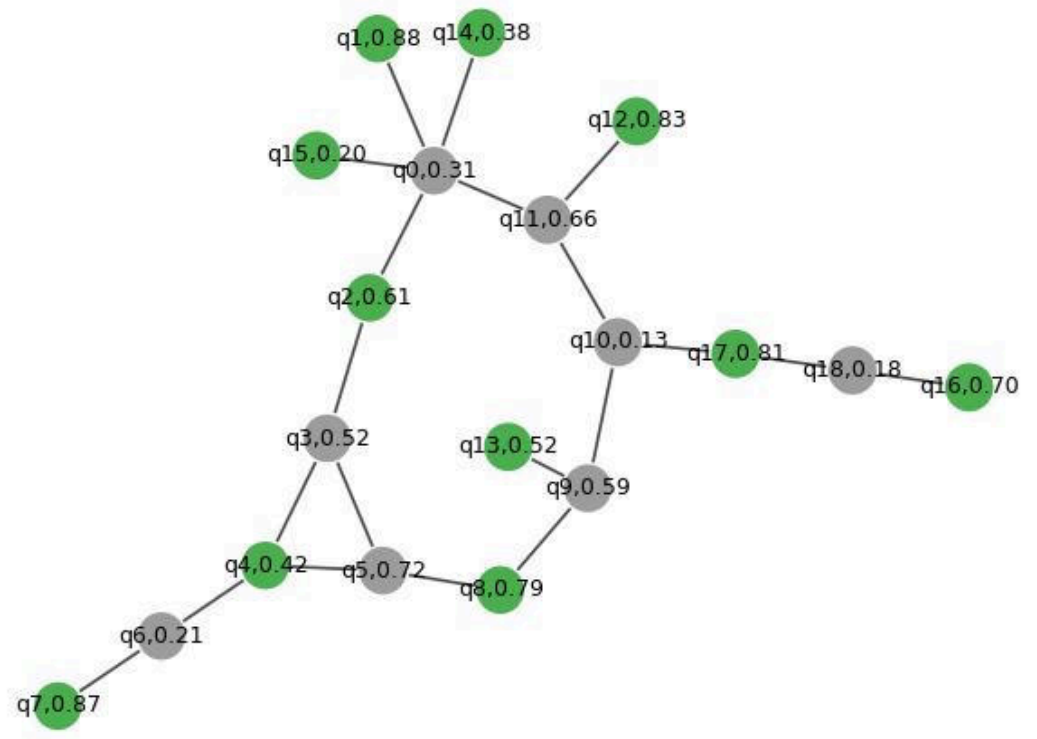

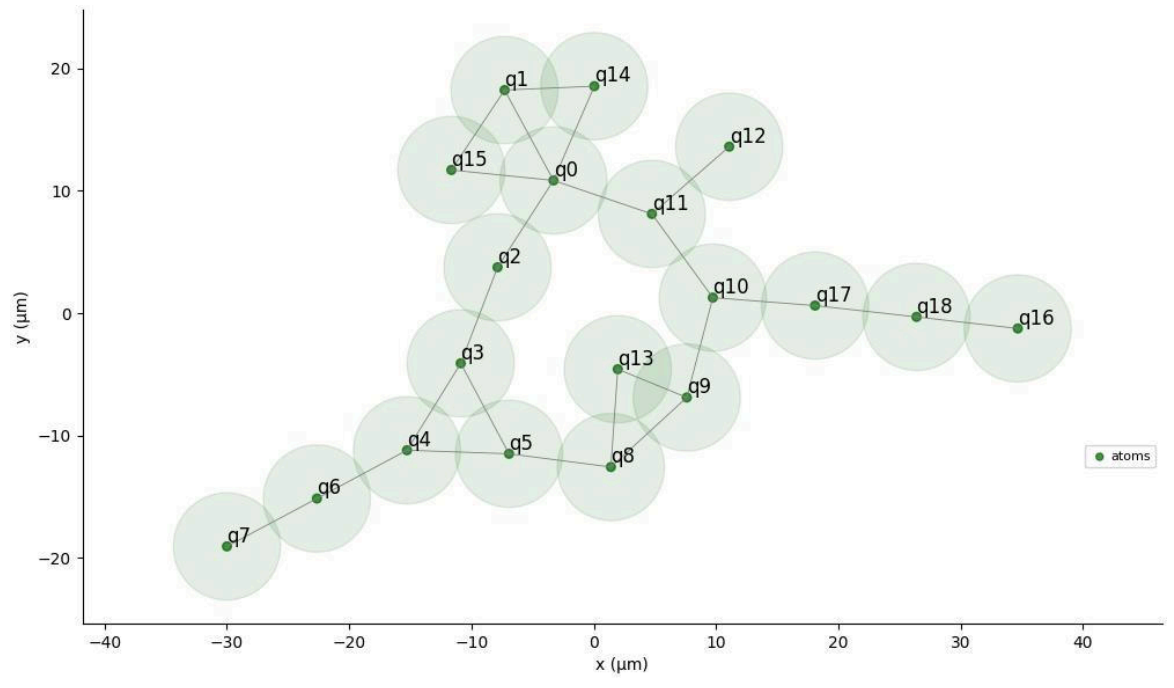

| Local Detuning | | DMM | |
|---|---|---|---|
| **Success probability** | **Optimality Ratio** | **Success probability** | **Optimality Ratio** |
| 0.139 | 0.9335 | 0.003 | 0.8406 |

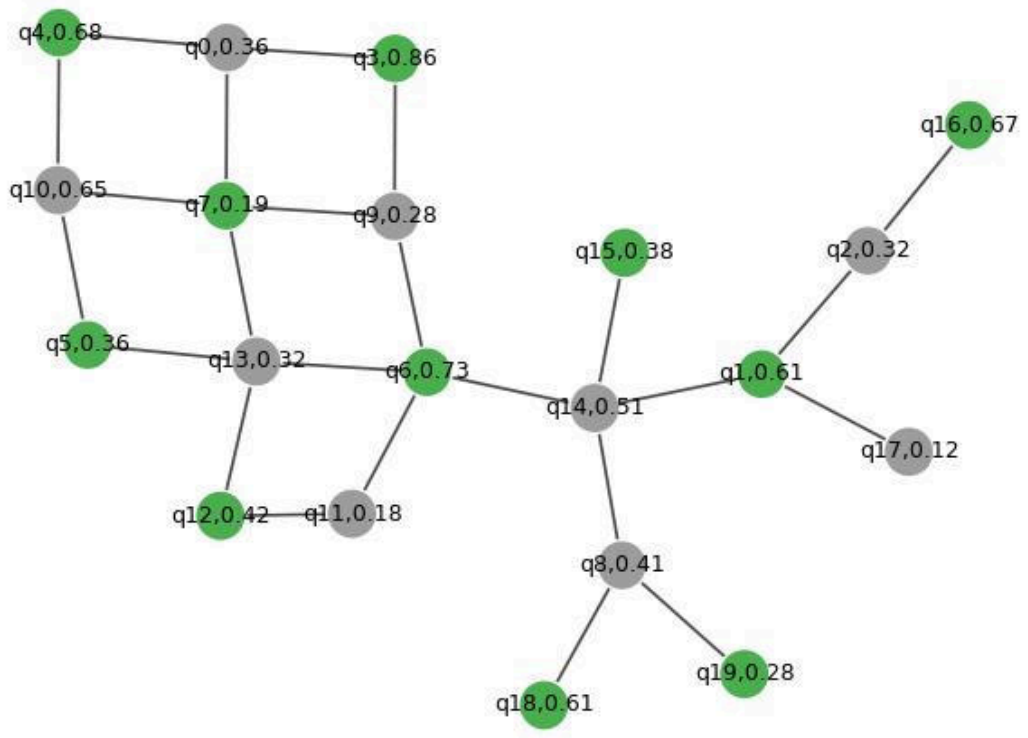

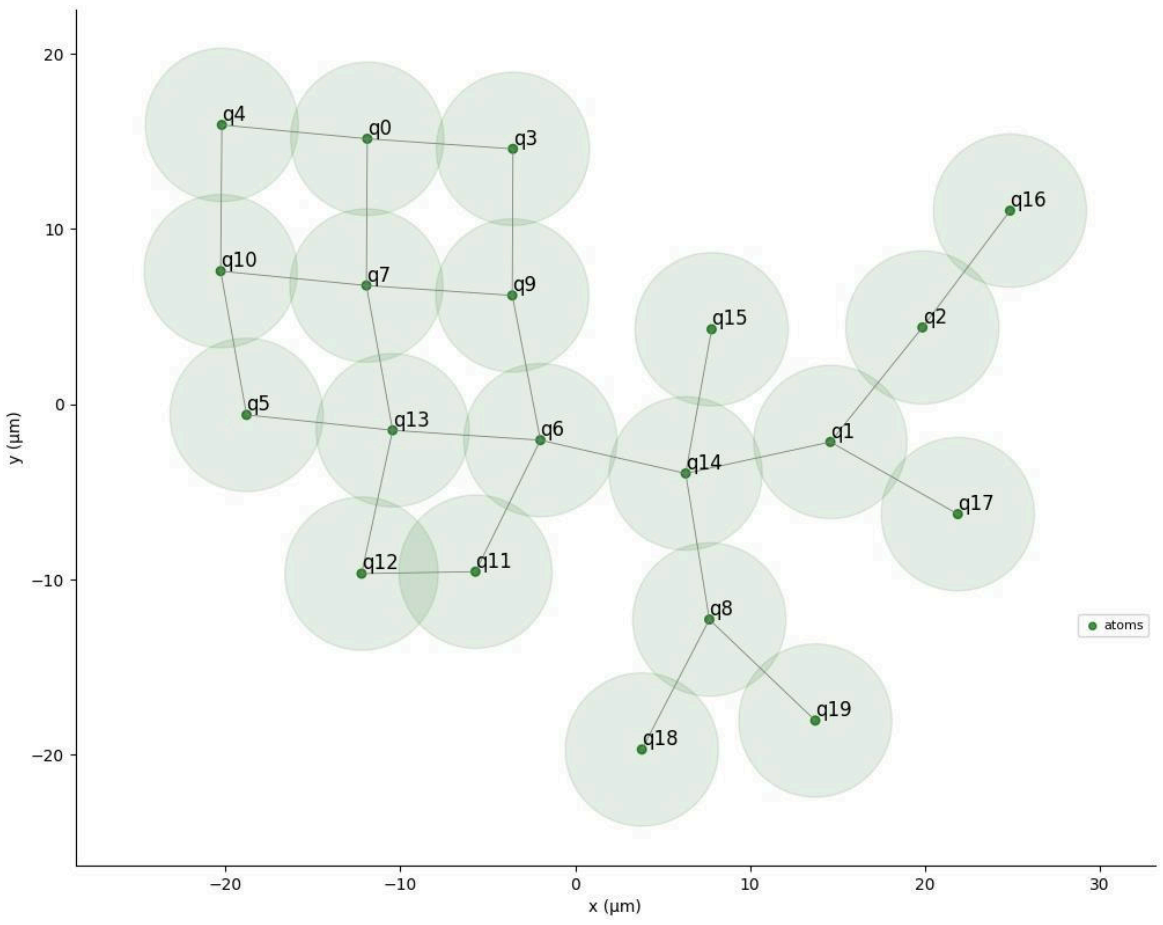

| Local Detuning | | DMM | |
|---|---|---|---|
| **Success probability** | **Optimality Ratio** | **Success probability** | **Optimality Ratio** |
| 0.392 | 0.9352 | 0.326 | 0.9276 |